\documentclass[sigconf, nonacm]{acmart}
\usepackage{enumitem}
\usepackage{algorithm}
\usepackage[noend]{algpseudocode}
\usepackage{subfig}
\usepackage{diagbox}
\usepackage{makecell}

\usepackage{dsfont}
\usepackage{relsize}
\usepackage{amsthm}
\usepackage{tabularx}
\usepackage{array} 

\theoremstyle{definition}
\newtheorem{definition}{Definition}[section]

\begin{document}
\title{A Survey on Efficient Processing of Similarity Queries over Neural Embeddings}

\author{Yifan Wang}
\affiliation{%
  \institution{University of Florida}
}
\email{wangyifan@ufl.edu}

\begin{abstract}
Similarity query is the family of queries based on some similarity metrics, like top-k queries where the results are ranked by their similarity to the queried object and similarity join queries where records from two datasets are joined by their similarities between each other instead of by specific keys. Unlike the traditional database queries which are mostly based on value equality, similarity queries aim to find targets "similar enough to" the given data objects, depending on some similarity metric, e.g., Euclidean distance, cosine similarity and so on. Similarity queries are applied widely in data integration (e.g., entity resolution and schema matching), information retrieval, question answering and many other areas of data science.

To measure the similarity between data objects, traditional methods normally work on low level or syntax features(e.g., basic visual features on images or bag-of-word features of text), which makes them weak to compute the semantic similarities between objects. So for measuring data similarities semantically, neural embedding is applied. Embedding techniques work by representing the raw data objects as vectors (so called "embeddings" or "neural embeddings" since they are mostly generated by neural network models) that expose the hidden semantics of the raw data, based on which embeddings do show outstanding effectiveness on capturing data similarities, making it one of the most widely used and studied techniques in the state-of-the-art similarity query processing research. But there are still many open challenges on the efficiency of embedding based similarity query processing, which are not so well-studied as the effectiveness.

In this survey, we first provide an overview of the “similarity query” and “similarity query processing” problems. Then we talk about recent approaches on designing the indexes and operators for highly efficient similarity query processing on top of embeddings (or more generally, high dimensional data). Finally, we investigate the specific solutions with and without using embeddings in selected application domains of similarity queries, including entity resolution and information retrieval. By comparing the solutions, we show how neural embeddings benefit those applications.
\end{abstract}

\maketitle

\section{Introduction}
Similarity queries are the queries based on approximate semantics of data instead of exact semantics of data~\cite{general-similarity-query-processing-1}. Or it can be defined as the queries involving similarity operators~\cite{general-similarity-query-processing-1, general-similarity-query-processing-2}, including similarity selection (e.g., k nearest neighbor search), similarity join, similarity groupby, etc. In short, any query looking for answers based on similarity between records instead of exact value match is a similarity query. Similarity queries are widely used on various types of data, e.g., multimedia data(video, audio, images)~\cite{general-similarity-query-multimedia-1, general-similarity-query-multimedia-2, general-similarity-query-multimedia-3, general-similarity-query-multimedia-4, general-similarity-query-multimedia-5}, text~\cite{general-similarity-query-text-1, general-similarity-query-text-2, general-similarity-query-text-3, general-similarity-query-text-4}, graph/network~\cite{general-similarity-query-graph-1, general-similarity-query-graph-2, general-similarity-query-graph-3}, etc. And accurately measuring the similarities between data objects is the key to processing those similarity queries. Traditional methods normally use low level features to determine whether objects are similar. For instance, in image processing area, visual features (like color, texture and shape) are widely used to measure image similarity~\cite{traditional-method-image-color-feature-1, traditional-method-image-color-feature-2, traditional-method-image-color-feature-3}; in natural language processing area, bag of words is a commonly used representation of the text document which represents the document as a multiset of its words. These features or representations are considered as "low level" since they do not extract the higher level semantic information from the raw data, which is the biggest problem of them. When humans are talking about "similar data objects", we want the two objects to have similar semantic content (e.g., images showing "a cat is running"). But such information is hard to learn by using low level features. Therefore as a powerful tool for capturing semantic similarities of data, embeddings are widely used in the state-of-the-art similarity query processing methods. In this survey we explore the exploitation of neural embeddings in three major categories of similarity queries: similarity search, similarity join and similarity group-by/grouping.

In general, an embedding is a multi-dimensional vector representation of the raw data object, where the vector exposes the semantic information of the raw data and the semantic similarity between raw data objects can be measured by computing some similarity/distance metric over the corresponding vectors. In recent years, embedding techniques are rapidly developing and showing significantly high performance on computing semantic similarities in multiple domains, including text~\cite{embedding-text-1, embedding-text-2, embedding-text-3}, image~\cite{embedding-image-1, embedding-image-2}, graph~\cite{embedding-graph-1, embedding-graph-2, embedding-graph-3, embedding-graph-4}, etc. With embeddings, many new methods are proposed and outperform the traditional methods for processing similarity queries.

Index is one of the most critical components in database and search systems as it provides fast access to the retrieval targets. Typical indexes include B tree, B+ tree, Log-structured merge-tree (LSM-tree), etc. Different index structures fit in different scenarios, e.g., KD-tree is suitable for low-dimensional spatial data search but performs weakly in high-dimensional spaces. For neural embeddings that are normally high-dimensional vectors, the proper indexes are majorly including 
four categories: hashing based~\cite{Index-survey-hashing-1, LSH-1, LSH-2, LSH-3, LSH-4, LSH-5, Learning-to-hash-1}, product quantization based~\cite{Product-quantization-1, Product-quantization-3, Product-quantization-5, Product-quantization-9, pq-distance-encoded-2014, pq-inverted-multi-index-2012, pq-ivfadc-2010, pq-locally-optimized-2014}, graph based~\cite{ Graph-based-index-2, Graph-based-index-5, Graph-based-index-7}, partition/tree based~\cite{Tree-based-index-2,  Tree-based-index-5, Tree-based-index-6, Tree-based-index-7}, 
where the state-of-the-art approaches normally belong to the first three categories  
in related research areas.

The operators involved in similarity query processing on high-dimensional data include but not limited to similarity search/selection, similarity join, similarity group-by, ranking/ordering, aggregation (e.g., sum and max), etc. In this survey we mainly investigate on the first three operators. 
Similarity search~\cite{Similarity-search-survey-1, Similarity-search-survey-2, Similarity-search-survey-3, Similarity-search-hashing-based-1, Similarity-search-hashing-based-2, Similarity-search-hashing-based-3,  Similarity-search-graph-based-1, Similarity-search-tree-based-1} mainly includes K-nearest-neighbor (KNN) search (i.e, search the top-k most similar neighbors to the query) and threshold search (also called range search, i.e., search all the objects within a given similarity or distance range centered at the query point). 
Similarity join also includes two categories like the similarity search, KNN join~\cite{KNN-join-2, KNN-join-3, KNN-join-5} and distance join~\cite{Condition-based-similarity-join-1, Condition-based-similarity-join-2, Condition-based-similarity-join-3,  Condition-based-similarity-join-5, Condition-based-similarity-join-8}. 
The major difference between the two joins are that KNN join is asymmetric while distance join is symmetric. More details about this difference and the consequence due to it are discussed in Section~\ref{sec:similarity-join}. The third similarity operator, similarity group-by~\cite{Similarity-group-by-1, Similarity-group-by-2, Similarity-group-by-3, Similarity-group-by-4, Similarity-group-by-5, Similarity-group-by-6}, refers to the process of collecting data into groups where similar data are within the same groups while different data are located in different groups. For high-dimensional data, the group-by operation is equivalent to the clustering operation which is well studied.

In addition to the software-level indexing and similarity operators, the new heterogeneous hardware (e.g., GPU and FPGA) also helps make advance on similarity query processing for high-dimensional data~\cite{Heterogeneous-hardware-1, Heterogeneous-hardware-2, Heterogeneous-hardware-3, Heterogeneous-hardware-4, Heterogeneous-hardware-5}. In some cases the GPU has outperformed CPU by hundreds of times of query processing speedup.

Finally in this survey we explore some common applications of similarity queries in data science community, and compare their traditional approaches (without embeddings) to their recent embedding-based approaches, showing the advantages of utilizing embeddings in similarity queries.

\begin{figure*}[!t]
  \centering
  \includegraphics[width=1.5\columnwidth]{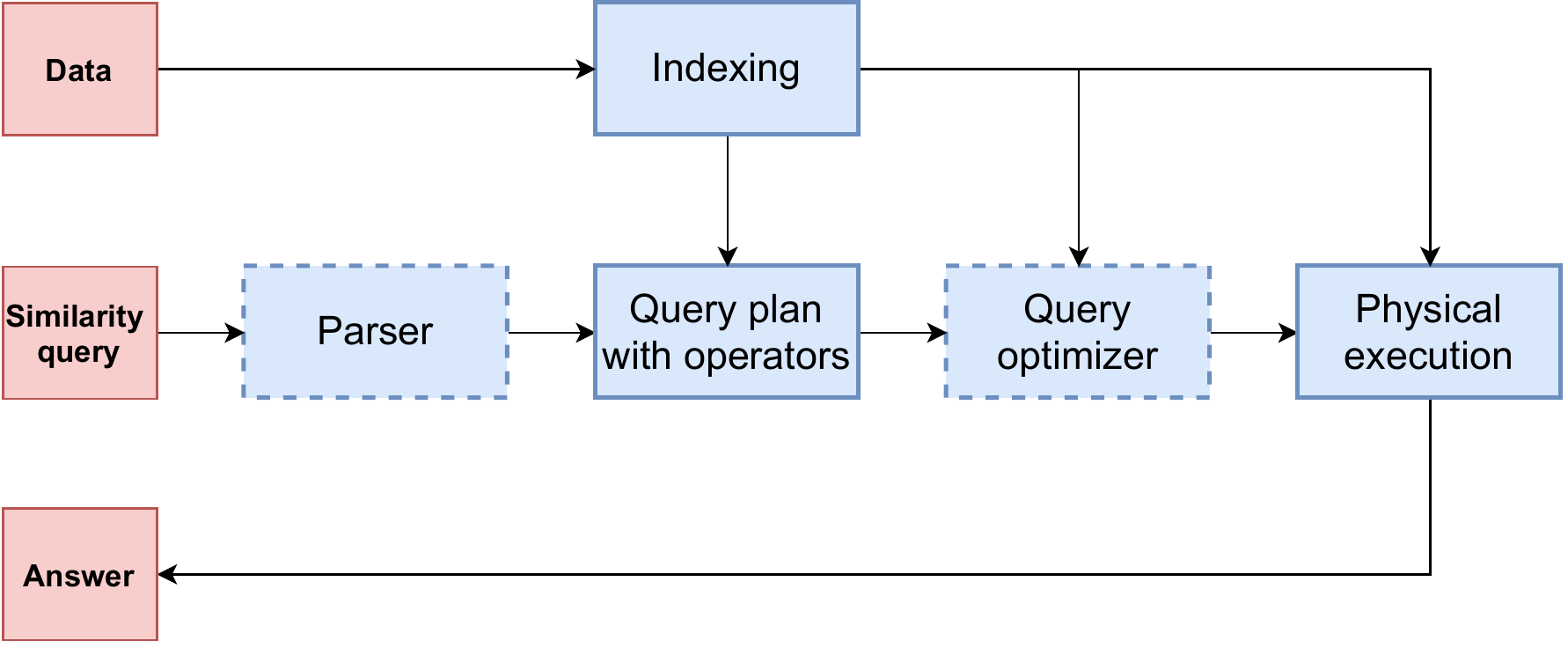}
  \caption{Similarity query processing workflow}
  \label{fig:similarity-query-processing-workflow}
\end{figure*}

\section{Background}
\subsection{Similarity query processing}
Similarity query is a very wide concept. Any query expecting answers based on similarity instead of exact value match is a similarity query, including but not limited to asking for similar text, similar faces, similar medical records, etc. And the corresponding data analytic operators become similarity operators, like similarity search/selection, similarity join, etc. As three of the most commonly used database operators, there have been many studies on similarity selection, join and group-by which we mainly investigate in this survey.

Figure~\ref{fig:similarity-query-processing-workflow} illustrates the workflow of similarity query processing. The pink boxes stand for the user end where the data and similarity queries are from and the answers are sent to. The blue boxes present major components during the processing. Similar to querying in a database system, the similarity queries go through query parser, query plan generator, query optimizer and finally are executed physically, while the data is indexed before querying and the index will be utilized in operators, optimization and execution. In some scenarios the query parsing and optimization can be skipped if the queries are simple, e.g., simply asking for the K nearest neighbors of a given query object. So the parser and optimizer boxes are drawn using dashed-line, and we do not discuss about them in this survey.

\subsection{Embeddings in similarity query processing}
As an important track of deep learning, neural embedding techniques have been more and more studied and applied in various domains. The applications of neural embeddings can be classified into two major categories: learning/fine-tuning task-specific embeddings and using pre-trained embeddings. The first category aims at learning new embeddings for specific tasks or tuning existing embeddings to better suit the tasks. By using the neural models designed and well tuned for the target tasks, these methods normally achieve the state-of-the-art quality, e.g.,~\cite{ER-embedding-3, example-learning-task-specific-embedding-for-Graph-1, example-learning-task-specific-embedding-for-text-1}. The second category tries to utilize existing neural models/embeddings pre-trained by previous researches (e.g., pre-trained BERT~\cite{bert} and ResNet~\cite{resnet}) and make slight adaption (like simply aggregating pre-trained word embeddings to get a sentence embedding) to fit their target tasks. These approaches are widely applied by the systems aiming at easing the use of deep learning techniques~\cite{example-pretrained-embedding-GNES} or studies with the goal of enhancing efficiency~\cite{example-pretrained-embedding-for-efficiency-1, example-pretrained-embedding-for-efficiency-2}.

However, no matter how the embeddings are generated, the workflows of using them to process similarity queries are similar: inferring the embeddings of target raw data and computing their distances to determine the similarities between the raw data during the similarity operations. For instance, recommender systems may execute a KNN search for the items whose embeddings have the smallest cosine distances to your favorite products, and recommend them to you as the similar products you may like. This survey will concentrate on efficient execution techniques of the following process after the embedding inference, without detailed discussion on generating the embeddings.

\section{Indexing neural embedding}
\label{sec:index}
Index is one of the most critical tools for fast data access which has been researched for decades. Traditional similarity/distance based indexes like KD-tree work well on low-dimensional data but perform weakly on high-dimensional data due to the curse of dimensionality. Since neural embeddings are normally high-dimensional vectors (like several hundreds of dimensions), they require high-dimensional similarity based indexing methods. And because most of the state-of-the-art embedding based methods~\cite{mikolov2013efficient,lacroix2018canonical,shi2019probabilistic,bert, embedding-based-work-using-euc-distance-1, embedding-based-work-using-euc-distance-2, embedding-based-work-using-euc-distance-3} rely on cosine similarity or Euclidean distance to accurately measure the embedding similarities, we focus on these two similarity metrics and the corresponding indexing approaches, and skip the discussion on other metrics and indexes.

To make it clearer, we present a summary of all the investigated indexing methods by Table~\ref{tab:summary-indexes} at the end of this section. The meaning of each abbreviation is given in Table~\ref{tab:abbreviations} which will be used in all the summary tables in the following sections.

\subsection{Hashing based}
Essentially, hashing based similarity index is built by hashing all the data objects in search space. For answering a query, it hashes the query, then looks up and verifies the candidates within the hash buckets where the hashed query is located. Hashing is widely used for efficient large-scale data retrieval since it maps high-dimensional data to low-dimensional spaces, which significantly reduces the computation. Furthermore, pair-wise comparison between most hash codes can be completed by light-weighted bit operations (e.g., computing Hamming distance of two hash codes by XOR), which further increases the processing speed.

There are two main categories of hashing based high-dimensional similarity indexing methods~\cite{Learning-to-hash-1, Index-survey-hashing-1}: locality-sensitive hashing (LSH) and learning to hash. 
\subsubsection{Locality-sensitive hashing} \hfill\\
Locality-sensitive hashing (LSH) is a hashing framework for efficient similarity search on high-dimensional data. Its basic idea is using a batch of specialized hashing functions $h_1(\cdot),...,h_m(\cdot)$ to hash the data records such that each record will correspond to one hashing value sequence. Given two records, the more element-wise overlaps (i.e., hashing values generated by the same hashing function $h_i(\cdot)$ are equal between the two sequences) there are between their hashing sequences, the more similar they are to each other. To increase precision or recall of the search, LSH amplification may be applied. There are two types of amplification, AND-amplification and OR-amplification. AND-amplification means only when all the corresponding hashing values are identical between two sequences (i.e., the two sequences are exactly the same), we think the two records are similar, while by OR-amplification we think they are similar as long as at least one pair of corresponding hashing values are identical. The specialized hashing function $h_i(\cdot)$ is so called \textit{LSH function} while the sequence of them ${G(\cdot) = (h_1(\cdot),...,h_m(\cdot))}$ is normally called \textit{compound LSH function}. LSH function is specially designed to achieve both of randomization and locality-preserving. Its formal definition is:

\begin{definition}[LSH function]
Given a metric space and its distance metric $d$, a threshold $t > 0$, and any two data points $\vec{u}$ and $\vec{v}$ in the metric space, an LSH function $h(\cdot)$ should satisfy the following two conditions:

(1) if $d(\vec{u}, \vec{v}) \leq t$, then $h(\vec{u}) = h(\vec{v})$ with a probability at least $p_1$

(2) if $d(\vec{u}, \vec{v}) \geq ct$, then $h(\vec{u}) = h(\vec{v})$ with probability at most $p_2$

where $c > 1$ is an approximation factor, and the probabilities $p_1 > p_2$.

\end{definition}
In short, such definition means an LSH function maps similar/close points into the same hash bucket with a higher probability than mapping dissimilar points into the same bucket. As a result, the points within the same bucket are likely to be similar to each other.

LSH functions are metric-dependent, i.e., different LSH function families work with different similarity metrics. Like $p$-stable distribution LSH~\cite{p-stable-lsh-datar2004locality} for $\ell_p$ distance, random projection LSH~\cite{random-projection-lsh} for cosine similarity, etc. More types of LSH functions for other distance metrics are well reviewed by~\cite{Index-survey-hashing-1} and we do not discuss about them here.

For Euclidean distance which is $\ell_2$ distance, the corresponding LSH is $p$-stable distribution LSH where $p = 2$. The $p$-stable LSH function is defined as
\begin{align}
    h(\vec{u}) = \lfloor \frac{\vec{a}\cdot \vec{u} + b}{w} \rfloor
\end{align}
where $\vec{a}$ is a vector (with the same dimension as data point $\vec{u}$) whose entries are chosen independently from a $p$-stable distribution, $w$ is a fixed window length set by users, $b$ is a real number sampled uniformly from the range $[0, w]$.

For cosine similarity, the random projection LSH function is shown below
\begin{equation}
\label{eq:hash-function}
h_{\vec{r}}(\vec{u}) = \left\{
 \begin{aligned}
             \begin{array}{lr}
             1 & if \  \vec{r} \cdot \vec{u} \ge 0 \\
             0 & if \  \vec{r} \cdot \vec{u} < 0
             \end{array}
\end{aligned}
\right.
\end{equation}
where $\vec{r}$ is a random vector (with the same dimension as the data point $\vec{u}$) of which each coordinate is drawn from an one-dimensional Gaussian distribution.

Based on the LSH functions, many index schemes are proposed. \cite{Multi-probe-LSH-lv2007multi} proposes multi-probe LSH which is one of the first LSH indexes using the LSH amplification techinques. \cite{Data-oriented-lsh-zhang2010data} focuses on the optimization on construction of the hash tables instead of the searching phase. Query adaptive LSH~\cite{Query-adaptative-lsh} does not use the pre-specific hashing functions set before construction, but selects proper LSH functions from a function pool online according to the given query. \cite{Posteriori-lsh} optimizes the multi-probe LSH by utilizing some prior knowledge about the queries and the datasets. \cite{distribuetd-lsh} designs a distributed LSH to fit on the emerging distributed systems. \cite{LSH-3} presents  SortingKeys-LSH (SK-LSH) which defines a specialized order on the hashing value of each data point such that the similar points will be placed together in the ordered sequence rather than in independent buckets, by which it can efficiently expand the search when the target buckets have no enough points to return. Super-Bit LSH~\cite{Super-Bit-lsh} orthogonalizes the random vectors used by random projection LSH to solve the problem that the variance is too large in the estimation of random projection LSH, which leads to large search error. Super-Bit LSH achieves significantly higher search accuracy compared to the original random projection LSH when length of the hash keys is fixed.

\subsubsection{Learning to hash}\hfill\\
Different from LSH whose hashing functions are data-independent (i.e., selection of the hashing functions is independent from data distribution), learning to hash is a family of data-dependent methods which learn the hashing functions best fitting the data distribution. A generalized form of hashing functions of learning to hash methods is
\begin{align}
    h_i(x) = \operatorname{sign}(f(\textbf{w}^T_ix + b_i))
\end{align}
where $f(\cdot)$ is normally a pre-defined function, including linear projection, kernel function, neural network, etc. Given a pre-specific loss function, the goal is to learn the best parameters $w_i$ and $b_i$ to minimize the loss. There is a significant number of related works on learning to hash in the past decades~\cite{l2h-multidim-features-supervised-kulis2009-binary-recons-embed, l2h-multidim-features-supervised-liu2012-supervised-hashing-with-kernels, l2h-multidim-features-supervised-norouzi2011-minimal-loss-hashing, l2h-multidim-features-unsupervised-gong2012-angular-quantization, l2h-multidim-features-unsupervised-liu2011-Anchor-Graph-hashing, l2h-multidim-features-unsupervised-weiss2008-spectral-hashing, l2h-multidim-features-supervised-gong2012-iterative-quantization}, including unsupervised and supervised learning approaches. Though most of them were evaluated on low-level feature vectors like image descriptor instead of semantic neural embedding, they can be applied to similarity search of neural embeddings, since by nature they are both high-dimensional numeric vectors. But we do not include their details in this survey as those studies have been well reviewed by several surveys~\cite{Learning-to-hash-1, Index-survey-hashing-1, learning-to-hash-good-survey-1, learning-to-hash-good-survey-2}. The interest of this survey on learning to hash is the more recent family of learning to hash methods, \textit{deep learning to hash}.

CNNH~\cite{deep-l2h-xia2014-supervised-hashing-image-retrieval} explores to simultaneously learn both of neural embeddings and hash functions in one learning run. It splits the learning process into two stages. In the first stage it learns approximate hash codes for the training samples by decomposing their similarity matrix. Then in the second stage, a convolution neural network (CNN) is used to learn both of the image embeddings and the hashing functions. Input of the CNN is raw images, while training targets/labels are the approximate hash codes for those images learned in stage 1. In the CNN, image embeddings are generated and input to the final output layer in order to train the output layer to approximate the target hash codes bit by bit. Later the trained output layer will act as the hashing functions to predict the hash codes of testing data. \cite{deep-l2h-xia2014-supervised-hashing-image-retrieval} also proposes an advanced approach (CNNH+) which uses the class labels of the images together with the approximate hash codes to train the output layer.

NINH~\cite{deep-l2h-lai2015-simultaneous-learning-embedding-and-hash-functions} also learns neural embeddings and hash functions simultaneously. It uses a similar deep neural network structure to CNNH~\cite{deep-l2h-xia2014-supervised-hashing-image-retrieval} including a CNN and a following hash coding module. Its major differences from CNNH is (1) its hash coding module is more complex than a simple fully-connected output layer, in order to reduce the redundancy among the hash bits, and (2) its input are triples including both of positive and negative training samples, and its training objective is to make generated hash code closer to that of the positive sample than to negative sample. Thus it does not need pre-computed target hash codes.

DPSH~\cite{deep-l2h-li2015-pairwise-labels},  DSH~\cite{deep-l2h-liu2016-dsh-for-image-retrieval}, DHN~\cite{deep-l2h-zhu2016-deep-hashing-net} and HashNet~\cite{deep-l2h-cao2017hashnet} use similar network structures to CNNH, i.e., a CNN with a hash code output layer. But they use pairwise labels instead of the approximate hash codes or the triple labels used by many existing works for training.

\cite{deep-l2h-erin2015-deep-hashing-supervised-DH} proposes Deep Hashing (DH) and Supervised Deep Hashing (SDH) based on multi-layer neural network whose input is raw data and output is hash code. DH trains the network by minimizing a specialized loss to minimize the difference between final output hash code and the real-number embedding right before the quantization, as well as maximize the variance of the bits in the output hash code to make them balanced. Based on DH, SDH further utilizes the labeled positive and negative samples to include supervision during the training.

Due to the difficulty to get large-scale labeled data, some weakly-supervised and unsupervised approaches have been proposed. WDHT \cite{deep-l2h-gattupalli2019-weakly-supervised} is a weakly-supervised approach which applies user-generated tags of online images instead of explicit image labels to learning the hashing. It generates word2vec based embeddings of the tags and uses them to constrain the learning. SADH~\cite{deep-lash-shen2018-unsupervised} is an unsupervised method that does not directly input the learned real-number embedding into the hashing layer, but uses it first to update the similarity graph matrix, then the matrix is used to improve the subsequent code optimization. It also proposes a discrete optimization algorithm with a general hashing loss.

In the works mentioned above, deep learning to hash techniques are mostly applied in image retrieval. In addition, they have been explored and exploited in many other domains, including knowledge graph similarity search~\cite{deep-l2h-wang2020-hash-for-KG-embed, deep-l2h-wang2019-hash-for-imcomplete-kg-search}, cross-modal retrieval~\cite{deep-l2h-cao2016-deep-visual-semantic-hash}, recommender systems~\cite{deep-l2h-shi2020-for-recommendation, deep-l2h-tan2020-with-GNN-for-recommendation}, and so on.

\subsection{Product quantization based}
Vector quantization (VQ)~\cite{Quantization-survey-gray1998} is a powerful tool to reduce the computation in high-dimensional nearest neighbor search. It works by encoding high-dimensional vectors into codewords using a many-to-one mapping, such that the cardinality of the representation space is reduced. Those codewords are named \textit{centroids}. The set of all the centroids forms the \textit{codebook} of such VQ. Then in similarity search, the distance between any two original vectors can be approximated by the distance between the corresponding codewords. Since the diversity of data values are reduced, the codewords are represented using much less bits/dimensions than the original vectors, which not only saves memory usage but also accelerates the computation of distances. Such advantages enable VQ to achieve state-of-the-art performance in similarity search.

But when VQ is applied on large-scale data, it still requires huge amount of memory occupation since the codebook size increases exponentially with length of each codeword (i.e., k-bit cordword leads to a codebook with up to $2^k$ centroids). Many modern similarity retrieval systems, like image retrieval, handle millions to billions of feature vectors, making it impossible to apply VQ. Therefore \textit{product quantization}
 (PQ)~\cite{Quantization-survey-gray1998, pq-ivfadc-2010, optimized-pq-kaiming2013} was proposed to address this issue. PQ partitions each original vector into M sub-vectors of lower dimensions, then quantizes each sub-vector individually, by which PQ actually decomposes the original vector space into the Cartesian product of M subspaces~\cite{optimized-pq-kaiming2013}. And each subspace corresponds to a much smaller codebook than that required by VQ. Finally the quantization of the original space is represented by the set of those subspace codebooks, which reduces the memory space requirement of VQ by several orders of magnitude.

Based on to what extent the search space is explored, there are two categories of production quantiztion indexing approaches: exhaustive and non-exhaustive methods, where the former means inspecting every data vector in the search space (i.e., brute force search) while the latter refers to only checking a fraction of the entire search space. Due to the recently rapid emergence of deep learning based product quantization, we investigate it separately in addition to the two major categories in this section.

\subsubsection{Non-exhaustive methods} \hfill\\
Non-exhaustive methods are normally facilitated by variant inverted index structures.
As one of the earliest studies for applying product quantization to similarity search, ~\cite{pq-ivfadc-2010} proposes an inverted file system with the asymmetric distance computation
(IVFADC). \cite{pq-ivfadc-2010} first introduces two types of distance computation, the Symmetric Distance Computation (SDC) and Asymmetric Distance Computations (ADC) in quantization based similarity search. Given a query vector and a candidate data vector, SDC approximates the distance between them as that between their codewords, while ADC uses the distance between the original query vector and codeword of the data vector to approximate it. \cite{pq-ivfadc-2010} proves SDC and ADC have similar time complexity and SDC only has a minor advantage over ADC that SDC stores the query using less space, while ADC obtains a lower distance distortion (i.e., the approximation error) than SDC, which is the reason for choosing ADC in the IVFADC approach. The inverted file system (IVF) is an inverted index on the codewords and the original data, where the codewords act as keys of the inverted lists. In the case of  non-exhaustive search, IVF enables fast access to a small fraction of candidate data records, which further enhances the computing efficiency by avoiding exhaustive comparisons between each data record and the query. Overall, IVFADC constructs an inverted index system with two quantizers, a \textit{coarse quantizer} to compute the first stage codeword of each original vector, and a \textit{locally defined product quantizer} to quantize the residues between the original vectors and their first stage codewords, which will be stored in the inverted list instead of directly storing the original vectors there. And the distance estimation during nearest neighbor search using such a system is also based on the quantized residues. This is because residues cost fewer bits to store than the original vectors, and encoding the residue is more precise than encoding the vector itself, resulting increased search accuracy.

Following the idea of ``inverted index + quantization'' from IVFADC, \cite{pq-inverted-multi-index-2012} proposes \textit{the inverted multi-index} which replaces the coarse vector quanziter with a product quantizer such that it is able to produce much finer subdivisions of the search space without
increasing the query time and the preprocessing time compared to the standard inverted index. Therefore the inverted multi-index achieves higher similarity search accuracy in similar time or higher speed under similar accuracy compared to inverted index. Specifically, unlike IVFADC which uses the codebook of the coarse vector quantizer as index keys, the inverted multi-index segments original vectors into equal-length sub-vectors and quantizes the subspaces using multiple quantizers respectively, then it uses the Cartesian product of the multiple codebooks as its index keys. During nearest neighbor search, compared to IVFADC which initially finds the top several nearest codewords (i.e., the index keys) to the query, the multi-index first looks up the top nearest codewords respectively in each codebook, then the Cartesian product of those found codewords refers to the index keys to inspect (in ascending order by the distance between the key and the query until enough data vectors have been inspected and returned).

Some works attempt to improve product quantization based indexing by combining multiple optimization techniques. LOPQ~\cite{pq-locally-optimized-2014} introduces the non-parametric and parametric optimization techniques of \cite{optimized-pq-kaiming2013, optimized-pq-kaiming2013-2} into IVFADC and the inverted multi-index approaches, to optimize the product quantizers used by the two index systems. By this, LOPQ optimizes the space and time overhead required by ordinary production quantization, and it is easy to be fit in existing search frameworks for the best performance with little overhead.
PQTable~\cite{pq-pqtable-2015} does not optimize the quantization itself but replaces the inverted index structure (like those in IVFADC and inverted multi-index) with hashtables, and improves the search efficiency using a similar table division and
merging method to the multi-index hashing~\cite{multi-index-hashing-norouzi2012}.

\subsubsection{Exhaustive methods} \hfill\\
Unlike non-exhaustive methods, exhaustive approaches mostly focus on optimizing the product quantization itself instead of designing a complex index system including several components, as they are required to execute in a brute force manner. The goal of optimizing product quantization is minimizing the quantization distortion (i.e., the error between the original vectors and the corresponding codewords).

\cite{optimized-pq-kaiming2013, optimized-pq-kaiming2013-2} formulate the optimization to be the process of finding the optimal codewords and space decomposition, i.e., how to learn the quantization centroids and how to segment the original vectors into sub-vectors such that the overall distortion is minimal. They propose two solutions, a non-parametric and a parametric approach. The non-parametric method is data-independent that works without knowledge about the data distribution. In this method, two sub-processes are iteratively and repeatedly executed until a maximal number of iterations is reached: (1) the decomposition is fixed and codewords are varied to minimize the objective, then (2) the codewords are fixed while decomposition is varied to further minimize it. The parametric solution is data-dependent in which the data distribution is assumed to be a parametric Gaussian distribution. Based on such assumption, a lower bound of the distortion is derived and reached when the decomposed subspaces are mutually independent and the variances of vectors in those subspaces are balanced. Then the \textit{Eigenvalue Allocation} algorithm using PCA and eigenvalues is designed for the space decomposition to achieve that lower bound.

\cite{pq-distance-encoded-2014} figures out that the IVFADC~\cite{pq-ivfadc-2010} and optimized PQ~\cite{optimized-pq-kaiming2013, optimized-pq-kaiming2013-2} approaches do not achieve a enough accuracy improvement in practice when sizes of the codebooks are increased, which is because they encode the original vectors/sub-vectors located in the same cell to the cell centroids, ignoring the distances between the original vectors and the corresponding centroids. In such a situation, the data vectors far from their cell centroids will face a relatively large distortion. If most of the vectors are located near the cell edges instead of the centers, the overall distortion will be significantly large. In summary, the quantization in IVFADC and the optimized PQ is too coarse. Therefore \cite{pq-distance-encoded-2014} presents an improved product quantization method, \textit{Distance-encoded Product Quantization} (DPQ), encoding a data vector/sub-vector using both of its assigned centroid and distance to the centroid. Specifically, DPQ further partitions each cell into several regions around the cell centroid with different distances to it, then concatenates a few extra bits (which are the least significant bits) to each codeword for encoding the region ID. For example, if there are 4 finer regions in each cell, two extra bits will be added to each quantization codeword. In another word, each codeword generated by DPQ is a concatenation of the cell centroid and the region bits. DPQ also designs two specific distance metrics, statistics and geometry based metrics, which are tailored for such an encoding schema for higher accuracy.

\subsubsection{Deep product quantization}\hfill\\
\cite{deep-pq-deep-quantization-net-2016} proposes \textit{Deep quantization network} (DQN), a deep neural network model for learning the optimal product quantizer as well as the data embeddings. The model is similar to those of deep learning to hash approaches. Specifically, it includes a CNN for learning the embeddings and a following full-connected layer for quantizing the embeddings. The model is trained with two losses, a pairwise cosine loss used by the embedding learning, and a product quantization loss to optimize the quantizer layer.

Product quantization network (PQN)~\cite{pq-pqnet-yu2018} is also a CNN based deep model but with a generalized product quantizer layer, so called the \textit{soft quantizer}. It is a generalized version of the standard product quantization which better fits in a neural network since its derivative is easier to be derived than the standard PQ. In addition, PQN introduces a novel asymmetric triplet loss to train the model for better performance.

Unlike many of the existing works using a fully-connected layer as quantizer, Deep Product Quantization (DPQ)~\cite{deep-pq-end2end-supervised-klein2019} designs a more complex quantizer module following the base CNN network, The quantizer includes several multi-layer perceptions (MLP) for learning on the embeddings from CNN and their sub-vectors, then generating the quantized vectors, so called soft and hard representations of the original embeddings, which will be used for Euclidean distance estimation during the later similarity search. DPQ also proposes a novel  \textit{Joint Central Loss} for training the quantizer module along with the softmax loss.

Deep Progressive Quantization~\cite{deep-pq-progressive-quantization-gao2019} is a end-to-end quantization code learning model that can simultaneously learn and generate quantized codes of different lengths for each input data embedding. It trains and optimizes multiple quantizer blocks at the same time where each  quantizer block is a layer of a CNN, corresponding to a output code of a different length. By this it avoids retraining the model when settings change and code with a different length is needed, which is unavoidable in standard product quantization and traditional deep product quantization.

\cite{deep-pq-module-liu2020} implements a plug-and-play module for deep product quantization that can be easily plugged onto any base deep learning model to quantize their learned embeddings. To solve the problem that quantization code is hardly derivable, it uses straight-through estimator and a modified MSE loss to optimize the hard quantization that assigns sub-vectors to exclusive codewords according to Euclidean distance, which has a lower distortion than the soft quantization (i.e., continuous relaxation).

Due to the difficulty on acquiring enough labeled training data, \cite{deep-pq-self-supervised-jang2021} presents a self-supervised deep product quantization model, Self-supervised Product Quantization (SPQ) network, which is label-free. Its main idea is namely \textit{Cross Quantized Contrastive learning}, i.e., transforming each input image by rotating, resizing, etc. and considering the transformed results of the same input image as correlated while those from different input images are uncorrelated, based on which the learning is executed to maximize the cross-similarity between the neural embeddings and the quantized codewords of the correlated transformed images. Such a learning strategy enables the model to learn discriminative quantization without any training label.

\subsection{Graph based}
Graph based indexes are emerging in recent years with the rapid development of GPU which is very suitable for accelerating graph computing. Formally, a graph based similarity index is such a (either directed or undirected) graph  $G(V, E)$ whose vertices $V$ is the set of all data points in the search space and the edge set $E$ is determined by distances between the data points. Specifically, if two data points are close enough to be thought as neighbors by some distance metric, there will be an edge connecting their corresponding vertices in $G$, otherwise that edge will not exist in $E$. \cite{graph-based-index-comprehensive-survey-vldb2021} provides a high-quality review on graph based indexes for similarity search. It classifies the indexes into four categories by the type of the base graph in the index: Delaunay Graph, Relative Neighborhood Graph, K-Nearest Neighbor Graph and Minimum Spanning Tree. We borrow those categories from their taxonomy, but do not exhaustively explore all the relevant works. Specifically, we classify the reviewed works in this section into two classes: graph construction which focuses on effectively building the base graphs and graph traversal which studies high-performance search on the existing graph index structures. And the publications in each of the two classes are organized by the categories borrowed from \cite{graph-based-index-comprehensive-survey-vldb2021}.

\subsubsection{Graph construction}\hfill\\
\noindent \textbf{K-Nearest Neighbor Graph}\hfill\\
K-Nearest Neighbor Graph (KNN Graph) is one of the most commonly used graph for similarity search indexing. It is a direct graph whose vertices are the data points in search space. Given a similarity metric, two arbitrary data points $p_u$, $p_v$ and their corresponding graph vertices $u$ and $v$, if $p_v$ is among the k nearest neighbors of $p_u$ under the similarity metric, there will be an edge from $u$ to $v$ in the KNN graph, otherwise the graph will not include such an edge.

\cite{Graph-based-index-NN-Descent-2011} proposes \textit{NN-Descent}, a simple but effective KNN graph construction algorithm. Its idea works under a intuitive principle: neighbor's neighbor is possibly also neighbor. So NN-Descent starts with a randomly initialized graph where each vertex's neighbors are randomly selected from the dataset. Then it repeatedly inspects neighbors' neighbors of each vertex and updates the vertex's neighbors if any neighbor's neighbor is closer than the vertex's current neighbors. The process terminates when the neighbor status stops changing. NN-Descent improves the KNN graph construction efficiency as it does not need to inspect the entire dataset. And it is easy to be implemented as distributed algorithm on MapReduce framework.

\cite{Graph-based-index-partition-construct-wang2012} constructs the KNN graph by recursive partitioning the search space, similar to the building of a KD-tree. The datasets are partitioned by hyperplanes recursively until each subspace is small enough. Then the data points within the same subspace is considered as neighbors and connected to each other. Such a random division process is repeated multiple times to reduce the cases that neighbors are partitioned into different subspaces. Finally the approach proposes a propagation strategy for spreading the neighborhood to wider area to further improve the construction accuracy.

Compared to many of the KNN graph construction algorithms that use Euclidean distance as the example similarity metric, L2Knng \cite{Graph-based-index-cosine-l2knng-anastasiu2015} studies on exact KNN graph construction with cosine similarity. It first builds an approximate KNN graph where each vertex's neighbors are not necessarily to be the exact k-nearest neighbors to it. During the building, it uses an inverted index of the data points to fast construct the initial approximate graph then follows the same idea of NN-Descent, i.e., neighbor's neighbor is likely to be also neighbor, to enhance the accuracy of the constructed graph using a similar greedy traversal method. Since NN-Descent results in an approximate instead of exact KNN graph, based on that approximate graph, L2Knng utilizes another inverted index to efficiently inspect the data points again and further update the vertices' neighbors, during which it does not traverse the entire search space but only inspects the most likely neighbors whose similarities to the current vertex satisfy some theoretic similarity bounds. Finally an exact KNN graph is acquired. Later \cite{Graph-based-index-cosine-l2knng-parallel-anastasiu2016} improves L2Knng to a parallel version, pL2Knng, which is more efficient.

\cite{Graph-based-index-kDNNG-xiao2018} constructs k-Diverse Nearest Neighbor (k-DNN) graph which achieves balance between precision and diversity. k-DNN graph is built in two steps. First an initial KNN graph is constructed using NN-Descent algorithm, then the initial graph is refined by re-ranking the neighbors of each vertex using a maximal marginal relevance (MMR) algorithm which obtains the k diverse nearest neighbors for each data point.

\cite{Graph-based-index-optimization-iwasaki2018} studies on optimization of the KNN graph for similarity search. It claims that the node degree, including both indegree and outdegree, is a critical factor that affects the search quality and efficiency. So it proposes three degree adjustment and one search path adjustment algorithms to optimize the node degrees both statically offline and dynamically during the search, such that the query time is reduced to the best effort with a high search quality.

\cite{Graph-based-index-NSG-2017} first presents comprehensive theoretical analysis on the factors/properties that affect the search efficiency the most. And based on the analysis they propose a novel Monotonic Relative Neighborhood Graph (MRNG) with high search efficiency on large-scale datasets. To further improve the efficiency, they also construct an approximate version of MRNG, namely Navigating Spreading-out Graph (NSG), by which they achieve the state-of-the-art similarity search performance with relatively small index sizes.

\subsubsection{Graph traversal}\hfill\\
\noindent \textbf{Search on K-Nearest Neighbor Graph}\hfill\\
GNNS~\cite{Graph-based-index-GNNS-hajebi2011} is a greedy search algorithm on KNN graph for the approximate KNN of a given query. The algorithm starts from a randomly chosen vertex in the KNN graph, then repeatedly replaces the current vertex by the neighbor that is closest to the query until no neighbor is closer than current vertex or a maximal number of iterations is reached. This repeated process will be restarted from different random starting vertices for several times to traverse enough candidate neighbors, during which all the traversed vertices are recorded and sorted by distance to the query. Finally the top-k closest vertices are returned as KNN of the query. Since GNNS is a greedy search algorithm focusing on local neighborhood, it faces a problem that it is likely to reach a local optimum rather than the globally optimal. To solve it, \cite{Graph-based-index-query-driven-iterated-wang2012} adapts the greedy neighborhood search of GNNS to the \textit{iterated local search} strategy which utilizes the search history to make the following restarted search processes more effective and avoid being stuck in local optima.

EFANNA~\cite{Graph-based-index-Efanna-2016} studies on both of efficient KNN graph and search on the graph. To construct the KNN graph, it follows the main idea of \cite{Graph-based-index-partition-construct-wang2012} to recursively partition the search space, then merge the close data points with a better filtering strategy to filter out the far data points and only inspect the points which are most likely to be neighbors during the merging stage. To search nearest neighbors, it first does a fast and coarse search to get initial KNN candidates in the hierarchical structures generated by the recursive partitioning, and then uses a similar greedy search on the constructed KNN graph as of NN-Descent to refine the initial candidates.

\noindent \textbf{Search on Delaunay Graph or Relative Neighborhood Graph}\hfill\\
As an approximation of Delaunay graph, Navigable small world (NSW) graph is a type of undirected graphs with high clustering coefficient and low diameter. A NSW graph for nearest neighbor search is constructed by one-to-one mapping all the data points to the graph vertices, and two vertices are connected by an edge if the corresponding data points are neighbors. Within an ideal NSW graph, any two vertices are connected by some path whose length is proportional to $logN$ where $N$ is the total number of vertices in the graph. Such a property guarantees that given a query point, its nearest neighbors can be found in a small number of traversal starting from any entry point of the NSW graph, as by up to $O(logN)$ ``skips'' we can reach any vertex in the graph.

\cite{Graph-based-index-nsw-malkov2014} proposes one of the first NSW graph based similarity search algorithms. It builds a base NSW graph with two types of edges, short-range and long-range links, where the former is to approximate the Delaunay graph~\cite{delaunay-graph-aurenhammer1991} while the latter is to guarantee the NSW graph property mentioned above. The construction process is simple, i.e., inserting the data point one by one and for each newly inserted point, connecting it to its closest neighbors in the current graph and vice versa. During this process some of the early created short-range links will gradually become long-range links. To search nearest neighbors on such a graph,\cite{Graph-based-index-nsw-malkov2014} uses a similar greedy algorithm to \cite{Graph-based-index-NN-Descent-2011}, i.e., repeatedly replacing current vertex with one of its neighbors which is the closest to the query, until no neighbors are closer than the current vertex, meaning that a local optima is reached. During this process the candidates of the k nearest neighbors are collected. Then such a process is repeated several times on the vertices that are never traversed in previous iterations to update the KNN candidates until no more change can be made on them, and they are returned as the final results. \cite{Graph-based-index-hnsw-malkov2016} improves \cite{Graph-based-index-nsw-malkov2014} to the Hierarchical Navigable Small World (HNSW) graph based algorithm. HNSW graph is essentially a multi-layer NSW graph where the links are separated according to their length scale into different layers, such that upper layers include longer range links while lower layers include shorter range links. The search starts at the upper layer where the same greedy algorithm as of \cite{Graph-based-index-nsw-malkov2014} is executed to reach the local optimal vertex on the current layer. Then the search goes down and starts at the next lower layer from the local optimal vertex of the previous layer. Since the lower layer includes more shorter range links, the search becomes finer. The algorithm repeats such a process until the lowest layer is searched where the finest search results are acquired. HNSW graph provides the capability of a finer search than NSW and achieves higher recall. In addition, its hierarchical structure is easy to be implemented in distributed systems for scaling up.

FANNG~\cite{Graph-based-index-fanng-2016} constructs a specialized approximation of Relative Neighborhood (RN) graph with an occlusion rule to avoid adding unnecessary edges into the graph, by which it reduces the redundant edges. Therefore the neighborhood traversal is reduced to the minimal during the greedy search, making the search more efficient. It also modifies the greedy search itself. Specifically, when no more update can be made on the KNN candidates, it backtracks to the previous closest vertex and traverse its unexplored neighbors as in depth-first-search, instead of just terminating the search.

\begin{table}[!h]
  \caption{Abbreviations used in the summary tables}
  \label{tab:abbreviations}
  \begin{tabular}{c|c}
    \toprule
     Full name & Abbreviation \\
    \midrule
    \ Locality-sensitive hashing & LSH \\
    \ Learning to hash  & L2H \\
    \ Euclidean distance & EUC \\ 
    \ Cosine similarity & COS \\
    \ Any common similarity metric & AnySim \\
    \ Product quantization & PQ \\
    \ Deep product quantization & DeepPQ \\
    \ Non-exhaustive & NE \\
    \ Exhaustive & EX \\
    \ Graph construction & GC \\
    \ Graph traversal & GT \\
    \bottomrule
  \end{tabular}
\end{table}

\begin{table*}[!h]
  \caption{Summary of high-dimensional indexes for similarity search}
  \label{tab:summary-indexes}
  \begin{tabular}{|c|c|cp{4cm}|}
    \toprule
     Category & Similarity metric & Method \\
    \Xhline{3\arrayrulewidth}
    \ Hashing(LSH) & COS & \multicolumn{1}{|m{5cm}|}{Random projection LSH~\cite{random-projection-lsh}, Super-Bit LSH~\cite{Super-Bit-lsh}} \\
    \hline
    \ Hashing(LSH) & EUC & \multicolumn{1}{|m{5cm}|}{Multi-Probe LSH~\cite{Multi-probe-LSH-lv2007multi}, Posteriori Multi-Probe LSH~\cite{Posteriori-lsh}, Data-Oriented LSH~\cite{Data-oriented-lsh-zhang2010data}, Query-adaptive LSH~\cite{Query-adaptative-lsh},  Distributed Layered LSH~\cite{distribuetd-lsh},  SortingKeys-LSH (SK-LSH)~\cite{LSH-3} } \\
    \hline
    \ Hashing(L2H) & EUC & \multicolumn{1}{|m{5cm}|}{CNNH/CNNH+~\cite{deep-l2h-xia2014-supervised-hashing-image-retrieval}, NINH~\cite{deep-l2h-lai2015-simultaneous-learning-embedding-and-hash-functions}, DPSH~\cite{deep-l2h-li2015-pairwise-labels}, DSH~\cite{deep-l2h-liu2016-dsh-for-image-retrieval}, DHN~\cite{deep-l2h-zhu2016-deep-hashing-net},  HashNet~\cite{deep-l2h-cao2017hashnet}, DH/SDH~\cite{deep-l2h-erin2015-deep-hashing-supervised-DH}, WDHT \cite{deep-l2h-gattupalli2019-weakly-supervised}, SADH~\cite{deep-lash-shen2018-unsupervised}} \\
    \Xhline{3\arrayrulewidth}
    \ PQ(NE) & EUC & \multicolumn{1}{|m{5cm}|}{IVFADC~\cite{pq-ivfadc-2010}, Inverted multi-index~\cite{pq-inverted-multi-index-2012}, LOPQ~\cite{pq-locally-optimized-2014}, PQTable~\cite{pq-pqtable-2015}} \\
    \hline
    \ PQ(EX) & EUC & \multicolumn{1}{|m{5cm}|}{\cite{optimized-pq-kaiming2013, optimized-pq-kaiming2013-2}, DPQ~\cite{pq-distance-encoded-2014}} \\
    \hline
    \ DeepPQ & EUC &  \multicolumn{1}{|m{5cm}|}{DQN~\cite{deep-pq-deep-quantization-net-2016}, PQN~\cite{pq-pqnet-yu2018}, DPQ~\cite{deep-pq-end2end-supervised-klein2019}, Deep Progressive Quantization~\cite{deep-pq-progressive-quantization-gao2019}, \cite{deep-pq-module-liu2020}, SPQ~\cite{deep-pq-self-supervised-jang2021}}  \\
    \Xhline{3\arrayrulewidth}
    \ Graph(GC) & AnySim &  \multicolumn{1}{|m{5cm}|}{NN-Descent~\cite{Graph-based-index-NN-Descent-2011}, \cite{Graph-based-index-partition-construct-wang2012}, k-DNN graph~\cite{Graph-based-index-kDNNG-xiao2018}, \cite{Graph-based-index-optimization-iwasaki2018}} \\
    \hline
    \ Graph(GC) & COS &  \multicolumn{1}{|m{5cm}|}{L2Knng \cite{Graph-based-index-cosine-l2knng-anastasiu2015}, pL2Knng~\cite{Graph-based-index-cosine-l2knng-parallel-anastasiu2016}} \\
    \hline
    \ Graph(GC) & EUC &  \multicolumn{1}{|m{5cm}|}{MRNG/NSG~\cite{Graph-based-index-NSG-2017}} \\
    \hline
    \ Graph(GT) & AnySim &  \multicolumn{1}{|m{5cm}|}{GNNS~\cite{Graph-based-index-GNNS-hajebi2011}, Query-driven iterated search~\cite{Graph-based-index-query-driven-iterated-wang2012}, EFANNA~\cite{Graph-based-index-Efanna-2016}, NSW~\cite{Graph-based-index-nsw-malkov2014}, HNSW~\cite{Graph-based-index-hnsw-malkov2016}, FANNG~\cite{Graph-based-index-fanng-2016}} \\
    \Xhline{3\arrayrulewidth}
    \ Tree & EUC &  \multicolumn{1}{|m{5cm}|}{\cite{Tree-based-index-metric-spill-trees-liu2004}, Randomized KD-tree~\cite{Tree-based-index-randomized-kd-tree-silpa2008}, Random project tree~\cite{Tree-based-index-random-projection-tree-dasgupta2008}, Randomly-oriented KD-tree~\cite{Tree-based-index-randomly-oriented-kd-tree-vempala2012}, Randomized partition tree~\cite{Tree-based-index-randomized-partition-trees-2013}, \cite{Tree-based-index-revisiting-kd-tree-2019}, \cite{Tree-based-index-which-tree-to-use-2013}, FLANN~\cite{simSearch-lib-flann, simSearch-lib-flann-2}, MRPT~\cite{Tree-based-index-MRPT-2016}, \cite{Tree-based-index-auxiliary-info-rptree-2018}} \\
    \hline
    \ Tree & AnySim &  \multicolumn{1}{|m{5cm}|}{rpForests~\cite{Tree-based-index-rpForests-2019}, Comparison tree~\cite{Tree-based-index-comparison-tree-2017}} \\
    \hline
    \bottomrule
  \end{tabular}
\end{table*}

\subsection{Partition / tree based}
Tree based (or space-partitioning based) indexes are one of the most commonly used indexes for similarity search, e.g., KD-tree. For low-dimensional data, they are able to achieve logarithmic time complexity. But due to the curse of dimensionality, their performance is diminished to be no better than brute force search on high-dimensional space. However, many studies have been conducted to propose variants of the classic tree based indexes to fit high-dimensional data.

A classic KD-tree partitions the space recursively into two subspaces at each step. The partitioning pivot is the median of the max variance along one of the coordinate axes. \cite{Tree-based-index-metric-spill-trees-liu2004} suggests to select the median along with a random direction rather than a coordinate axis in each split, in another word, using a randomly rotated orthogonal basis for the space to build a KD-tree. It also suggests to use \textit{random projection} to reduce the data dimension by projecting the dataset to a lower-dimensional space, which is based on the statements of \cite{random-project-high-to-low-dim-lindenstrauss1984}: a dataset of n points can be embedded in a subspace of dimension O(log n) with little distortion on the pair-wise distances. Following the suggestion of \cite{Tree-based-index-metric-spill-trees-liu2004} that introduces randomness into the space partitioning of KD-tree, \cite{Tree-based-index-random-projection-tree-dasgupta2008} proposes \textit{random project tree} which introduces another more randomness. In addition to the random direction for the splitting pivot, the random project tree no longer splits on the median but randomly selects a cutting point in an interval around the median. \cite{Tree-based-index-randomly-oriented-kd-tree-vempala2012,  Tree-based-index-randomized-partition-trees-2013, Tree-based-index-revisiting-kd-tree-2019} make some further refinement on the random projection tree with the same core idea (i.e., random rotated basis and random cutting point selection on an interval).  \cite{Tree-based-index-random-projection-tree-dasgupta2008, Tree-based-index-randomly-oriented-kd-tree-vempala2012} also provide analysis on the reason why these randomness can make kd-tree fit high-dimensional data: though the data lies in high-dimensional space, its \textit{intrinsic dimension} is often very low, and the introduced randomness adapts KD-tree to the low intrinsic dimension rather than the high ambient dimension. In short words, random projection tree captures a small number of the most significant dimensions which are deterministic, so it works well no matter how many ``redundant'' dimensions exist since they have only minor effects on its partitioning and searching.

\cite{Tree-based-index-which-tree-to-use-2013} provides further theoretic analysis on the similarity search performance of commonly used binary space-partitioning trees (BSP-trees), e.g., KD-tree and random projection tree. They conclude that the search performance is proportional to the performance when that tree is used in vector quantization, and margins of the partitions are another important factor affecting the search performance. In summary, they claim and prove that best BSP-tree for search should satisfy two properties: low quantization error and large partition margins.

\cite{simSearch-lib-flann, simSearch-lib-flann-2} evaluate many nearest neighbor search algorithm on high-dimensional data and find the two best of them, the\textit{ priority search k-means tree } and the \textit{multiple randomized KD-trees}, where the former is a novel similarity search index structure proposed by them and the latter is implemented according to the \textit{randomized KD-tree}~\cite{Tree-based-index-randomized-kd-tree-silpa2008}. Basically, randomized KD-tree is a variant of KD-tree which selects the partitioning direction randomly among the first several dimensions with the largest variances instead of always cutting on exactly the dimension with the maximal variance. To achieve better search accuracy, the index normally includes multiple randomized KD-trees (which is also called randomized KD-tree forest). The search on the forest is executed parallelly in each tree, and a priority queue is used to maintain the results from all the trees ascendingly ordered by their distances to the corresponding decision boundaries. This guarantees the closest leaves from all the trees will be explored first. Another chosen index in \cite{simSearch-lib-flann, simSearch-lib-flann-2}, the priority search k-means tree, is built also by partitioning the space recursively. The major difference between it and KD-tree is that it splits data points of each subspace into k finer subspaces by k-means clustering based on the distance computed using all dimensions instead of cutting on only one dimension. When searching in a k-means tree, at each level, the closest cluster center to the query will be chosen and the search moves down to the corresponding subspace. All the research achievements of \cite{simSearch-lib-flann, simSearch-lib-flann-2} are integrated and released in a library, namely \textit{fast library for approximate nearest neighbors} (FLANN), with the capability of automatically selecting and configuring the optimal nearest neighbor algorithm for the given datasets.

\textit{Multiple random projection trees} (MRPT)~\cite{Tree-based-index-MRPT-2016} and \textit{random projection forests} (rpForests)~\cite{Tree-based-index-rpForests-2019} are two similar variants of random projection tree~\cite{Tree-based-index-random-projection-tree-dasgupta2008}. To overcome the shortcoming of random projection tree that it requires heavy computation and large memory occupation for the construction and search, MRPT makes several adaptions: (1) it uses sparse vector instead of dense vector to determine the splitting direction in each subspace, (2) it uses the same direction vector for all subspaces at the same level (while random projection tree selects a different random vector for each new subspace), and (3) it rollbacks to splitting on the median instead of randomly cutting in the interval around the median. MRPT builds several such modified random projection trees and uses their votes on each candidate data point to filter out the points  less likely to be a nearest neighbor, by which the final accuracy and efficiency are further improved. Similarly, rpForests also improves random projection tree by using ensemble of multiple random projection trees to form forests. And it optimizes the forests by detecting the best random direction vector for each split. Specifically, before each split, it randomly generates a partitioning vector and computes the variance of data points projected on the vector's direction, then repeats such generating-and-computing step several times and selects the vector along which the variance is maximal. This strategy improves the tree construction efficiency and also makes it more balanced, which leads to a higher search quality and speed. Unlike MRPT and rpForests, instead of using multiple trees, \cite{Tree-based-index-auxiliary-info-rptree-2018} studies optimizing the search on single random projection tree. It proposes the method that stores auxiliary information in internal nodes of random projection tree and utilizes the information during search to achieve higher quality and efficiency. It also proposes two priority functions to retrieve data points efficiently under a given computational budget in single tree.

\cite{Tree-based-index-comparison-tree-2017} proposes the \textit{comparison tree}, which is a generalized comparison based tree index on metric space where the distance metric is unknown and only the \textit{triplet comparisons} are accessible to the users. \textit{triplet comparisons} means given a triple of three points $(x, y, z)$, the users know whether $d(x, y) < d(x, z)$ is true or not, where $d(x, y)$ stands for the distance between points $x$ and $y$ under the unknown distance metric. Such a setting is common in the crowd sourcing literature. The comparison tree is also constructed by splitting space into two subspaces recursively like KD-tree, but due to the constraint above, it selects two random pivots in each subspace, and each point in the subspace will be assigned to its closer pivot between the two. By such a strategy the points will be grouped into two finer subspaces without knowing the distance metric. Then the search is straightforward: in each subspace, checking which pivot the query is closer to and searching into the corresponding finer subspace recursively.

\section{Similarity operators}
\subsection{Similarity search}
Similarity search normally includes two categories, k-nearest neighbor search and range search (also called radius search or $\epsilon$-similarity search). k-nearest neighbor (KNN) search is defined as such: given a query point and some distance metric, searching the top-k closest data points to the query in the whole dataset/search space. And the range search is defined as the search of all data points whose distance to the query is smaller than a given threshold under some distance metric. Formally, we formulate these two types of search below:

\begin{definition}[k-nearest neighbor search]
    Given a dataset $P = \{p_i|i = 1, 2, ..., n\}$, a distance metric $d(\cdot, \cdot)$ and a query point $q$, k-nearest neighbor search tries to find a set of k data points $P^\ast = \{p^\ast_i | i = 1, 2, ..., k\}$ such that for any $p_i \in P - P^\ast$ and $p^\ast_j \in P^\ast$, $d(q, p_i) \geq d(q, p^\ast_j)$.
\end{definition}

\begin{definition}[range search]
    Given a dataset $P = \{p_i|i = 1, 2, ..., n\}$, a distance metric $d(\cdot, \cdot)$, a query point $q$ and a radius/threshold $\epsilon$, range search tries to find a set of data points $P^\ast$ such that for any $p^\ast_i \in P^\ast$, $d(q, p^\ast_i) \leq \epsilon$.
\end{definition}

For neural embeddings, range search normally makes no sense since it is hard to determine the distance threshold without a thorough inspection of the target datasets. For example, if using Euclidean distance as the similarity metric, it may varies from 0 to infinity, while if using cosine similarity, even though its range is only from -1 to 1, it is still hard to tell which number can be an accurate boundary between ``similar'' and ``unsimilar''. In addition, tree-based indexes and methods are the mainstream in solving range search problem~\cite{range-search-using-tree-bohm2000cost, range-search-using-tree-bohm2001searching, range-search-using-tree-Kim2021, range-search-using-tree-white1996similarity}, which are less actively studied in recent years. Though there are some new research works using hashing-based methods on range search~\cite{range-search-using-hashing-ahle2017parameter, range-search-using-hashing-gao2015selective, range-search-using-hashing-Tuxen2016RangeQO} recently, the progress is not as significant as the advances on the research of KNN search problem. Again, on high dimensional data like neural embeddings, searching based on similarity threshold is not practical in real-world applications, due to the difficulty on selecting a proper threshold. Therefore, we will not discuss more on range search over embeddings, but focus on the KNN search.

The definition of KNN search mentioned above is about the exact KNN search, which returns exactly the top-k closest neighbors to the query. However, in real-world applications, especially in the scenarios of big data, an accurate exact KNN search usually requires unaffordable computing time. To balance the computing time and accuracy, \textit{approximate k-nearest neighbor} (ANN) search is more and more studied today. ANN search returns k points that are not necessarily the 100\% accurate top-k closest points and thus the search accuracy of ANN algorithms is normally measured by some ratio between the returned results and the true answers, e.g., the number of correct results over $k$ or the average distance of the true answers to the query over that of the returned points to the query, and so on.

In most cases, KNN and ANN search are done by simply using some specialized index to retrieve the results efficiently and effectively. And actually we have already explored them in Section~\ref{sec:index}. So please refer to Section~\ref{sec:index} for the existing KNN and ANN methods. In this section, we present a summary of the mainstream real-world KNN/ANN software libraries that are facilitated by those indexing methods in Table~\ref{tab:summary-KNN-libs}.    

\begin{table*}[!h]
  \caption{Summary of mainstream real-world KNN/ANN search libraries}
  \label{tab:summary-KNN-libs}
  \begin{tabular}{|c|c|c|} 
    \toprule
     Library name & Index category & Index method\\
    \Xhline{3\arrayrulewidth}
    \ NearPy~\cite{simSearch-lib-NearPy} & Hashing &  LSH \\
    \hline
    \ FALCONN~\cite{simSearch-lib-FALCONN} & Hashing & LSH~\cite{simSearch-lib-falconn-basedOn-lsh} \\
    \hline
    \ PUFFINN~\cite{simSearch-lib-PUFFINN} & Hashing & parameterless LSH~\cite{LSH-parameterless-PUFFINN-aumuller2019} \\
    \hline
    \ FAISS~\cite{simSearch-lib-FAISS} & PQ & IVFADC~\cite{simSearch-lib-fb-faiss-basedOn-pq-plus-inverted-index} \\
    \hline
    \ ScaNN~\cite{simSearch-lib-ScaNN} & PQ & anisotropic vector quantization~\cite{simSearch-lib-google-scann-basedOn-pq} \\
    \hline
    \ KGraph~\cite{simSearch-lib-KGraph} & Graph &  KNN graph \\
    \hline
    \ PyNNDescent~\cite{simSearch-lib-PyNNDescent} & Graph &  NN-Descent~\cite{Graph-based-index-NN-Descent-2011} \\
    \hline
    \ NSG~\cite{simSearch-lib-NSG} & Graph & Navigating Spreading-out Graph~\cite{Graph-based-index-NSG-2017} \\
    \hline
    \ EFANNA~\cite{simSearch-lib-EFANNA} & Graph & KNN graph~\cite{Graph-based-index-Efanna-2016} \\
    \hline
    \ NMSLIB~\cite{simSearch-lib-NMSLIB}, Hnswlib~\cite{simSearch-lib-Hnswlib} & Graph &  Hierarchical Navigable Small World graph~\cite{Graph-based-index-hnsw-malkov2016} \\
    \hline
    \ FLANN & Tree &  k-means trees + randomized KD-trees~\cite{simSearch-lib-flann, simSearch-lib-flann-2} \\
    \hline
    \ MRPT~\cite{simSearch-lib-MRPT}, Annoy~\cite{simSearch-lib-annoy} & Tree &  multiple random projection trees~\cite{Tree-based-index-MRPT-2016} \\
    \hline
    \ rpforest~\cite{simSearch-lib-rpforest} & Tree &  random projection forests~\cite{Tree-based-index-rpForests-2019} \\
    \hline
    \ NGT~\cite{simSearch-lib-NGT} & Graph + Tree & \cite{Graph-based-index-optimization-iwasaki2018} \\ 
    \hline
    \ SPTAG~\cite{simSearch-lib-SPTAG} & Graph + Tree & KNN graph~\cite{Graph-based-index-partition-construct-wang2012, Tree-based-index-trinary-projection-trees} + Balanced k-means trees \\
    \hline
    \bottomrule
  \end{tabular}
\end{table*}

\subsection{Similarity join}
\label{sec:similarity-join}
Similar to the similarity search, similarity join also includes two major categories, KNN join and distance join (a.k.a, similarity range join). KNN join is an asymmetric join where each data point in the the left dataset joins with its k nearest neighbors in the right dataset. In contrast, distance join is a symmetric join that returns all pairs of the data points between the left and right datasets where the two points in a pair have a smaller distance than the given threshold. Formally, we define them as such

\begin{definition}[KNN join]
    Given two datasets $R$ and $S$, an integer $k$, the KNN join between $R$ and $S$ is denoted by $R \ltimes_{KNN} S$, abbreviated as $R \ltimes S$, which combines each data point $r \in R$ with its k nearest neighbors from $S$. Formally
    \begin{align}
        R \ltimes_{KNN} S = \{(r, s) | \forall r \in R, \forall s\ where\ s \in KNN(r, S, k)\}
    \end{align}
    where $KNN(r, S, k)$ stands for the k nearest neighbors from dataset $S$ to point $r$. 
\end{definition}

\begin{definition}[Distance join]
   Given two datasets $R$ and $S$, a distance threshold $t$, and a distance metric $d(\cdot, \cdot)$, the distance join between $R$ and $S$ is denoted by $R \bowtie_t S$, which combines each point $r \in R$ with each point $s \in S$ that is close/similar enough to $r$ (i.e., with distance smaller than or equal to $t$). Formally
    \begin{align}
        R \bowtie_t S = \{(r, s) | \forall r \in R, \forall s \in S\ where\ d(r,\ s) \leq t\}
    \label{eq:sim-join}
    \end{align}
\end{definition}

\subsubsection{KNN join}\hfill\\
There are tons of studies on KNN join for low-dimensional data, especially spatial data, since looking for the 2D or 3D closest locations to some given places is one of the most common scenarios in spatial applications, e.g., recommending a couple of closest restaurants to the users' current locations on map. But on high-dimensional data those approaches normally degrade due to the curse of dimensionality. So more and more high-dimensional KNN join methods are proposed. We organize them according to whether their target datasets are dynamic or static and present two categories here: static KNN join and dynamic KNN join. Some algorithms or systems are claimed to be for KNN search, but their inputs are a batch of multiple query points instead of single query, which are actually also KNN join if we see the multiple query points as the left dataset. So we also include such approaches in this section.

\noindent \textbf{Static KNN join} \hfill \\
High-dimensional KNN join on static datasets (i.e., the datasets do not change with time) is well studied in the past decades. Gorder KNN-join~\cite{KNN-join-Gorder-2004} is an efficient and IO optimized KNN join method. It optimizes the disk IO efficiency during KNN join by sorting the two datasets based on the \textit{Grid Order} and then executing a block nested loop join on top of the ordered data. Specifically, the ordering, namely \textit{G-ordering}, includes two steps, firstly principal component analysis (PCA) is applied on the high-dimensional data, then the dimension-reduced space is partitioned to be a grid of multiple hyper-rectangle cells and the data vectors are sorted according to their located cells, such that the vectors in the same cell are grouped together in the sorted datasets. After the G-ordering, a block nested loop join is executed where each block contains data from multiple disk pages in order to reduce the disk IO time. iJoin~\cite{KNN-join-iJoin-yu2007} is a high-dimensional KNN join based on a specialized index, \textit{iDistance}, which partitions the space and selects a proper reference point for each partition, then indexes each data point using its distance to the closest reference point, by which the high-dimensional data point is represented as a one-dimensional scalar value that can be retrieved using existing index structures like B+ tree. Then basically, the iJoin itself is executed by searching the k nearest neighbors of each left dataset point in the right dataset, utilizing the index. \cite{KNN-join-z-value-2010, KNN-join-z-value-2012} both map multidimensional data into one-dimensional space using \textit{z-value}, a space filling curve to transform the high-dimensional data into one-dimensional space with preserving the data locality. After the data transformation, the kNN join is transformed to a sequence of one-dimensional range search, which can be efficiently executed using some existing indexes, which is similar to the idea of iDistance. But Gorder join and iJoin, as well as other similar space partitioning based methods like \cite{KNN-join-z-value-2010,KNN-join-z-value-2012}, focus on data with less than 100 dimensions, sometimes even less than 10 dimensions, though their approaches are claimed to be for high-dimensional data.

With the hashing based index emerging in high-dimensional KNN search, it is also used in KNN join approaches. RankReduce~\cite{KNN-join-Rankreduce-stupar2010} implements a distributed LSH based KNN search system using MapReduce. The hashing tables of LSH are distributed across multiple computing nodes in RankReduce. When processing KNN queries, a list of multiple query vectors are input and kept in each mapper. In a mapper, a portion of the data vectors is loaded and the mapper computes the distances between each of the data vectors to each of the query vectors and maintains the current k closest neighbors for each query. Finally the reducer collects those closest neighbors from all mappers and re-ranks them to get the final KNN for each query. Hashing based index is suitable for significantly high dimensional data (at least more than 100 dimensions), so following RankReduce, many hashing based studies are conducted for high-dimensional KNN join (including KNN search methods whose input is a batch of queries, as we mention above).  \cite{KNN-join-inverted-index-LSH-2015} proposes a distributed LSH-Based Inverted Index, namely LBI, where each key is a distinct hash value of data vectors and value is a list of the data vector IDs corresponding to that hash value. Comparing to RankReduce which stores the data vectors in memory, LBI only maintains their IDs and the vectors are only loaded into memory whenever necessary. By this it reduces the space overhead significantly. \cite{KNN-join-LSH-zhou2013hdkv} presents a LSH based index for efficient value-based search on high-dimensional data. It solves the problem that most key-value (KV) stores (1) does not provide fast value-based search or (2) provide value-based search using tree-based indexes that are inefficient on high-dimensional data. \cite{KNN-join-parallel-LHS-teixeira2013scalable} parallelizes LSH index based on the dataflow programming paradigm. Specifically, it decomposes the LSH building and KNN search processes into multiple computing stages organized in several conceptual pipelines, which makes indexing and searching using LSH parallel in distributed or multi-core processor systems. It successfully applied such a technique to Content-Based Multimedia Retrieval (CBMR).

There are also studies of KNN join on top of other types of indexes, e.g., distributed KNN join based on tree index~\cite{KNN-join-tree-based-2014}, distributed KNN join on parallel product quantization~\cite{KNN-join-parallel-PQ-ANDRADE201981}, localized KNN join based on the multi-index with product quantization for word embeddings stored in relational databases~\cite{KNN-join-PQ-multi-index-based-on-word2vec-gunther2019fast}, distributed pivot-based KNN join which partitions search space into Voronoi cells\cite{KNN-join-pivot-based-Voronoi-cells-2020}, etc. Since the traditional localized KNN join has been pretty well studied, the mainstream KNN join research has turned to the distributed/parallel scenarios.    

\noindent \textbf{Dynamic KNN join} \hfill \\
In addition to the distributed static KNN join, another most active research direction in KNN join problem is that on dynamic data, i.e., the datasets with frequent updating (e.g., inserting/replacing/deleting data), such as stream data.

KNNJoin+\cite{KNN-join-dynamic-incremental-Yu2010HighdimensionalKJ} is a KNN join method supporting efficient incremental update. Basically it maintains a KNN join result table and iDistance index for the initial datasets, and updates them as soon as the left or right dataset is updated, by which it does not need to re-compute the KNN join on the whole datasets but only incrementally updates the KNN join table and the index for each change in the datasets.

\cite{KNN-join-dynamic-HDR-tree-2014} proposes KNN join methods supporting real-time recommendation in the scenario where content is updated frequently. They proposes two R-tree based indexes for high-dimensional feature vectors of the users and content, HDR-tree and HDR*-tree, by which they can fast retrieve the users who are affected by the content update, and then re-compute the KNN join results between the affected users and the updated content (which is only a small fraction of the whole user and content datasets). The two methods also utilize PCA and random projection respectively for dimension reduction to reduce the distance computation cost. Later, based on the two methods, \cite{KNN-join-dynamic-LSH-hu2019efficient} proposes another improved version method by using a LSH index to replace the HDR-tree and HDR*-tree. Due to the simple structure of LSH index, such a method further improves the retrieval speed of the affected users and also possesses much less index construction and maintenance cost than the two tree-based indexes.

Other KNN join studies on stream data includes those on sets~\cite{KNN-join-on-sets-2007}, on spatial data~\cite{KNN-join-on-spatial-2021}, etc. Since we focus on neural embedding data, we will not discuss their details in this section. 

\subsubsection{Distance join}\hfill\\
Distance join is very commonly used in many domains, including data mining, data integration and cleaning (like near duplicate detection), and so on. But many applications of distance join in those domains are for string, set, or other types of data instead of high-dimensional embeddings. For example, text token based near duplicate detection for Web pages~\cite{Condition-based-join-text-token-xiao2011efficient}, string similarity joins with
edit distance~\cite{Condition-based-join-string-sim-join-edit-dist-2011}, or \cite{Condition-based-join-string-sim-join-edit-dist-2017} whose ``embedding'' is used with edit distance and is actually not a neural embedding generated by deep models as we refer to. etc. And similar to the range search, distance join on neural embeddings normally makes no sense due to the vague distance boundary between ``similar'' and ``unsimilar''. Therefore we will not investigate more about distance join in this survey. 

\subsection{Similarity group-by}
Group-by is an important operator in RDBMS that needs specialized designs. But in terms of high-dimensional embeddings, the similarity based group-by operator for them is actually equivalent to the clustering operation. Clustering approaches for high-dimensional data has been well reviewed by many surveys~\cite{clustering-survey-arora2014survey, clustering-survey-babu2011clustering, clustering-survey-kriegel2009clustering, clustering-survey-pandove2018systematic, clustering-survey-pavithra2017survey, clustering-survey-xu2015comprehensive, clustering-survey-zimek2018clustering}. So we will skip the discussion on them in this section.

\section{New hardware for similarity query processing}
Modern hardware, especially the heterogeneous hardware acceleration techniques have been widely applied on high-dimensional similarity query processing. In addition to optimizing similarity query processing for CPU, many publications also attempt to utilize other kinds of processors to accelerate the processing of similarity queries. The representatives of them are GPU and FPGA.

\subsection{GPU accelerated similarity query processing}
GPU is designed for high-performance parallel computing, especially on matrix manipulation. So they are suitable to help construct and speed up parallel algorithms. Unlike CPU, GPU requires that the parallelization degree is high enough (i.e., there should be a large number of parallel subtasks), each parallel subtask should reduce the branches to the best effort, and the shared memory size should be properly assigned to achieve efficient memory access, etc. Such new requirements lead to new designs of algorithms and indexes that are different from the CPU-based versions.

\cite{gpu-lsh-2011} builds a GPU based parallel LSH index for fast KNN search on high-dimensional data. The index includes two levels. The first level is a  random projection tree that roughly partition the search space, while the second level includes multiple LSH built on each of the subspaces. Different from the traditional LSH, here each second level LSH index is a hierarchical structure built based on  a Morton curve, which can help balance the workload among GPU threads. Specifically, the hierarchical LSH can choose different search strategies depending on the data density of each subspace.

\cite{gpu-kernel-lsh-lukavc2015gpu} designs a kernel based LSH index and uses it in fast KNN search for high-dimensional satellite image retrieval, with a mixed CPU+GPU settings. It first preprocesses the dataset and generates the kernel matrix for training on CPU to build the index, then the components of index are stored in different parts of the GPU cache/memory separately (for the best access speed), which are used later to parallelly search the KNN of multiple query points on GPU.

FLASH (Fast LSH Algorithm for Similarity search accelerated with HPC)~\cite{gpu-fast-lsh-ultra-high-dim-Wang2018RandomizedAA} is an improved LSH which well fits GPU computing on ultra-high dimensional data (millions of dimensions) where the traditional indexes like general LSH and product quantization cannot be used. It solves several shortcomings of LSH via multiple state-of-the-art techniques, e.g., by using Densied One Permutation Hashes (DOPH), FLASH can compute hundreds of minwise hashes in one pass rapidly and once the data is indexed completely, only the index needs to be maintained while the raw data is no longer needed (in contrast, LSH must keep both of the hash tables and the raw data for the distance computations), such that the space overhead is significantly reduced; by using fixed sized Reservoir sampling, the data skews among LSH hash buckets are fixed and the resulting sample arrays of fixed length are easier to be accessed and maintained by GPU, also making the load better balanced among GPU threads; etc.

\cite{gpu-PQ-Wieschollek_2016_CVPR} focuses on GPU accelerated product quantization for similarity search. It designs the \textit{product quantization tree} (PQT) in which each part of product quantization is not a flat codebook but a hierarchical tree structure. Since tree structure has better access efficiency (i.e., requiring fewer times of distance computation and comparison) than linear scan, the PQT speed up the overall query processing than traditional product quantization. And either the construction or the utilization of such an index structure is easy to be highly parallelized on GPU.

FAISS~\cite{simSearch-lib-fb-faiss-basedOn-pq-plus-inverted-index} is one of the most popular and state-of-the-art product quantization based similarity search libraries today. It has been widely and successfully applied in industry. By utilizing the registers, \cite{simSearch-lib-fb-faiss-basedOn-pq-plus-inverted-index} proposes an extreme fast k-selection algorithm on GPU. Then they design a near-optimal computation layout for parallelizing product quantization indexing algorithms on GPU and demonstrate it by improving the original IVFADC~\cite{pq-ivfadc-2010} to a parallel version on GPU.

\cite{gpu-PQ-RobustiQ-2019} presents a hierarchical inverted indexing structure based on Vector and Bilayer Line Quantization (VBLQ). The index includes three levels of different types of quantizers, by which they solve the problem that a single subspace resulted from quantization may include too many points. In addition, by FAISS, the increasing number of subspaces will increase search accuracy but at the same time require a longer search time as the subspaces becomes finer and denser. The hierarchical index of \cite{gpu-PQ-RobustiQ-2019} successfully solve this problem. The index raises only little memory overhead when increasing the number of subspaces significantly to achieve higher search accuracy.

There are also many approaches based on other types of indexes for high-dimensional similarity queries on GPU, including space-filling curves based index~\cite{teodoro2012approximate}, permutation based index~\cite{Permutation-based-indexing-gpu}, inverted index~\cite{inverted-index-gpu}, grid based index~\cite{grid-based-index-gpu}, and so on.

\subsection{FPGA accelerated similarity query processing}
FPGA is another powerful hardware for high-performance parallel computing. Instead of being installed as a computing unit on computers, FPGA is normally applied in embedded systems and devices. So unlike GPU, it has fewer chances to interact with those commonly used large-scale databases. Therefore the applications and publications of FPGA in traditional similarity query processing are not as many as GPU. But we still explore some related works in this survey since FPGA is being applied in wider and wider range for high-performance computing, including but not limited to IOT (Internet-of-Thing), stock market, smart home, and so on. It may be also widely integrated into personal computers and servers in the future.

\cite{fpga-jun2015large} uses FPGA with flash memory to speed up large scale high-dimensional KNN search. The flash memory is to hold the large scale datasets that cannot fit in DRAM, while FPGA is used as auxiliary accelerator for parallel distance comparison during the search.

\cite{fpga-faiss-Danopoulos2019FPGAAO} speeds up FAISS by implementing the inverted file index (IVF) part of FAISS on FPGA, since the index creation takes a significant long time, which is therefore worthy to be further accelerated. Finally the integration of FAISS onto FPGA achieves much higher performance than on GPU under the same power limit.

\cite{fpga-PQ-Zhang_2018_CVPR} further improves FAISS by implementing and optimizing product quantization based KNN search on FPGA. Specifically, it designs a compression method to reduce the codebook sizes in FAISS, which could be significant large but FAISS never attempts to reduce. By such a compression, it successfully fits the codebooks into the extreme fast but also very small on-chip memory of FPGA. Similar to FAISS, it also proposes an optimized k-selection algorithm for FPGA (while the one proposed by FAISS is for GPU) to accelerate KNN search.

\cite{fpga-multi-index-hashing-2019} focuses on accelerating the multi-index hashing in KNN search. Because multi-index hashing needs to lookup many times among several hash tables for distance computation and sort the initially retrieved candidates of nearest neighbors, it is considered as the most data-intensive part of the whole KNN search algorithm. Therefore, to apply such a KNN search algorithm in embedded devices like IOT, the multi-index hashing has to be accelerated by specialized hardware such as FPGA. \cite{fpga-multi-index-hashing-2019} carefully designs the distance computation and sorting unit in logical circuit level and implements it in FPGA.

More studies are conducted to improve the usability of FPGA-accelerated KNN search.  \cite{fpga-open-source-knn} presents a fully open-source FPGA accelerator for vector similarity searching, in order to make such a hardware acceleration technique available to more users instead of letting it stay in research institutes. \cite{fpga-configurable-knn} implements a configurable FPGA-based KNN accelerator aiming at solving the problem that many of the existing FPGA-accelerated KNN methods only support fixed parameters such as fixed feature dimensions and similarity metrics.

\section{Applications of similarity queries}
To better demonstrate the effect of neural embedding in similarity query processing, we select two typical application scenarios of similarity queries, entity solution and information retrieval, then compare the related works with and without applying neural embeddings in those scenarios. 

\subsection{Entity resolution}
In data science, multiple data records may refer to the same real-world entity. The task of recognizing/matching them for deduplication on the raw datasets is called \textit{entity resolution} (ER) (a.k.a., entity matching (EM), near-duplicate detection, etc.). Obviously, ER applications have to often answer similarity queries, including KNN search, KNN join and so on, to find similar records to the queried records.

Traditional ER approaches normally match the records by some or all of their attributes~\cite{ER-traditional-1, ER-traditional-2, ER-traditional-3} with some token based similarity metric such as cosine similarity for token sequences or edit distance for strings. Such kind of matching algorithms have a major shortcoming that they are weak to extract the semantic information from the data records. This sometimes significantly affects their matching quality since an entity can be represented using different words in various ways. And their core matching modules often need to be highly task-specific, like \cite{ER-traditional-3} utilizes different similarity functions for different attributes in its target data tables, making the methods hard to be generalized.

Because of the advantage of embeddings on semantic encoding, more recent ER studies are conducted with neural embeddings. \cite{ER-embedding-1} uses word embeddings to generate attribute embeddings and summarizes the attribute embeddings to form the record embeddings, which are used as representations of the data records for similarity comparison. \cite{ER-embedding-2} also starts with word embeddings to get attribute embeddings but it directly computes attribute-wise similarities between two records which results in a similarity vector rather than a scalar similarity value for each two records. Then the similarity vectors are input to a binary classifier to determine whether the corresponding two records are matched or not. \cite{ER-embedding-8} develops the method from \cite{ER-embedding-2} by using the state-of-the-art pre-trained embedding, BERT, to generate the record embeddings. It also proposes several optimization approaches on the embedding generation like utilizing domain knowledge to find-tune the pre-trained BERT model. \cite{ER-embedding-3} directly learns the embeddings on the data records without using pre-trained deep models, and then uses nearest neighbor search to find the most similar record to each given record, and vice versa. If the two records are most similar to each other, then they are likely to refer to the same entity.

Those embedding based ER approaches show state-of-the-art performance. And utilization of embeddings also makes the methods more generalized, as the embedding is unstructured and schema-less, meaning that it can be used more flexibly than the original record attributes.

\subsection{Information retrieval}
Information retrieval (IR) includes a very wide range of applications, e.g., image retrieval, text retrieval, video retrieval, etc. Among them there are two special kinds, multimodal information retrieval (MMIR) and cross-modal information retrieval (CMIR). MMIR refers to the process of retrieving information using multiple retrieval models like searching for online posts with a query consisting of both image and text. CMIR refers to the retrieval of information in one modality using query in another modality, like searching images using text query.

For MMIR and CMIR, one major category of the mainstream methods is to map different modalities into an isomorphic latent space and then do similarity search or join in that space. Such methods have been successfully applied before the emergence of embeddings by maximizing inter-modal correlations~\cite{IR-traditional-unified, IR-traditional-joint-1, IR-traditional-joint-2}. Since deep neural models have shown a significantly better capability of capturing the hidden semantic information from data, they gradually replace the traditional linear algebra and statistics based space transformation methods. Most of the recent works on MMIR and CMIR are learning joint embeddings over data from different modalities directly via deep neural networks and transformers~\cite{IR-embedding-joint-mithun2018learning, IR-embedding-joint-sadeh2019joint, IR-embedding-joint-2016}. By using deep models and embeddings, they achieve much better retrieval quality than the traditional methods.

\section{Conclusion}
In this survey, we first review the workflow of neural embedding based similarity query processing which includes index, query parser, query plan generator, query optimizer and plan executor. Then we investigate the techniques of the most critical components in the processing workflow, index (including hashing, product quantization, graph and tree based indexing algorithms) and similarity operators (including majorly similarity search, join and group-by), from both software and hardware levels. Finally we select some application domains of similarity queries, and present the differences between the solutions in those domains with and without embeddings, in order to better show the strength of neural embeddings in facilitating similarity query processing.

\bibliographystyle{ACM-Reference-Format}
\bibliography{survey}


\begin{thebibliography}{242}


\ifx \showCODEN    \undefined \def \showCODEN     #1{\unskip}     \fi
\ifx \showDOI      \undefined \def \showDOI       #1{#1}\fi
\ifx \showISBNx    \undefined \def \showISBNx     #1{\unskip}     \fi
\ifx \showISBNxiii \undefined \def \showISBNxiii  #1{\unskip}     \fi
\ifx \showISSN     \undefined \def \showISSN      #1{\unskip}     \fi
\ifx \showLCCN     \undefined \def \showLCCN      #1{\unskip}     \fi
\ifx \shownote     \undefined \def \shownote      #1{#1}          \fi
\ifx \showarticletitle \undefined \def \showarticletitle #1{#1}   \fi
\ifx \showURL      \undefined \def \showURL       {\relax}        \fi
\providecommand\bibfield[2]{#2}
\providecommand\bibinfo[2]{#2}
\providecommand\natexlab[1]{#1}
\providecommand\showeprint[2][]{arXiv:#2}

\bibitem[\protect\citeauthoryear{??}{sim}{[n.d.]a}]%
        {simSearch-lib-annoy}
 \bibinfo{year}{[n.d.]}\natexlab{a}.
\newblock \bibinfo{title}{Annoy: Approximate Nearest Neighbors Oh Yeah}.
\newblock
\newblock
\urldef\tempurl%
\url{https://github.com/spotify/annoy}
\showURL{%
\tempurl}


\bibitem[\protect\citeauthoryear{??}{sim}{[n.d.]b}]%
        {simSearch-lib-EFANNA}
 \bibinfo{year}{[n.d.]}\natexlab{b}.
\newblock \bibinfo{title}{Efanna: An extremely fast approximate nearest
  neighbor search algorithm based on knn graph.}
\newblock
\newblock
\urldef\tempurl%
\url{https://github.com/ZJULearning/efanna}
\showURL{%
\tempurl}


\bibitem[\protect\citeauthoryear{??}{sim}{[n.d.]c}]%
        {simSearch-lib-FAISS}
 \bibinfo{year}{[n.d.]}\natexlab{c}.
\newblock \bibinfo{title}{FAISS - a library for efficient similarity search and
  clustering of dense vectors}.
\newblock
\newblock
\urldef\tempurl%
\url{https://github.com/facebookresearch/faiss}
\showURL{%
\tempurl}


\bibitem[\protect\citeauthoryear{??}{sim}{[n.d.]d}]%
        {simSearch-lib-FALCONN}
 \bibinfo{year}{[n.d.]}\natexlab{d}.
\newblock \bibinfo{title}{FALCONN - FAst Lookups of Cosine and Other Nearest
  Neighbors}.
\newblock
\newblock
\urldef\tempurl%
\url{https://github.com/FALCONN-LIB/FALCONN}
\showURL{%
\tempurl}


\bibitem[\protect\citeauthoryear{??}{exa}{[n.d.]}]%
        {example-pretrained-embedding-GNES}
 \bibinfo{year}{[n.d.]}\natexlab{}.
\newblock \bibinfo{title}{Generic Neural Elastic Search: From bert-as-service
  and Go Way Beyond}.
\newblock
  \bibinfo{howpublished}{\url{https://hanxiao.io/2019/07/29/Generic-Neural-Elastic-Search-From-bert-as-service-and-Go-Way-Beyond}}.
\newblock


\bibitem[\protect\citeauthoryear{??}{sim}{[n.d.]e}]%
        {simSearch-lib-Hnswlib}
 \bibinfo{year}{[n.d.]}\natexlab{e}.
\newblock \bibinfo{title}{Hnswlib - fast approximate nearest neighbor search}.
\newblock
\newblock
\urldef\tempurl%
\url{https://github.com/nmslib/hnswlib}
\showURL{%
\tempurl}


\bibitem[\protect\citeauthoryear{??}{sim}{[n.d.]f}]%
        {simSearch-lib-KGraph}
 \bibinfo{year}{[n.d.]}\natexlab{f}.
\newblock \bibinfo{title}{KGraph: A Library for Approximate Nearest Neighbor
  Search}.
\newblock
\newblock
\urldef\tempurl%
\url{https://github.com/aaalgo/kgraph}
\showURL{%
\tempurl}


\bibitem[\protect\citeauthoryear{??}{lea}{[n.d.]}]%
        {learning-to-hash-good-survey-1}
 \bibinfo{year}{[n.d.]}\natexlab{}.
\newblock \bibinfo{title}{Learning to Hash for Big Data: A Tutorial.}
\newblock
  \bibinfo{howpublished}{\url{https://cs.nju.edu.cn/lwj/slides/L2H.pdf}}.
\newblock


\bibitem[\protect\citeauthoryear{??}{sim}{[n.d.]g}]%
        {simSearch-lib-MRPT}
 \bibinfo{year}{[n.d.]}\natexlab{g}.
\newblock \bibinfo{title}{MRPT - fast nearest neighbor search with random
  projection}.
\newblock
\newblock
\urldef\tempurl%
\url{https://github.com/vioshyvo/mrpt}
\showURL{%
\tempurl}


\bibitem[\protect\citeauthoryear{??}{sim}{[n.d.]h}]%
        {simSearch-lib-NSG}
 \bibinfo{year}{[n.d.]}\natexlab{h}.
\newblock \bibinfo{title}{Navigating Spread-out Graph For Approximate Nearest
  Neighbor Search}.
\newblock
\newblock
\urldef\tempurl%
\url{https://github.com/ZJULearning/nsg}
\showURL{%
\tempurl}


\bibitem[\protect\citeauthoryear{??}{sim}{[n.d.]i}]%
        {simSearch-lib-NearPy}
 \bibinfo{year}{[n.d.]}\natexlab{i}.
\newblock \bibinfo{title}{NearPy: ANN search in large, high-dimensional data
  sets (in python)}.
\newblock
\newblock
\urldef\tempurl%
\url{http://pixelogik.github.io/NearPy/}
\showURL{%
\tempurl}


\bibitem[\protect\citeauthoryear{??}{sim}{[n.d.]j}]%
        {simSearch-lib-NGT}
 \bibinfo{year}{[n.d.]}\natexlab{j}.
\newblock \bibinfo{title}{NGT: Neighborhood Graph and Tree for Indexing
  High-dimensional Data}.
\newblock
\newblock
\urldef\tempurl%
\url{https://github.com/yahoojapan/NGT}
\showURL{%
\tempurl}


\bibitem[\protect\citeauthoryear{??}{sim}{[n.d.]k}]%
        {simSearch-lib-NMSLIB}
 \bibinfo{year}{[n.d.]}\natexlab{k}.
\newblock \bibinfo{title}{Non-Metric Space Library (NMSLIB)}.
\newblock
\newblock
\urldef\tempurl%
\url{https://github.com/nmslib/nmslib}
\showURL{%
\tempurl}


\bibitem[\protect\citeauthoryear{??}{sim}{[n.d.]l}]%
        {simSearch-lib-PUFFINN}
 \bibinfo{year}{[n.d.]}\natexlab{l}.
\newblock \bibinfo{title}{PUFFINN: Parameterless and Universal Fast FInding of
  Nearest Neighbors}.
\newblock
\newblock
\urldef\tempurl%
\url{https://github.com/puffinn/puffinn}
\showURL{%
\tempurl}


\bibitem[\protect\citeauthoryear{??}{sim}{[n.d.]m}]%
        {simSearch-lib-PyNNDescent}
 \bibinfo{year}{[n.d.]}\natexlab{m}.
\newblock \bibinfo{title}{PyNNDescent: a Python nearest neighbor descent for
  approximate nearest neighbors.}
\newblock
\newblock
\urldef\tempurl%
\url{https://github.com/lmcinnes/pynndescent}
\showURL{%
\tempurl}


\bibitem[\protect\citeauthoryear{??}{sim}{[n.d.]n}]%
        {simSearch-lib-rpforest}
 \bibinfo{year}{[n.d.]}\natexlab{n}.
\newblock \bibinfo{title}{rpforest}.
\newblock
\newblock
\urldef\tempurl%
\url{https://github.com/lyst/rpforest}
\showURL{%
\tempurl}


\bibitem[\protect\citeauthoryear{??}{sim}{[n.d.]o}]%
        {simSearch-lib-ScaNN}
 \bibinfo{year}{[n.d.]}\natexlab{o}.
\newblock \bibinfo{title}{ScaNN: Scalable Nearest Neighbors}.
\newblock
\newblock
\urldef\tempurl%
\url{https://github.com/google-research/google-research/tree/master/scann}
\showURL{%
\tempurl}


\bibitem[\protect\citeauthoryear{??}{sim}{[n.d.]p}]%
        {simSearch-lib-SPTAG}
 \bibinfo{year}{[n.d.]}\natexlab{p}.
\newblock \bibinfo{title}{SPTAG: A library for fast approximate nearest
  neighbor search}.
\newblock
\newblock
\urldef\tempurl%
\url{https://github.com/microsoft/SPTAG}
\showURL{%
\tempurl}


\bibitem[\protect\citeauthoryear{??}{IR-}{2013}]%
        {IR-traditional-unified}
 \bibinfo{year}{2013}\natexlab{}.
\newblock \showarticletitle{A unified framework for multimodal retrieval}.
\newblock \bibinfo{journal}{\emph{Pattern Recognition}} \bibinfo{volume}{46},
  \bibinfo{number}{12} (\bibinfo{year}{2013}), \bibinfo{pages}{3358--3370}.
\newblock
\showISSN{0031-3203}
\urldef\tempurl%
\url{https://doi.org/10.1016/j.patcog.2013.05.023}
\showDOI{\tempurl}


\bibitem[\protect\citeauthoryear{Abbasifard, Ghahremani, and Naderi}{Abbasifard
  et~al\mbox{.}}{2014}]%
        {Similarity-search-survey-2}
\bibfield{author}{\bibinfo{person}{Mohammad~Reza Abbasifard},
  \bibinfo{person}{Bijan Ghahremani}, {and} \bibinfo{person}{Hassan Naderi}.}
  \bibinfo{year}{2014}\natexlab{}.
\newblock \showarticletitle{A survey on nearest neighbor search methods}.
\newblock \bibinfo{journal}{\emph{International Journal of Computer
  Applications}} \bibinfo{volume}{95}, \bibinfo{number}{25}
  (\bibinfo{year}{2014}).
\newblock


\bibitem[\protect\citeauthoryear{Abdelsadek and Hefeeda}{Abdelsadek and
  Hefeeda}{2014}]%
        {KNN-join-tree-based-2014}
\bibfield{author}{\bibinfo{person}{Ahmed Abdelsadek} {and}
  \bibinfo{person}{Mohamed Hefeeda}.} \bibinfo{year}{2014}\natexlab{}.
\newblock \showarticletitle{DIMO: Distributed Index for Matching Multimedia
  Objects Using MapReduce}. In \bibinfo{booktitle}{\emph{Proceedings of the 5th
  ACM Multimedia Systems Conference}} (Singapore, Singapore)
  \emph{(\bibinfo{series}{MMSys '14})}. \bibinfo{publisher}{Association for
  Computing Machinery}, \bibinfo{address}{New York, NY, USA},
  \bibinfo{pages}{115–126}.
\newblock
\showISBNx{9781450327053}
\urldef\tempurl%
\url{https://doi.org/10.1145/2557642.2557650}
\showDOI{\tempurl}


\bibitem[\protect\citeauthoryear{Ahle, Aum{\"u}ller, and Pagh}{Ahle
  et~al\mbox{.}}{2017}]%
        {range-search-using-hashing-ahle2017parameter}
\bibfield{author}{\bibinfo{person}{Thomas~D Ahle}, \bibinfo{person}{Martin
  Aum{\"u}ller}, {and} \bibinfo{person}{Rasmus Pagh}.}
  \bibinfo{year}{2017}\natexlab{}.
\newblock \showarticletitle{Parameter-free locality sensitive hashing for
  spherical range reporting}. In \bibinfo{booktitle}{\emph{Proceedings of the
  Twenty-Eighth Annual ACM-SIAM Symposium on Discrete Algorithms}}. SIAM,
  \bibinfo{pages}{239--256}.
\newblock


\bibitem[\protect\citeauthoryear{Akbik, Blythe, and Vollgraf}{Akbik
  et~al\mbox{.}}{2018}]%
        {example-learning-task-specific-embedding-for-text-1}
\bibfield{author}{\bibinfo{person}{Alan Akbik}, \bibinfo{person}{Duncan
  Blythe}, {and} \bibinfo{person}{Roland Vollgraf}.}
  \bibinfo{year}{2018}\natexlab{}.
\newblock \showarticletitle{Contextual string embeddings for sequence
  labeling}. In \bibinfo{booktitle}{\emph{Proceedings of the 27th international
  conference on computational linguistics}}. \bibinfo{pages}{1638--1649}.
\newblock


\bibitem[\protect\citeauthoryear{Altowim, Kalashnikov, and Mehrotra}{Altowim
  et~al\mbox{.}}{2014}]%
        {ER-traditional-3}
\bibfield{author}{\bibinfo{person}{Yasser Altowim}, \bibinfo{person}{Dmitri~V
  Kalashnikov}, {and} \bibinfo{person}{Sharad Mehrotra}.}
  \bibinfo{year}{2014}\natexlab{}.
\newblock \showarticletitle{Progressive approach to relational entity
  resolution}.
\newblock \bibinfo{journal}{\emph{Proceedings of the VLDB Endowment}}
  \bibinfo{volume}{7}, \bibinfo{number}{11} (\bibinfo{year}{2014}),
  \bibinfo{pages}{999--1010}.
\newblock


\bibitem[\protect\citeauthoryear{Amagata, Hara, and Xiao}{Amagata
  et~al\mbox{.}}{2019}]%
        {KNN-join-on-sets-2007}
\bibfield{author}{\bibinfo{person}{Daichi Amagata}, \bibinfo{person}{Takahiro
  Hara}, {and} \bibinfo{person}{Chuan Xiao}.} \bibinfo{year}{2019}\natexlab{}.
\newblock \showarticletitle{Dynamic Set kNN Self-Join}. In
  \bibinfo{booktitle}{\emph{2019 IEEE 35th International Conference on Data
  Engineering (ICDE)}}. \bibinfo{pages}{818--829}.
\newblock
\urldef\tempurl%
\url{https://doi.org/10.1109/ICDE.2019.00078}
\showDOI{\tempurl}


\bibitem[\protect\citeauthoryear{Anastasiu and Karypis}{Anastasiu and
  Karypis}{2015}]%
        {Graph-based-index-cosine-l2knng-anastasiu2015}
\bibfield{author}{\bibinfo{person}{David~C Anastasiu} {and}
  \bibinfo{person}{George Karypis}.} \bibinfo{year}{2015}\natexlab{}.
\newblock \showarticletitle{L2knng: Fast exact k-nearest neighbor graph
  construction with l2-norm pruning}. In \bibinfo{booktitle}{\emph{Proceedings
  of the 24th ACM International on Conference on Information and Knowledge
  Management}}. \bibinfo{pages}{791--800}.
\newblock


\bibitem[\protect\citeauthoryear{Anastasiu and Karypis}{Anastasiu and
  Karypis}{2016}]%
        {Graph-based-index-cosine-l2knng-parallel-anastasiu2016}
\bibfield{author}{\bibinfo{person}{David~C Anastasiu} {and}
  \bibinfo{person}{George Karypis}.} \bibinfo{year}{2016}\natexlab{}.
\newblock \showarticletitle{Fast parallel cosine k-nearest neighbor graph
  construction}. In \bibinfo{booktitle}{\emph{2016 6th Workshop on Irregular
  Applications: Architecture and Algorithms (IA3)}}. IEEE,
  \bibinfo{pages}{50--53}.
\newblock


\bibitem[\protect\citeauthoryear{Andoni, Indyk, Laarhoven, Razenshteyn, and
  Schmidt}{Andoni et~al\mbox{.}}{2015}]%
        {simSearch-lib-falconn-basedOn-lsh}
\bibfield{author}{\bibinfo{person}{Alexandr Andoni}, \bibinfo{person}{Piotr
  Indyk}, \bibinfo{person}{Thijs Laarhoven}, \bibinfo{person}{Ilya
  Razenshteyn}, {and} \bibinfo{person}{Ludwig Schmidt}.}
  \bibinfo{year}{2015}\natexlab{}.
\newblock \bibinfo{title}{Practical and Optimal LSH for Angular Distance}.
\newblock
\newblock
\showeprint[arxiv]{1509.02897}~[cs.DS]


\bibitem[\protect\citeauthoryear{Andrade, Fernandes, Gomes, Ferreira, and
  Teodoro}{Andrade et~al\mbox{.}}{2019}]%
        {KNN-join-parallel-PQ-ANDRADE201981}
\bibfield{author}{\bibinfo{person}{Guilherme Andrade}, \bibinfo{person}{André
  Fernandes}, \bibinfo{person}{Jeremias~M. Gomes}, \bibinfo{person}{Renato
  Ferreira}, {and} \bibinfo{person}{George Teodoro}.}
  \bibinfo{year}{2019}\natexlab{}.
\newblock \showarticletitle{Large-scale parallel similarity search with Product
  Quantization for online multimedia services}.
\newblock \bibinfo{journal}{\emph{J. Parallel and Distrib. Comput.}}
  \bibinfo{volume}{125} (\bibinfo{year}{2019}), \bibinfo{pages}{81--92}.
\newblock
\showISSN{0743-7315}
\urldef\tempurl%
\url{https://doi.org/10.1016/j.jpdc.2018.11.009}
\showDOI{\tempurl}


\bibitem[\protect\citeauthoryear{Arora and Chana}{Arora and Chana}{2014}]%
        {clustering-survey-arora2014survey}
\bibfield{author}{\bibinfo{person}{Saurabh Arora} {and}
  \bibinfo{person}{Inderveer Chana}.} \bibinfo{year}{2014}\natexlab{}.
\newblock \showarticletitle{A survey of clustering techniques for big data
  analysis}. In \bibinfo{booktitle}{\emph{2014 5th International
  Conference-Confluence The Next Generation Information Technology Summit
  (Confluence)}}. IEEE, \bibinfo{pages}{59--65}.
\newblock


\bibitem[\protect\citeauthoryear{Aum{\"u}ller, Christiani, Pagh, and
  Vesterli}{Aum{\"u}ller et~al\mbox{.}}{2019}]%
        {LSH-parameterless-PUFFINN-aumuller2019}
\bibfield{author}{\bibinfo{person}{Martin Aum{\"u}ller},
  \bibinfo{person}{Tobias Christiani}, \bibinfo{person}{Rasmus Pagh}, {and}
  \bibinfo{person}{Michael Vesterli}.} \bibinfo{year}{2019}\natexlab{}.
\newblock \showarticletitle{PUFFINN: parameterless and universally fast finding
  of nearest neighbors}.
\newblock \bibinfo{journal}{\emph{arXiv preprint arXiv:1906.12211}}
  (\bibinfo{year}{2019}).
\newblock


\bibitem[\protect\citeauthoryear{Aurenhammer}{Aurenhammer}{1991}]%
        {delaunay-graph-aurenhammer1991}
\bibfield{author}{\bibinfo{person}{Franz Aurenhammer}.}
  \bibinfo{year}{1991}\natexlab{}.
\newblock \showarticletitle{Voronoi diagrams—a survey of a fundamental
  geometric data structure}.
\newblock \bibinfo{journal}{\emph{ACM Computing Surveys (CSUR)}}
  \bibinfo{volume}{23}, \bibinfo{number}{3} (\bibinfo{year}{1991}),
  \bibinfo{pages}{345--405}.
\newblock


\bibitem[\protect\citeauthoryear{Babu, Chandra, and Gopal}{Babu
  et~al\mbox{.}}{2011}]%
        {clustering-survey-babu2011clustering}
\bibfield{author}{\bibinfo{person}{B~Hari Babu}, \bibinfo{person}{N~Subash
  Chandra}, {and} \bibinfo{person}{T~Venu Gopal}.}
  \bibinfo{year}{2011}\natexlab{}.
\newblock \showarticletitle{Clustering Algorithms For High Dimensional Data--A
  Survey Of Issues And Existing Approaches}.
\newblock \bibinfo{journal}{\emph{Special Issue of International Journal of
  Computer Science \& Informatics}} \bibinfo{volume}{2}, \bibinfo{number}{1}
  (\bibinfo{year}{2011}), \bibinfo{pages}{2}.
\newblock


\bibitem[\protect\citeauthoryear{Bahmani, Goel, and Shinde}{Bahmani
  et~al\mbox{.}}{2012a}]%
        {LSH-1}
\bibfield{author}{\bibinfo{person}{Bahman Bahmani}, \bibinfo{person}{Ashish
  Goel}, {and} \bibinfo{person}{Rajendra Shinde}.}
  \bibinfo{year}{2012}\natexlab{a}.
\newblock \showarticletitle{Efficient distributed locality sensitive hashing}.
  In \bibinfo{booktitle}{\emph{Proceedings of the 21st ACM international
  conference on Information and knowledge management}}.
  \bibinfo{pages}{2174--2178}.
\newblock


\bibitem[\protect\citeauthoryear{Bahmani, Goel, and Shinde}{Bahmani
  et~al\mbox{.}}{2012b}]%
        {distribuetd-lsh}
\bibfield{author}{\bibinfo{person}{Bahman Bahmani}, \bibinfo{person}{Ashish
  Goel}, {and} \bibinfo{person}{Rajendra Shinde}.}
  \bibinfo{year}{2012}\natexlab{b}.
\newblock \showarticletitle{Efficient Distributed Locality Sensitive Hashing}.
  In \bibinfo{booktitle}{\emph{Proceedings of the 21st ACM International
  Conference on Information and Knowledge Management}} (Maui, Hawaii, USA)
  \emph{(\bibinfo{series}{CIKM '12})}. \bibinfo{publisher}{Association for
  Computing Machinery}, \bibinfo{address}{New York, NY, USA},
  \bibinfo{pages}{2174–2178}.
\newblock
\showISBNx{9781450311564}
\urldef\tempurl%
\url{https://doi.org/10.1145/2396761.2398596}
\showDOI{\tempurl}


\bibitem[\protect\citeauthoryear{Barrientos, G{\'o}mez, Tenllado, Matias, and
  Marin}{Barrientos et~al\mbox{.}}{2012}]%
        {Heterogeneous-hardware-3}
\bibfield{author}{\bibinfo{person}{Ricardo~J Barrientos},
  \bibinfo{person}{Jos{\'e}~I G{\'o}mez}, \bibinfo{person}{Christian Tenllado},
  \bibinfo{person}{Manuel~Prieto Matias}, {and} \bibinfo{person}{Mauricio
  Marin}.} \bibinfo{year}{2012}\natexlab{}.
\newblock \showarticletitle{Range query processing in a multi-GPU environment}.
  In \bibinfo{booktitle}{\emph{2012 IEEE 10th International Symposium on
  Parallel and Distributed Processing with Applications}}. IEEE,
  \bibinfo{pages}{419--426}.
\newblock


\bibitem[\protect\citeauthoryear{Bartolini, Ciaccia, and Waas}{Bartolini
  et~al\mbox{.}}{2001}]%
        {general-similarity-query-multimedia-1}
\bibfield{author}{\bibinfo{person}{Ilaria Bartolini}, \bibinfo{person}{Paolo
  Ciaccia}, {and} \bibinfo{person}{Florian Waas}.}
  \bibinfo{year}{2001}\natexlab{}.
\newblock \showarticletitle{FeedbackBypass: A new approach to interactive
  similarity query processing}. In \bibinfo{booktitle}{\emph{VLDB}}.
  \bibinfo{pages}{201--210}.
\newblock


\bibitem[\protect\citeauthoryear{Beecks and Berrendorf}{Beecks and
  Berrendorf}{2018}]%
        {general-similarity-query-multimedia-5}
\bibfield{author}{\bibinfo{person}{Christian Beecks} {and} \bibinfo{person}{Max
  Berrendorf}.} \bibinfo{year}{2018}\natexlab{}.
\newblock \showarticletitle{Optimal k-nearest-neighbor query processing via
  multiple lower bound approximations}. In \bibinfo{booktitle}{\emph{2018 IEEE
  International Conference on Big Data (Big Data)}}. IEEE,
  \bibinfo{pages}{614--623}.
\newblock


\bibitem[\protect\citeauthoryear{Bhatia et~al\mbox{.}}{Bhatia
  et~al\mbox{.}}{2010}]%
        {Similarity-search-survey-1}
\bibfield{author}{\bibinfo{person}{Nitin Bhatia} {et~al\mbox{.}}}
  \bibinfo{year}{2010}\natexlab{}.
\newblock \showarticletitle{Survey of nearest neighbor techniques}.
\newblock \bibinfo{journal}{\emph{arXiv preprint arXiv:1007.0085}}
  (\bibinfo{year}{2010}).
\newblock


\bibitem[\protect\citeauthoryear{B{\"o}hm}{B{\"o}hm}{2000a}]%
        {general-similarity-query-multimedia-4}
\bibfield{author}{\bibinfo{person}{Christian B{\"o}hm}.}
  \bibinfo{year}{2000}\natexlab{a}.
\newblock \showarticletitle{A cost model for query processing in high
  dimensional data spaces}.
\newblock \bibinfo{journal}{\emph{ACM Transactions on Database Systems (TODS)}}
  \bibinfo{volume}{25}, \bibinfo{number}{2} (\bibinfo{year}{2000}),
  \bibinfo{pages}{129--178}.
\newblock


\bibitem[\protect\citeauthoryear{B{\"o}hm}{B{\"o}hm}{2000b}]%
        {range-search-using-tree-bohm2000cost}
\bibfield{author}{\bibinfo{person}{Christian B{\"o}hm}.}
  \bibinfo{year}{2000}\natexlab{b}.
\newblock \showarticletitle{A cost model for query processing in high
  dimensional data spaces}.
\newblock \bibinfo{journal}{\emph{ACM Transactions on Database Systems (TODS)}}
  \bibinfo{volume}{25}, \bibinfo{number}{2} (\bibinfo{year}{2000}),
  \bibinfo{pages}{129--178}.
\newblock


\bibitem[\protect\citeauthoryear{B{\"o}hm, Berchtold, and Keim}{B{\"o}hm
  et~al\mbox{.}}{2001}]%
        {range-search-using-tree-bohm2001searching}
\bibfield{author}{\bibinfo{person}{Christian B{\"o}hm}, \bibinfo{person}{Stefan
  Berchtold}, {and} \bibinfo{person}{Daniel~A Keim}.}
  \bibinfo{year}{2001}\natexlab{}.
\newblock \showarticletitle{Searching in high-dimensional spaces: Index
  structures for improving the performance of multimedia databases}.
\newblock \bibinfo{journal}{\emph{ACM Computing Surveys (CSUR)}}
  \bibinfo{volume}{33}, \bibinfo{number}{3} (\bibinfo{year}{2001}),
  \bibinfo{pages}{322--373}.
\newblock


\bibitem[\protect\citeauthoryear{Bronstein, Bronstein, Michel, and
  Paragios}{Bronstein et~al\mbox{.}}{2010}]%
        {IR-traditional-joint-2}
\bibfield{author}{\bibinfo{person}{Michael~M. Bronstein},
  \bibinfo{person}{Alexander~M. Bronstein}, \bibinfo{person}{Fabrice Michel},
  {and} \bibinfo{person}{Nikos Paragios}.} \bibinfo{year}{2010}\natexlab{}.
\newblock \showarticletitle{Data fusion through cross-modality metric learning
  using similarity-sensitive hashing}. In \bibinfo{booktitle}{\emph{2010 IEEE
  Computer Society Conference on Computer Vision and Pattern Recognition}}.
  \bibinfo{pages}{3594--3601}.
\newblock
\urldef\tempurl%
\url{https://doi.org/10.1109/CVPR.2010.5539928}
\showDOI{\tempurl}


\bibitem[\protect\citeauthoryear{Cao, Long, Wang, Yang, and Yu}{Cao
  et~al\mbox{.}}{2016a}]%
        {deep-l2h-cao2016-deep-visual-semantic-hash}
\bibfield{author}{\bibinfo{person}{Yue Cao}, \bibinfo{person}{Mingsheng Long},
  \bibinfo{person}{Jianmin Wang}, \bibinfo{person}{Qiang Yang}, {and}
  \bibinfo{person}{Philip~S Yu}.} \bibinfo{year}{2016}\natexlab{a}.
\newblock \showarticletitle{Deep visual-semantic hashing for cross-modal
  retrieval}. In \bibinfo{booktitle}{\emph{Proceedings of the 22nd ACM SIGKDD
  International Conference on Knowledge Discovery and Data Mining}}.
  \bibinfo{pages}{1445--1454}.
\newblock


\bibitem[\protect\citeauthoryear{Cao, Long, Wang, Zhu, and Wen}{Cao
  et~al\mbox{.}}{2016b}]%
        {deep-pq-deep-quantization-net-2016}
\bibfield{author}{\bibinfo{person}{Yue Cao}, \bibinfo{person}{Mingsheng Long},
  \bibinfo{person}{Jianmin Wang}, \bibinfo{person}{Han Zhu}, {and}
  \bibinfo{person}{Qingfu Wen}.} \bibinfo{year}{2016}\natexlab{b}.
\newblock \bibinfo{title}{Deep Quantization Network for Efficient Image
  Retrieval}.
\newblock
\newblock
\urldef\tempurl%
\url{https://www.aaai.org/ocs/index.php/AAAI/AAAI16/paper/view/12040}
\showURL{%
\tempurl}


\bibitem[\protect\citeauthoryear{Cao, Long, Wang, and Yu}{Cao
  et~al\mbox{.}}{2017}]%
        {deep-l2h-cao2017hashnet}
\bibfield{author}{\bibinfo{person}{Zhangjie Cao}, \bibinfo{person}{Mingsheng
  Long}, \bibinfo{person}{Jianmin Wang}, {and} \bibinfo{person}{Philip~S Yu}.}
  \bibinfo{year}{2017}\natexlab{}.
\newblock \showarticletitle{Hashnet: Deep learning to hash by continuation}. In
  \bibinfo{booktitle}{\emph{Proceedings of the IEEE international conference on
  computer vision}}. \bibinfo{pages}{5608--5617}.
\newblock


\bibitem[\protect\citeauthoryear{Cappuzzo, Papotti, and
  Thirumuruganathan}{Cappuzzo et~al\mbox{.}}{2020}]%
        {ER-embedding-3}
\bibfield{author}{\bibinfo{person}{Riccardo Cappuzzo}, \bibinfo{person}{Paolo
  Papotti}, {and} \bibinfo{person}{Saravanan Thirumuruganathan}.}
  \bibinfo{year}{2020}\natexlab{}.
\newblock \showarticletitle{Creating embeddings of heterogeneous relational
  datasets for data integration tasks}. In
  \bibinfo{booktitle}{\emph{Proceedings of the 2020 ACM SIGMOD International
  Conference on Management of Data}}. \bibinfo{pages}{1335--1349}.
\newblock


\bibitem[\protect\citeauthoryear{{\v{C}}ech, Loko{\v{c}}, and Silva}{{\v{C}}ech
  et~al\mbox{.}}{2020}]%
        {KNN-join-pivot-based-Voronoi-cells-2020}
\bibfield{author}{\bibinfo{person}{P{\v{r}}emysl {\v{C}}ech},
  \bibinfo{person}{Jakub Loko{\v{c}}}, {and} \bibinfo{person}{Yasin~N Silva}.}
  \bibinfo{year}{2020}\natexlab{}.
\newblock \showarticletitle{Pivot-based approximate k-NN similarity joins for
  big high-dimensional data}.
\newblock \bibinfo{journal}{\emph{Information Systems}}  \bibinfo{volume}{87}
  (\bibinfo{year}{2020}), \bibinfo{pages}{101410}.
\newblock


\bibitem[\protect\citeauthoryear{Charikar}{Charikar}{2002}]%
        {random-projection-lsh}
\bibfield{author}{\bibinfo{person}{Moses~S Charikar}.}
  \bibinfo{year}{2002}\natexlab{}.
\newblock \showarticletitle{Similarity estimation techniques from rounding
  algorithms}. In \bibinfo{booktitle}{\emph{Proceedings of the thirty-fourth
  annual ACM symposium on Theory of computing}}. \bibinfo{pages}{380--388}.
\newblock


\bibitem[\protect\citeauthoryear{Chen, Chen, Zou, Li, Lu, and Zhao}{Chen
  et~al\mbox{.}}{2019}]%
        {gpu-PQ-RobustiQ-2019}
\bibfield{author}{\bibinfo{person}{Wei Chen}, \bibinfo{person}{Jincai Chen},
  \bibinfo{person}{Fuhao Zou}, \bibinfo{person}{Yuan-Fang Li},
  \bibinfo{person}{Ping Lu}, {and} \bibinfo{person}{Wei Zhao}.}
  \bibinfo{year}{2019}\natexlab{}.
\newblock \showarticletitle{RobustiQ: A Robust ANN Search Method for
  Billion-Scale Similarity Search on GPUs}. In
  \bibinfo{booktitle}{\emph{Proceedings of the 2019 on International Conference
  on Multimedia Retrieval}} (Ottawa ON, Canada) \emph{(\bibinfo{series}{ICMR
  '19})}. \bibinfo{publisher}{Association for Computing Machinery},
  \bibinfo{address}{New York, NY, USA}, \bibinfo{pages}{132–140}.
\newblock
\showISBNx{9781450367653}
\urldef\tempurl%
\url{https://doi.org/10.1145/3323873.3325018}
\showDOI{\tempurl}


\bibitem[\protect\citeauthoryear{Danopoulos, Kachris, and Soudris}{Danopoulos
  et~al\mbox{.}}{2019}]%
        {fpga-faiss-Danopoulos2019FPGAAO}
\bibfield{author}{\bibinfo{person}{Dimitrios Danopoulos},
  \bibinfo{person}{Christoforos Kachris}, {and} \bibinfo{person}{Dimitrios
  Soudris}.} \bibinfo{year}{2019}\natexlab{}.
\newblock \showarticletitle{FPGA Acceleration of Approximate KNN Indexing on
  High- Dimensional Vectors}.
\newblock \bibinfo{journal}{\emph{2019 14th International Symposium on
  Reconfigurable Communication-centric Systems-on-Chip (ReCoSoC)}}
  (\bibinfo{year}{2019}), \bibinfo{pages}{59--65}.
\newblock


\bibitem[\protect\citeauthoryear{Dasgupta and Freund}{Dasgupta and
  Freund}{2008}]%
        {Tree-based-index-random-projection-tree-dasgupta2008}
\bibfield{author}{\bibinfo{person}{Sanjoy Dasgupta} {and} \bibinfo{person}{Yoav
  Freund}.} \bibinfo{year}{2008}\natexlab{}.
\newblock \showarticletitle{Random projection trees and low dimensional
  manifolds}. In \bibinfo{booktitle}{\emph{Proceedings of the fortieth annual
  ACM symposium on Theory of computing}}. \bibinfo{pages}{537--546}.
\newblock


\bibitem[\protect\citeauthoryear{Dasgupta and Sinha}{Dasgupta and
  Sinha}{2013}]%
        {Tree-based-index-randomized-partition-trees-2013}
\bibfield{author}{\bibinfo{person}{Sanjoy Dasgupta} {and}
  \bibinfo{person}{Kaushik Sinha}.} \bibinfo{year}{2013}\natexlab{}.
\newblock \showarticletitle{Randomized partition trees for exact nearest
  neighbor search}. In \bibinfo{booktitle}{\emph{Conference on Learning
  Theory}}. PMLR, \bibinfo{pages}{317--337}.
\newblock


\bibitem[\protect\citeauthoryear{Datar, Immorlica, Indyk, and Mirrokni}{Datar
  et~al\mbox{.}}{2004}]%
        {p-stable-lsh-datar2004locality}
\bibfield{author}{\bibinfo{person}{Mayur Datar}, \bibinfo{person}{Nicole
  Immorlica}, \bibinfo{person}{Piotr Indyk}, {and} \bibinfo{person}{Vahab~S
  Mirrokni}.} \bibinfo{year}{2004}\natexlab{}.
\newblock \showarticletitle{Locality-sensitive hashing scheme based on p-stable
  distributions}. In \bibinfo{booktitle}{\emph{Proceedings of the twentieth
  annual symposium on Computational geometry}}. \bibinfo{pages}{253--262}.
\newblock


\bibitem[\protect\citeauthoryear{Devlin, Chang, Lee, and Toutanova}{Devlin
  et~al\mbox{.}}{2018a}]%
        {embedding-text-3}
\bibfield{author}{\bibinfo{person}{Jacob Devlin}, \bibinfo{person}{Ming-Wei
  Chang}, \bibinfo{person}{Kenton Lee}, {and} \bibinfo{person}{Kristina
  Toutanova}.} \bibinfo{year}{2018}\natexlab{a}.
\newblock \showarticletitle{Bert: Pre-training of deep bidirectional
  transformers for language understanding}.
\newblock \bibinfo{journal}{\emph{arXiv preprint arXiv:1810.04805}}
  (\bibinfo{year}{2018}).
\newblock


\bibitem[\protect\citeauthoryear{Devlin, Chang, Lee, and Toutanova}{Devlin
  et~al\mbox{.}}{2018b}]%
        {bert}
\bibfield{author}{\bibinfo{person}{Jacob Devlin}, \bibinfo{person}{Ming-Wei
  Chang}, \bibinfo{person}{Kenton Lee}, {and} \bibinfo{person}{Kristina
  Toutanova}.} \bibinfo{year}{2018}\natexlab{b}.
\newblock \bibinfo{title}{BERT: Pre-training of Deep Bidirectional Transformers
  for Language Understanding}.
\newblock
\newblock
\showeprint[arxiv]{1810.04805}~[cs.CL]


\bibitem[\protect\citeauthoryear{Dong, Moses, and Li}{Dong
  et~al\mbox{.}}{2011}]%
        {Graph-based-index-NN-Descent-2011}
\bibfield{author}{\bibinfo{person}{Wei Dong}, \bibinfo{person}{Charikar Moses},
  {and} \bibinfo{person}{Kai Li}.} \bibinfo{year}{2011}\natexlab{}.
\newblock \showarticletitle{Efficient k-nearest neighbor graph construction for
  generic similarity measures}. In \bibinfo{booktitle}{\emph{Proceedings of the
  20th international conference on World wide web}}. \bibinfo{pages}{577--586}.
\newblock


\bibitem[\protect\citeauthoryear{Dong, Indyk, Razenshteyn, and Wagner}{Dong
  et~al\mbox{.}}{2019}]%
        {Graph-based-index-7}
\bibfield{author}{\bibinfo{person}{Yihe Dong}, \bibinfo{person}{Piotr Indyk},
  \bibinfo{person}{Ilya Razenshteyn}, {and} \bibinfo{person}{Tal Wagner}.}
  \bibinfo{year}{2019}\natexlab{}.
\newblock \showarticletitle{Learning space partitions for nearest neighbor
  search}.
\newblock \bibinfo{journal}{\emph{arXiv preprint arXiv:1901.08544}}
  (\bibinfo{year}{2019}).
\newblock


\bibitem[\protect\citeauthoryear{Ebraheem, Thirumuruganathan, Joty, Ouzzani,
  and Tang}{Ebraheem et~al\mbox{.}}{2018}]%
        {ER-embedding-2}
\bibfield{author}{\bibinfo{person}{Muhammad Ebraheem},
  \bibinfo{person}{Saravanan Thirumuruganathan}, \bibinfo{person}{Shafiq Joty},
  \bibinfo{person}{Mourad Ouzzani}, {and} \bibinfo{person}{Nan Tang}.}
  \bibinfo{year}{2018}\natexlab{}.
\newblock \showarticletitle{Distributed representations of tuples for entity
  resolution}.
\newblock \bibinfo{journal}{\emph{Proceedings of the VLDB Endowment}}
  \bibinfo{volume}{11}, \bibinfo{number}{11} (\bibinfo{year}{2018}),
  \bibinfo{pages}{1454--1467}.
\newblock


\bibitem[\protect\citeauthoryear{Echihabi, Zoumpatianos, and Palpanas}{Echihabi
  et~al\mbox{.}}{2021}]%
        {Similarity-search-survey-3}
\bibfield{author}{\bibinfo{person}{Karima Echihabi}, \bibinfo{person}{Kostas
  Zoumpatianos}, {and} \bibinfo{person}{Themis Palpanas}.}
  \bibinfo{year}{2021}\natexlab{}.
\newblock \showarticletitle{High-Dimensional Similarity Search for Scalable
  Data Science}. In \bibinfo{booktitle}{\emph{2021 IEEE 37th International
  Conference on Data Engineering (ICDE)}}. IEEE, \bibinfo{pages}{2369--2372}.
\newblock


\bibitem[\protect\citeauthoryear{Eghbali, Ashtiani, and Tahvildari}{Eghbali
  et~al\mbox{.}}{2017}]%
        {Tree-based-index-7}
\bibfield{author}{\bibinfo{person}{Sepehr Eghbali}, \bibinfo{person}{Hassan
  Ashtiani}, {and} \bibinfo{person}{Ladan Tahvildari}.}
  \bibinfo{year}{2017}\natexlab{}.
\newblock \showarticletitle{Online nearest neighbor search in binary space}. In
  \bibinfo{booktitle}{\emph{2017 IEEE International Conference on Data Mining
  (ICDM)}}. IEEE, \bibinfo{pages}{853--858}.
\newblock


\bibitem[\protect\citeauthoryear{Erin~Liong, Lu, Wang, Moulin, and
  Zhou}{Erin~Liong et~al\mbox{.}}{2015}]%
        {deep-l2h-erin2015-deep-hashing-supervised-DH}
\bibfield{author}{\bibinfo{person}{Venice Erin~Liong}, \bibinfo{person}{Jiwen
  Lu}, \bibinfo{person}{Gang Wang}, \bibinfo{person}{Pierre Moulin}, {and}
  \bibinfo{person}{Jie Zhou}.} \bibinfo{year}{2015}\natexlab{}.
\newblock \showarticletitle{Deep hashing for compact binary codes learning}. In
  \bibinfo{booktitle}{\emph{Proceedings of the IEEE conference on computer
  vision and pattern recognition}}. \bibinfo{pages}{2475--2483}.
\newblock


\bibitem[\protect\citeauthoryear{Fan, Kong, Zhang, Liu, Pan, and Lu}{Fan
  et~al\mbox{.}}{2020}]%
        {Tree-based-index-6}
\bibfield{author}{\bibinfo{person}{Bin Fan}, \bibinfo{person}{Qingqun Kong},
  \bibinfo{person}{Baoqian Zhang}, \bibinfo{person}{Hongmin Liu},
  \bibinfo{person}{Chunhong Pan}, {and} \bibinfo{person}{Jiwen Lu}.}
  \bibinfo{year}{2020}\natexlab{}.
\newblock \showarticletitle{Efficient nearest neighbor search in high
  dimensional hamming space}.
\newblock \bibinfo{journal}{\emph{Pattern Recognition}}  \bibinfo{volume}{99}
  (\bibinfo{year}{2020}), \bibinfo{pages}{107082}.
\newblock


\bibitem[\protect\citeauthoryear{Frome, Corrado, Shlens, Bengio, Dean, Ranzato,
  and Mikolov}{Frome et~al\mbox{.}}{2013}]%
        {embedding-image-2}
\bibfield{author}{\bibinfo{person}{Andrea Frome}, \bibinfo{person}{Greg
  Corrado}, \bibinfo{person}{Jonathon Shlens}, \bibinfo{person}{Samy Bengio},
  \bibinfo{person}{Jeffrey Dean}, \bibinfo{person}{Marc’Aurelio Ranzato},
  {and} \bibinfo{person}{Tomas Mikolov}.} \bibinfo{year}{2013}\natexlab{}.
\newblock \showarticletitle{Devise: A deep visual-semantic embedding model}.
\newblock  (\bibinfo{year}{2013}).
\newblock


\bibitem[\protect\citeauthoryear{Fu and Cai}{Fu and Cai}{2016}]%
        {Graph-based-index-Efanna-2016}
\bibfield{author}{\bibinfo{person}{Cong Fu} {and} \bibinfo{person}{Deng Cai}.}
  \bibinfo{year}{2016}\natexlab{}.
\newblock \showarticletitle{Efanna: An extremely fast approximate nearest
  neighbor search algorithm based on knn graph}.
\newblock \bibinfo{journal}{\emph{arXiv preprint arXiv:1609.07228}}
  (\bibinfo{year}{2016}).
\newblock


\bibitem[\protect\citeauthoryear{Fu, Xiang, Wang, and Cai}{Fu
  et~al\mbox{.}}{2017}]%
        {Graph-based-index-NSG-2017}
\bibfield{author}{\bibinfo{person}{Cong Fu}, \bibinfo{person}{Chao Xiang},
  \bibinfo{person}{Changxu Wang}, {and} \bibinfo{person}{Deng Cai}.}
  \bibinfo{year}{2017}\natexlab{}.
\newblock \showarticletitle{Fast approximate nearest neighbor search with the
  navigating spreading-out graph}.
\newblock \bibinfo{journal}{\emph{arXiv preprint arXiv:1707.00143}}
  (\bibinfo{year}{2017}).
\newblock


\bibitem[\protect\citeauthoryear{Gao, Jagadish, Ooi, and Wang}{Gao
  et~al\mbox{.}}{2015}]%
        {range-search-using-hashing-gao2015selective}
\bibfield{author}{\bibinfo{person}{Jinyang Gao}, \bibinfo{person}{HV Jagadish},
  \bibinfo{person}{Beng~Chin Ooi}, {and} \bibinfo{person}{Sheng Wang}.}
  \bibinfo{year}{2015}\natexlab{}.
\newblock \showarticletitle{Selective hashing: Closing the gap between radius
  search and k-nn search}. In \bibinfo{booktitle}{\emph{Proceedings of the 21th
  ACM SIGKDD International Conference on Knowledge Discovery and Data Mining}}.
  \bibinfo{pages}{349--358}.
\newblock


\bibitem[\protect\citeauthoryear{Gao, Zhu, Song, Zhao, and Shen}{Gao
  et~al\mbox{.}}{2019}]%
        {deep-pq-progressive-quantization-gao2019}
\bibfield{author}{\bibinfo{person}{Lianli Gao}, \bibinfo{person}{Xiaosu Zhu},
  \bibinfo{person}{Jingkuan Song}, \bibinfo{person}{Zhou Zhao}, {and}
  \bibinfo{person}{Heng~Tao Shen}.} \bibinfo{year}{2019}\natexlab{}.
\newblock \showarticletitle{Beyond product quantization: Deep progressive
  quantization for image retrieval}.
\newblock \bibinfo{journal}{\emph{arXiv preprint arXiv:1906.06698}}
  (\bibinfo{year}{2019}).
\newblock


\bibitem[\protect\citeauthoryear{Gattupalli, Zhuo, and Li}{Gattupalli
  et~al\mbox{.}}{2019}]%
        {deep-l2h-gattupalli2019-weakly-supervised}
\bibfield{author}{\bibinfo{person}{Vijetha Gattupalli}, \bibinfo{person}{Yaoxin
  Zhuo}, {and} \bibinfo{person}{Baoxin Li}.} \bibinfo{year}{2019}\natexlab{}.
\newblock \showarticletitle{Weakly supervised deep image hashing through tag
  embeddings}. In \bibinfo{booktitle}{\emph{Proceedings of the IEEE/CVF
  Conference on Computer Vision and Pattern Recognition}}.
  \bibinfo{pages}{10375--10384}.
\newblock


\bibitem[\protect\citeauthoryear{Ge, He, Ke, and Sun}{Ge
  et~al\mbox{.}}{2013a}]%
        {optimized-pq-kaiming2013}
\bibfield{author}{\bibinfo{person}{Tiezheng Ge}, \bibinfo{person}{Kaiming He},
  \bibinfo{person}{Qifa Ke}, {and} \bibinfo{person}{Jian Sun}.}
  \bibinfo{year}{2013}\natexlab{a}.
\newblock \showarticletitle{Optimized product quantization}.
\newblock \bibinfo{journal}{\emph{IEEE transactions on pattern analysis and
  machine intelligence}} \bibinfo{volume}{36}, \bibinfo{number}{4}
  (\bibinfo{year}{2013}), \bibinfo{pages}{744--755}.
\newblock


\bibitem[\protect\citeauthoryear{Ge, He, Ke, and Sun}{Ge
  et~al\mbox{.}}{2013b}]%
        {optimized-pq-kaiming2013-2}
\bibfield{author}{\bibinfo{person}{Tiezheng Ge}, \bibinfo{person}{Kaiming He},
  \bibinfo{person}{Qifa Ke}, {and} \bibinfo{person}{Jian Sun}.}
  \bibinfo{year}{2013}\natexlab{b}.
\newblock \showarticletitle{Optimized product quantization for approximate
  nearest neighbor search}. In \bibinfo{booktitle}{\emph{Proceedings of the
  IEEE Conference on Computer Vision and Pattern Recognition}}.
  \bibinfo{pages}{2946--2953}.
\newblock


\bibitem[\protect\citeauthoryear{Gionis, Indyk, Motwani, et~al\mbox{.}}{Gionis
  et~al\mbox{.}}{1999}]%
        {Similarity-search-hashing-based-1}
\bibfield{author}{\bibinfo{person}{Aristides Gionis}, \bibinfo{person}{Piotr
  Indyk}, \bibinfo{person}{Rajeev Motwani}, {et~al\mbox{.}}}
  \bibinfo{year}{1999}\natexlab{}.
\newblock \showarticletitle{Similarity search in high dimensions via hashing}.
  In \bibinfo{booktitle}{\emph{Vldb}}, Vol.~\bibinfo{volume}{99}.
  \bibinfo{pages}{518--529}.
\newblock


\bibitem[\protect\citeauthoryear{Gong, Kumar, Verma, and Lazebnik}{Gong
  et~al\mbox{.}}{2012a}]%
        {l2h-multidim-features-unsupervised-gong2012-angular-quantization}
\bibfield{author}{\bibinfo{person}{Yunchao Gong}, \bibinfo{person}{Sanjiv
  Kumar}, \bibinfo{person}{Vishal Verma}, {and} \bibinfo{person}{Svetlana
  Lazebnik}.} \bibinfo{year}{2012}\natexlab{a}.
\newblock \showarticletitle{Angular quantization-based binary codes for fast
  similarity search}.
\newblock  (\bibinfo{year}{2012}).
\newblock


\bibitem[\protect\citeauthoryear{Gong, Lazebnik, Gordo, and Perronnin}{Gong
  et~al\mbox{.}}{2012b}]%
        {l2h-multidim-features-supervised-gong2012-iterative-quantization}
\bibfield{author}{\bibinfo{person}{Yunchao Gong}, \bibinfo{person}{Svetlana
  Lazebnik}, \bibinfo{person}{Albert Gordo}, {and} \bibinfo{person}{Florent
  Perronnin}.} \bibinfo{year}{2012}\natexlab{b}.
\newblock \showarticletitle{Iterative quantization: A procrustean approach to
  learning binary codes for large-scale image retrieval}.
\newblock \bibinfo{journal}{\emph{IEEE transactions on pattern analysis and
  machine intelligence}} \bibinfo{volume}{35}, \bibinfo{number}{12}
  (\bibinfo{year}{2012}), \bibinfo{pages}{2916--2929}.
\newblock


\bibitem[\protect\citeauthoryear{Gowanlock and Karsin}{Gowanlock and
  Karsin}{2019a}]%
        {Heterogeneous-hardware-2}
\bibfield{author}{\bibinfo{person}{Michael Gowanlock} {and}
  \bibinfo{person}{Ben Karsin}.} \bibinfo{year}{2019}\natexlab{a}.
\newblock \showarticletitle{Accelerating the similarity self-join using the
  GPU}.
\newblock \bibinfo{journal}{\emph{Journal of parallel and distributed
  computing}}  \bibinfo{volume}{133} (\bibinfo{year}{2019}),
  \bibinfo{pages}{107--123}.
\newblock


\bibitem[\protect\citeauthoryear{Gowanlock and Karsin}{Gowanlock and
  Karsin}{2019b}]%
        {Heterogeneous-hardware-1}
\bibfield{author}{\bibinfo{person}{Michael Gowanlock} {and}
  \bibinfo{person}{Ben Karsin}.} \bibinfo{year}{2019}\natexlab{b}.
\newblock \showarticletitle{GPU-accelerated similarity self-join for
  multi-dimensional data}. In \bibinfo{booktitle}{\emph{Proceedings of the 15th
  International Workshop on Data Management on New Hardware}}.
  \bibinfo{pages}{1--9}.
\newblock


\bibitem[\protect\citeauthoryear{Gowanlock and Karsin}{Gowanlock and
  Karsin}{2019c}]%
        {grid-based-index-gpu}
\bibfield{author}{\bibinfo{person}{Michael Gowanlock} {and}
  \bibinfo{person}{Ben Karsin}.} \bibinfo{year}{2019}\natexlab{c}.
\newblock \showarticletitle{GPU-Accelerated Similarity Self-Join for
  Multi-Dimensional Data}. In \bibinfo{booktitle}{\emph{Proceedings of the 15th
  International Workshop on Data Management on New Hardware}} (Amsterdam,
  Netherlands) \emph{(\bibinfo{series}{DaMoN'19})}.
  \bibinfo{publisher}{Association for Computing Machinery},
  \bibinfo{address}{New York, NY, USA}, Article \bibinfo{articleno}{6},
  \bibinfo{numpages}{9}~pages.
\newblock
\showISBNx{9781450368018}
\urldef\tempurl%
\url{https://doi.org/10.1145/3329785.3329920}
\showDOI{\tempurl}


\bibitem[\protect\citeauthoryear{Goyal and Ferrara}{Goyal and Ferrara}{2018}]%
        {embedding-graph-3}
\bibfield{author}{\bibinfo{person}{Palash Goyal} {and} \bibinfo{person}{Emilio
  Ferrara}.} \bibinfo{year}{2018}\natexlab{}.
\newblock \showarticletitle{Graph embedding techniques, applications, and
  performance: A survey}.
\newblock \bibinfo{journal}{\emph{Knowledge-Based Systems}}
  \bibinfo{volume}{151} (\bibinfo{year}{2018}), \bibinfo{pages}{78--94}.
\newblock


\bibitem[\protect\citeauthoryear{Gray and Neuhoff}{Gray and Neuhoff}{1998}]%
        {Quantization-survey-gray1998}
\bibfield{author}{\bibinfo{person}{Robert~M. Gray} {and}
  \bibinfo{person}{David~L. Neuhoff}.} \bibinfo{year}{1998}\natexlab{}.
\newblock \showarticletitle{Quantization}.
\newblock \bibinfo{journal}{\emph{IEEE transactions on information theory}}
  \bibinfo{volume}{44}, \bibinfo{number}{6} (\bibinfo{year}{1998}),
  \bibinfo{pages}{2325--2383}.
\newblock


\bibitem[\protect\citeauthoryear{Grover and Leskovec}{Grover and
  Leskovec}{2016}]%
        {embedding-graph-2}
\bibfield{author}{\bibinfo{person}{Aditya Grover} {and} \bibinfo{person}{Jure
  Leskovec}.} \bibinfo{year}{2016}\natexlab{}.
\newblock \showarticletitle{node2vec: Scalable feature learning for networks}.
  In \bibinfo{booktitle}{\emph{Proceedings of the 22nd ACM SIGKDD international
  conference on Knowledge discovery and data mining}}.
  \bibinfo{pages}{855--864}.
\newblock


\bibitem[\protect\citeauthoryear{G{\"u}nther, Thiele, and Lehner}{G{\"u}nther
  et~al\mbox{.}}{2019}]%
        {KNN-join-PQ-multi-index-based-on-word2vec-gunther2019fast}
\bibfield{author}{\bibinfo{person}{Michael G{\"u}nther}, \bibinfo{person}{Maik
  Thiele}, {and} \bibinfo{person}{Wolfgang Lehner}.}
  \bibinfo{year}{2019}\natexlab{}.
\newblock \showarticletitle{Fast Approximated Nearest Neighbor Joins For
  Relational Database Systems}.
\newblock \bibinfo{journal}{\emph{BTW 2019}} (\bibinfo{year}{2019}).
\newblock


\bibitem[\protect\citeauthoryear{Guo, Sun, Lindgren, Geng, Simcha, Chern, and
  Kumar}{Guo et~al\mbox{.}}{2020}]%
        {simSearch-lib-google-scann-basedOn-pq}
\bibfield{author}{\bibinfo{person}{Ruiqi Guo}, \bibinfo{person}{Philip Sun},
  \bibinfo{person}{Erik Lindgren}, \bibinfo{person}{Quan Geng},
  \bibinfo{person}{David Simcha}, \bibinfo{person}{Felix Chern}, {and}
  \bibinfo{person}{Sanjiv Kumar}.} \bibinfo{year}{2020}\natexlab{}.
\newblock \showarticletitle{Accelerating large-scale inference with anisotropic
  vector quantization}. In \bibinfo{booktitle}{\emph{International Conference
  on Machine Learning}}. PMLR, \bibinfo{pages}{3887--3896}.
\newblock


\bibitem[\protect\citeauthoryear{Haghiri, Ghoshdastidar, and von
  Luxburg}{Haghiri et~al\mbox{.}}{2017}]%
        {Tree-based-index-comparison-tree-2017}
\bibfield{author}{\bibinfo{person}{Siavash Haghiri}, \bibinfo{person}{Debarghya
  Ghoshdastidar}, {and} \bibinfo{person}{Ulrike von Luxburg}.}
  \bibinfo{year}{2017}\natexlab{}.
\newblock \showarticletitle{Comparison-based nearest neighbor search}. In
  \bibinfo{booktitle}{\emph{Artificial Intelligence and Statistics}}. PMLR,
  \bibinfo{pages}{851--859}.
\newblock


\bibitem[\protect\citeauthoryear{Hajebi, Abbasi-Yadkori, Shahbazi, and
  Zhang}{Hajebi et~al\mbox{.}}{2011}]%
        {Graph-based-index-GNNS-hajebi2011}
\bibfield{author}{\bibinfo{person}{Kiana Hajebi}, \bibinfo{person}{Yasin
  Abbasi-Yadkori}, \bibinfo{person}{Hossein Shahbazi}, {and}
  \bibinfo{person}{Hong Zhang}.} \bibinfo{year}{2011}\natexlab{}.
\newblock \showarticletitle{Fast approximate nearest-neighbor search with
  k-nearest neighbor graph}. In \bibinfo{booktitle}{\emph{Twenty-Second
  International Joint Conference on Artificial Intelligence}}.
\newblock


\bibitem[\protect\citeauthoryear{Hamilton, Ying, and Leskovec}{Hamilton
  et~al\mbox{.}}{2017}]%
        {embedding-graph-4}
\bibfield{author}{\bibinfo{person}{William~L Hamilton}, \bibinfo{person}{Rex
  Ying}, {and} \bibinfo{person}{Jure Leskovec}.}
  \bibinfo{year}{2017}\natexlab{}.
\newblock \showarticletitle{Representation learning on graphs: Methods and
  applications}.
\newblock \bibinfo{journal}{\emph{arXiv preprint arXiv:1709.05584}}
  (\bibinfo{year}{2017}).
\newblock


\bibitem[\protect\citeauthoryear{Han and Ma}{Han and Ma}{2002}]%
        {traditional-method-image-color-feature-1}
\bibfield{author}{\bibinfo{person}{Ju Han} {and} \bibinfo{person}{Kai-Kuang
  Ma}.} \bibinfo{year}{2002}\natexlab{}.
\newblock \showarticletitle{Fuzzy color histogram and its use in color image
  retrieval}.
\newblock \bibinfo{journal}{\emph{IEEE Transactions on Image Processing}}
  \bibinfo{volume}{11}, \bibinfo{number}{8} (\bibinfo{year}{2002}),
  \bibinfo{pages}{944--952}.
\newblock
\urldef\tempurl%
\url{https://doi.org/10.1109/TIP.2002.801585}
\showDOI{\tempurl}


\bibitem[\protect\citeauthoryear{Harwood and Drummond}{Harwood and
  Drummond}{2016}]%
        {Graph-based-index-fanng-2016}
\bibfield{author}{\bibinfo{person}{Ben Harwood} {and} \bibinfo{person}{Tom
  Drummond}.} \bibinfo{year}{2016}\natexlab{}.
\newblock \showarticletitle{Fanng: Fast approximate nearest neighbour graphs}.
  In \bibinfo{booktitle}{\emph{Proceedings of the IEEE Conference on Computer
  Vision and Pattern Recognition}}. \bibinfo{pages}{5713--5722}.
\newblock


\bibitem[\protect\citeauthoryear{He, Zhang, Ren, and Sun}{He
  et~al\mbox{.}}{2016}]%
        {resnet}
\bibfield{author}{\bibinfo{person}{Kaiming He}, \bibinfo{person}{X. Zhang},
  \bibinfo{person}{Shaoqing Ren}, {and} \bibinfo{person}{Jian Sun}.}
  \bibinfo{year}{2016}\natexlab{}.
\newblock \showarticletitle{Deep Residual Learning for Image Recognition}.
\newblock \bibinfo{journal}{\emph{2016 IEEE Conference on Computer Vision and
  Pattern Recognition (CVPR)}} (\bibinfo{year}{2016}),
  \bibinfo{pages}{770--778}.
\newblock


\bibitem[\protect\citeauthoryear{Hennequin, Khlif, Voituret, and
  Moussallam}{Hennequin et~al\mbox{.}}{2020}]%
        {example-pretrained-embedding-for-efficiency-1}
\bibfield{author}{\bibinfo{person}{Romain Hennequin}, \bibinfo{person}{Anis
  Khlif}, \bibinfo{person}{Felix Voituret}, {and} \bibinfo{person}{Manuel
  Moussallam}.} \bibinfo{year}{2020}\natexlab{}.
\newblock \showarticletitle{Spleeter: a fast and efficient music source
  separation tool with pre-trained models}.
\newblock \bibinfo{journal}{\emph{Journal of Open Source Software}}
  \bibinfo{volume}{5}, \bibinfo{number}{50} (\bibinfo{year}{2020}),
  \bibinfo{pages}{2154}.
\newblock


\bibitem[\protect\citeauthoryear{Heo, Lin, and Yoon}{Heo et~al\mbox{.}}{2014}]%
        {pq-distance-encoded-2014}
\bibfield{author}{\bibinfo{person}{Jae-Pil Heo}, \bibinfo{person}{Zhe Lin},
  {and} \bibinfo{person}{Sung-Eui Yoon}.} \bibinfo{year}{2014}\natexlab{}.
\newblock \showarticletitle{Distance encoded product quantization}. In
  \bibinfo{booktitle}{\emph{Proceedings of the IEEE Conference on Computer
  Vision and Pattern Recognition}}. \bibinfo{pages}{2131--2138}.
\newblock


\bibitem[\protect\citeauthoryear{Houle and Nett}{Houle and Nett}{2013}]%
        {Tree-based-index-5}
\bibfield{author}{\bibinfo{person}{Michael~E Houle} {and}
  \bibinfo{person}{Michael Nett}.} \bibinfo{year}{2013}\natexlab{}.
\newblock \showarticletitle{Rank cover trees for nearest neighbor search}. In
  \bibinfo{booktitle}{\emph{International Conference on Similarity Search and
  Applications}}. Springer, \bibinfo{pages}{16--29}.
\newblock


\bibitem[\protect\citeauthoryear{Hu, Yang, Zhan, Zhao, Li, and Li}{Hu
  et~al\mbox{.}}{2019}]%
        {KNN-join-dynamic-LSH-hu2019efficient}
\bibfield{author}{\bibinfo{person}{Yupeng Hu}, \bibinfo{person}{Chong Yang},
  \bibinfo{person}{Peng Zhan}, \bibinfo{person}{Jia Zhao},
  \bibinfo{person}{Yujun Li}, {and} \bibinfo{person}{Xueqing Li}.}
  \bibinfo{year}{2019}\natexlab{}.
\newblock \showarticletitle{Efficient continuous KNN join processing for
  real-time recommendation}.
\newblock \bibinfo{journal}{\emph{Personal and Ubiquitous Computing}}
  (\bibinfo{year}{2019}), \bibinfo{pages}{1--11}.
\newblock


\bibitem[\protect\citeauthoryear{Huang, Loy, and Tang}{Huang
  et~al\mbox{.}}{2016}]%
        {embedding-based-work-using-euc-distance-3}
\bibfield{author}{\bibinfo{person}{Chen Huang}, \bibinfo{person}{Chen~Change
  Loy}, {and} \bibinfo{person}{Xiaoou Tang}.} \bibinfo{year}{2016}\natexlab{}.
\newblock \showarticletitle{Local similarity-aware deep feature embedding}.
\newblock \bibinfo{journal}{\emph{Advances in neural information processing
  systems}}  \bibinfo{volume}{29} (\bibinfo{year}{2016}),
  \bibinfo{pages}{1262--1270}.
\newblock


\bibitem[\protect\citeauthoryear{Hyv{\"o}nen, Pitk{\"a}nen, Tasoulis,
  J{\"a}{\"a}saari, Tuomainen, Wang, Corander, and Roos}{Hyv{\"o}nen
  et~al\mbox{.}}{2015}]%
        {Tree-based-index-2}
\bibfield{author}{\bibinfo{person}{Ville Hyv{\"o}nen}, \bibinfo{person}{Teemu
  Pitk{\"a}nen}, \bibinfo{person}{Sotiris Tasoulis}, \bibinfo{person}{Elias
  J{\"a}{\"a}saari}, \bibinfo{person}{Risto Tuomainen}, \bibinfo{person}{Liang
  Wang}, \bibinfo{person}{Jukka Corander}, {and} \bibinfo{person}{Teemu Roos}.}
  \bibinfo{year}{2015}\natexlab{}.
\newblock \showarticletitle{Fast k-nn search}.
\newblock \bibinfo{journal}{\emph{arXiv preprint arXiv:1509.06957}}
  (\bibinfo{year}{2015}).
\newblock


\bibitem[\protect\citeauthoryear{Hyv{\"o}nen, Pitk{\"a}nen, Tasoulis,
  J{\"a}{\"a}saari, Tuomainen, Wang, Corander, and Roos}{Hyv{\"o}nen
  et~al\mbox{.}}{2016}]%
        {Tree-based-index-MRPT-2016}
\bibfield{author}{\bibinfo{person}{Ville Hyv{\"o}nen}, \bibinfo{person}{Teemu
  Pitk{\"a}nen}, \bibinfo{person}{Sotiris Tasoulis}, \bibinfo{person}{Elias
  J{\"a}{\"a}saari}, \bibinfo{person}{Risto Tuomainen}, \bibinfo{person}{Liang
  Wang}, \bibinfo{person}{Jukka Corander}, {and} \bibinfo{person}{Teemu Roos}.}
  \bibinfo{year}{2016}\natexlab{}.
\newblock \showarticletitle{Fast nearest neighbor search through sparse random
  projections and voting}. In \bibinfo{booktitle}{\emph{2016 IEEE International
  Conference on Big Data (Big Data)}}. IEEE, \bibinfo{pages}{881--888}.
\newblock


\bibitem[\protect\citeauthoryear{Iwasaki and Miyazaki}{Iwasaki and
  Miyazaki}{2018}]%
        {Graph-based-index-optimization-iwasaki2018}
\bibfield{author}{\bibinfo{person}{Masajiro Iwasaki} {and}
  \bibinfo{person}{Daisuke Miyazaki}.} \bibinfo{year}{2018}\natexlab{}.
\newblock \showarticletitle{Optimization of indexing based on k-nearest
  neighbor graph for proximity search in high-dimensional data}.
\newblock \bibinfo{journal}{\emph{arXiv preprint arXiv:1810.07355}}
  (\bibinfo{year}{2018}).
\newblock


\bibitem[\protect\citeauthoryear{Jang and Cho}{Jang and Cho}{2021}]%
        {deep-pq-self-supervised-jang2021}
\bibfield{author}{\bibinfo{person}{Young~Kyun Jang} {and}
  \bibinfo{person}{Nam~Ik Cho}.} \bibinfo{year}{2021}\natexlab{}.
\newblock \showarticletitle{Self-supervised Product Quantization for Deep
  Unsupervised Image Retrieval}. In \bibinfo{booktitle}{\emph{Proceedings of
  the IEEE/CVF International Conference on Computer Vision}}.
  \bibinfo{pages}{12085--12094}.
\newblock


\bibitem[\protect\citeauthoryear{Jegou, Amsaleg, Schmid, and Gros}{Jegou
  et~al\mbox{.}}{2008}]%
        {Query-adaptative-lsh}
\bibfield{author}{\bibinfo{person}{Herve Jegou}, \bibinfo{person}{Laurent
  Amsaleg}, \bibinfo{person}{Cordelia Schmid}, {and} \bibinfo{person}{Patrick
  Gros}.} \bibinfo{year}{2008}\natexlab{}.
\newblock \showarticletitle{Query adaptative locality sensitive hashing}. In
  \bibinfo{booktitle}{\emph{2008 IEEE International Conference on Acoustics,
  Speech and Signal Processing}}. \bibinfo{pages}{825--828}.
\newblock
\urldef\tempurl%
\url{https://doi.org/10.1109/ICASSP.2008.4517737}
\showDOI{\tempurl}


\bibitem[\protect\citeauthoryear{Jegou, Douze, and Schmid}{Jegou
  et~al\mbox{.}}{2010}]%
        {pq-ivfadc-2010}
\bibfield{author}{\bibinfo{person}{Herve Jegou}, \bibinfo{person}{Matthijs
  Douze}, {and} \bibinfo{person}{Cordelia Schmid}.}
  \bibinfo{year}{2010}\natexlab{}.
\newblock \showarticletitle{Product quantization for nearest neighbor search}.
\newblock \bibinfo{journal}{\emph{IEEE transactions on pattern analysis and
  machine intelligence}} \bibinfo{volume}{33}, \bibinfo{number}{1}
  (\bibinfo{year}{2010}), \bibinfo{pages}{117--128}.
\newblock


\bibitem[\protect\citeauthoryear{Ji, Li, Yan, Zhang, and Tian}{Ji
  et~al\mbox{.}}{2012}]%
        {Super-Bit-lsh}
\bibfield{author}{\bibinfo{person}{Jianqiu Ji}, \bibinfo{person}{Jianmin Li},
  \bibinfo{person}{Shuicheng Yan}, \bibinfo{person}{Bo Zhang}, {and}
  \bibinfo{person}{Qi Tian}.} \bibinfo{year}{2012}\natexlab{}.
\newblock \showarticletitle{Super-Bit Locality-Sensitive Hashing}. In
  \bibinfo{booktitle}{\emph{Proceedings of the 25th International Conference on
  Neural Information Processing Systems - Volume 1}} (Lake Tahoe, Nevada)
  \emph{(\bibinfo{series}{NIPS'12})}. \bibinfo{publisher}{Curran Associates
  Inc.}, \bibinfo{address}{Red Hook, NY, USA}, \bibinfo{pages}{108–116}.
\newblock


\bibitem[\protect\citeauthoryear{Johansson and Pina}{Johansson and
  Pina}{2015}]%
        {embedding-based-work-using-euc-distance-1}
\bibfield{author}{\bibinfo{person}{Richard Johansson} {and}
  \bibinfo{person}{Luis~Nieto Pina}.} \bibinfo{year}{2015}\natexlab{}.
\newblock \showarticletitle{Embedding a semantic network in a word space}. In
  \bibinfo{booktitle}{\emph{Proceedings of the 2015 Conference of the North
  American Chapter of the Association for Computational Linguistics: Human
  Language Technologies}}. \bibinfo{pages}{1428--1433}.
\newblock


\bibitem[\protect\citeauthoryear{Johnson, Douze, and J{\'e}gou}{Johnson
  et~al\mbox{.}}{2019}]%
        {simSearch-lib-fb-faiss-basedOn-pq-plus-inverted-index}
\bibfield{author}{\bibinfo{person}{Jeff Johnson}, \bibinfo{person}{Matthijs
  Douze}, {and} \bibinfo{person}{Herv{\'e} J{\'e}gou}.}
  \bibinfo{year}{2019}\natexlab{}.
\newblock \showarticletitle{Billion-scale similarity search with gpus}.
\newblock \bibinfo{journal}{\emph{IEEE Transactions on Big Data}}
  (\bibinfo{year}{2019}).
\newblock


\bibitem[\protect\citeauthoryear{Joly and Buisson}{Joly and Buisson}{2008}]%
        {Posteriori-lsh}
\bibfield{author}{\bibinfo{person}{Alexis Joly} {and} \bibinfo{person}{Olivier
  Buisson}.} \bibinfo{year}{2008}\natexlab{}.
\newblock \showarticletitle{A Posteriori Multi-Probe Locality Sensitive
  Hashing}. In \bibinfo{booktitle}{\emph{Proceedings of the 16th ACM
  International Conference on Multimedia}} (Vancouver, British Columbia,
  Canada) \emph{(\bibinfo{series}{MM '08})}. \bibinfo{publisher}{Association
  for Computing Machinery}, \bibinfo{address}{New York, NY, USA},
  \bibinfo{pages}{209–218}.
\newblock
\showISBNx{9781605583037}
\urldef\tempurl%
\url{https://doi.org/10.1145/1459359.1459388}
\showDOI{\tempurl}


\bibitem[\protect\citeauthoryear{Jun, Chung, et~al\mbox{.}}{Jun
  et~al\mbox{.}}{2015}]%
        {fpga-jun2015large}
\bibfield{author}{\bibinfo{person}{Sang-Woo Jun}, \bibinfo{person}{Chanwoo
  Chung}, {et~al\mbox{.}}} \bibinfo{year}{2015}\natexlab{}.
\newblock \showarticletitle{Large-scale high-dimensional nearest neighbor
  search using flash memory with in-store processing}. In
  \bibinfo{booktitle}{\emph{2015 International Conference on ReConFigurable
  Computing and FPGAs (ReConFig)}}. IEEE, \bibinfo{pages}{1--8}.
\newblock


\bibitem[\protect\citeauthoryear{Kalantidis and Avrithis}{Kalantidis and
  Avrithis}{2014}]%
        {pq-locally-optimized-2014}
\bibfield{author}{\bibinfo{person}{Yannis Kalantidis} {and}
  \bibinfo{person}{Yannis Avrithis}.} \bibinfo{year}{2014}\natexlab{}.
\newblock \showarticletitle{Locally optimized product quantization for
  approximate nearest neighbor search}. In
  \bibinfo{booktitle}{\emph{Proceedings of the IEEE Conference on Computer
  Vision and Pattern Recognition}}. \bibinfo{pages}{2321--2328}.
\newblock


\bibitem[\protect\citeauthoryear{Keivani and Sinha}{Keivani and Sinha}{2018}]%
        {Tree-based-index-auxiliary-info-rptree-2018}
\bibfield{author}{\bibinfo{person}{Omid Keivani} {and} \bibinfo{person}{Kaushik
  Sinha}.} \bibinfo{year}{2018}\natexlab{}.
\newblock \showarticletitle{Improved nearest neighbor search using auxiliary
  information and priority functions}. In
  \bibinfo{booktitle}{\emph{International Conference on Machine Learning}}.
  PMLR, \bibinfo{pages}{2573--2581}.
\newblock


\bibitem[\protect\citeauthoryear{Khandelwal, Clark, Jurafsky, and
  Kaiser}{Khandelwal et~al\mbox{.}}{2019}]%
        {example-pretrained-embedding-for-efficiency-2}
\bibfield{author}{\bibinfo{person}{Urvashi Khandelwal}, \bibinfo{person}{Kevin
  Clark}, \bibinfo{person}{Dan Jurafsky}, {and} \bibinfo{person}{Lukasz
  Kaiser}.} \bibinfo{year}{2019}\natexlab{}.
\newblock \showarticletitle{Sample efficient text summarization using a single
  pre-trained transformer}.
\newblock \bibinfo{journal}{\emph{arXiv preprint arXiv:1905.08836}}
  (\bibinfo{year}{2019}).
\newblock


\bibitem[\protect\citeauthoryear{Kim, Liu, and Choi}{Kim et~al\mbox{.}}{2021}]%
        {range-search-using-tree-Kim2021}
\bibfield{author}{\bibinfo{person}{Mincheol Kim}, \bibinfo{person}{Ling Liu},
  {and} \bibinfo{person}{Woink Choi}.} \bibinfo{year}{2021}\natexlab{}.
\newblock \showarticletitle{Multi-GPU Efficient Indexing for Maximizing
  Parallelism of High Dimensional Range Query Services}.
\newblock \bibinfo{journal}{\emph{IEEE Transactions on Services Computing}}
  (\bibinfo{year}{2021}), \bibinfo{pages}{1--1}.
\newblock
\urldef\tempurl%
\url{https://doi.org/10.1109/TSC.2021.3079580}
\showDOI{\tempurl}


\bibitem[\protect\citeauthoryear{Klein and Wolf}{Klein and Wolf}{2019a}]%
        {Product-quantization-5}
\bibfield{author}{\bibinfo{person}{Benjamin Klein} {and} \bibinfo{person}{Lior
  Wolf}.} \bibinfo{year}{2019}\natexlab{a}.
\newblock \showarticletitle{End-to-end supervised product quantization for
  image search and retrieval}. In \bibinfo{booktitle}{\emph{Proceedings of the
  IEEE/CVF Conference on Computer Vision and Pattern Recognition}}.
  \bibinfo{pages}{5041--5050}.
\newblock


\bibitem[\protect\citeauthoryear{Klein and Wolf}{Klein and Wolf}{2019b}]%
        {deep-pq-end2end-supervised-klein2019}
\bibfield{author}{\bibinfo{person}{Benjamin Klein} {and} \bibinfo{person}{Lior
  Wolf}.} \bibinfo{year}{2019}\natexlab{b}.
\newblock \showarticletitle{End-to-end supervised product quantization for
  image search and retrieval}. In \bibinfo{booktitle}{\emph{Proceedings of the
  IEEE/CVF Conference on Computer Vision and Pattern Recognition}}.
  \bibinfo{pages}{5041--5050}.
\newblock


\bibitem[\protect\citeauthoryear{K\"{o}pcke, Thor, and Rahm}{K\"{o}pcke
  et~al\mbox{.}}{2010}]%
        {ER-traditional-2}
\bibfield{author}{\bibinfo{person}{Hanna K\"{o}pcke}, \bibinfo{person}{Andreas
  Thor}, {and} \bibinfo{person}{Erhard Rahm}.} \bibinfo{year}{2010}\natexlab{}.
\newblock \showarticletitle{Evaluation of Entity Resolution Approaches on
  Real-World Match Problems}.
\newblock \bibinfo{journal}{\emph{Proc. VLDB Endow.}} \bibinfo{volume}{3},
  \bibinfo{number}{1–2} (\bibinfo{date}{Sept.} \bibinfo{year}{2010}),
  \bibinfo{pages}{484–493}.
\newblock
\showISSN{2150-8097}
\urldef\tempurl%
\url{https://doi.org/10.14778/1920841.1920904}
\showDOI{\tempurl}


\bibitem[\protect\citeauthoryear{Kriegel, Kr{\"o}ger, and Zimek}{Kriegel
  et~al\mbox{.}}{2009}]%
        {clustering-survey-kriegel2009clustering}
\bibfield{author}{\bibinfo{person}{Hans-Peter Kriegel}, \bibinfo{person}{Peer
  Kr{\"o}ger}, {and} \bibinfo{person}{Arthur Zimek}.}
  \bibinfo{year}{2009}\natexlab{}.
\newblock \showarticletitle{Clustering high-dimensional data: A survey on
  subspace clustering, pattern-based clustering, and correlation clustering}.
\newblock \bibinfo{journal}{\emph{Acm transactions on knowledge discovery from
  data (tkdd)}} \bibinfo{volume}{3}, \bibinfo{number}{1}
  (\bibinfo{year}{2009}), \bibinfo{pages}{1--58}.
\newblock


\bibitem[\protect\citeauthoryear{Kruli{\v{s}}, Osipyan, and
  Marchand-Maillet}{Kruli{\v{s}} et~al\mbox{.}}{2015}]%
        {Heterogeneous-hardware-5}
\bibfield{author}{\bibinfo{person}{Martin Kruli{\v{s}}},
  \bibinfo{person}{Hasmik Osipyan}, {and} \bibinfo{person}{St{\'e}phane
  Marchand-Maillet}.} \bibinfo{year}{2015}\natexlab{}.
\newblock \showarticletitle{Optimizing sorting and top-k selection steps in
  permutation based indexing on gpus}. In \bibinfo{booktitle}{\emph{East
  European Conference on Advances in Databases and Information Systems}}.
  Springer, \bibinfo{pages}{305--317}.
\newblock


\bibitem[\protect\citeauthoryear{Kruliš, Osipyan, and
  Marchand-Maillet}{Kruliš et~al\mbox{.}}{2015}]%
        {Permutation-based-indexing-gpu}
\bibfield{author}{\bibinfo{person}{Martin Kruliš}, \bibinfo{person}{Hasmik
  Osipyan}, {and} \bibinfo{person}{Stéphane Marchand-Maillet}.}
  \bibinfo{year}{2015}\natexlab{}.
\newblock \showarticletitle{Permutation based indexing for high dimensional
  data on GPU architectures}. In \bibinfo{booktitle}{\emph{2015 13th
  International Workshop on Content-Based Multimedia Indexing (CBMI)}}.
  \bibinfo{pages}{1--6}.
\newblock
\urldef\tempurl%
\url{https://doi.org/10.1109/CBMI.2015.7153619}
\showDOI{\tempurl}


\bibitem[\protect\citeauthoryear{Kulis and Darrell}{Kulis and Darrell}{2009}]%
        {l2h-multidim-features-supervised-kulis2009-binary-recons-embed}
\bibfield{author}{\bibinfo{person}{Brian Kulis} {and} \bibinfo{person}{Trevor
  Darrell}.} \bibinfo{year}{2009}\natexlab{}.
\newblock \showarticletitle{Learning to Hash with Binary Reconstructive
  Embeddings.}. In \bibinfo{booktitle}{\emph{NIPS}}, Vol.~\bibinfo{volume}{22}.
  Citeseer, \bibinfo{pages}{1042--1050}.
\newblock


\bibitem[\protect\citeauthoryear{Kumar and Udupa}{Kumar and Udupa}{2011}]%
        {IR-traditional-joint-1}
\bibfield{author}{\bibinfo{person}{Shaishav Kumar} {and}
  \bibinfo{person}{Raghavendra Udupa}.} \bibinfo{year}{2011}\natexlab{}.
\newblock \showarticletitle{Learning Hash Functions for Cross-View Similarity
  Search}. In \bibinfo{booktitle}{\emph{Proceedings of the Twenty-Second
  International Joint Conference on Artificial Intelligence - Volume Volume
  Two}} (Barcelona, Catalonia, Spain) \emph{(\bibinfo{series}{IJCAI'11})}.
  \bibinfo{publisher}{AAAI Press}, \bibinfo{pages}{1360–1365}.
\newblock
\showISBNx{9781577355144}


\bibitem[\protect\citeauthoryear{Lacroix, Usunier, and Obozinski}{Lacroix
  et~al\mbox{.}}{2018}]%
        {lacroix2018canonical}
\bibfield{author}{\bibinfo{person}{Timoth{\'e}e Lacroix},
  \bibinfo{person}{Nicolas Usunier}, {and} \bibinfo{person}{Guillaume
  Obozinski}.} \bibinfo{year}{2018}\natexlab{}.
\newblock \showarticletitle{Canonical tensor decomposition for knowledge base
  completion}.
\newblock \bibinfo{journal}{\emph{arXiv preprint arXiv:1806.07297}}
  (\bibinfo{year}{2018}).
\newblock


\bibitem[\protect\citeauthoryear{Lai, Pan, Liu, and Yan}{Lai
  et~al\mbox{.}}{2015}]%
        {deep-l2h-lai2015-simultaneous-learning-embedding-and-hash-functions}
\bibfield{author}{\bibinfo{person}{Hanjiang Lai}, \bibinfo{person}{Yan Pan},
  \bibinfo{person}{Ye Liu}, {and} \bibinfo{person}{Shuicheng Yan}.}
  \bibinfo{year}{2015}\natexlab{}.
\newblock \showarticletitle{Simultaneous feature learning and hash coding with
  deep neural networks}. In \bibinfo{booktitle}{\emph{Proceedings of the IEEE
  conference on computer vision and pattern recognition}}.
  \bibinfo{pages}{3270--3278}.
\newblock


\bibitem[\protect\citeauthoryear{Lempitsky and Babenko}{Lempitsky and
  Babenko}{2012}]%
        {pq-inverted-multi-index-2012}
\bibfield{author}{\bibinfo{person}{Victor Lempitsky} {and} \bibinfo{person}{A
  Babenko}.} \bibinfo{year}{2012}\natexlab{}.
\newblock \showarticletitle{The inverted multi-index}. In
  \bibinfo{booktitle}{\emph{2012 IEEE Conference on Computer Vision and Pattern
  Recognition}}. IEEE Computer Society, \bibinfo{pages}{3069--3076}.
\newblock


\bibitem[\protect\citeauthoryear{Li, Deng, Wang, and Feng}{Li
  et~al\mbox{.}}{2011}]%
        {Condition-based-join-string-sim-join-edit-dist-2011}
\bibfield{author}{\bibinfo{person}{Guoliang Li}, \bibinfo{person}{Dong Deng},
  \bibinfo{person}{Jiannan Wang}, {and} \bibinfo{person}{Jianhua Feng}.}
  \bibinfo{year}{2011}\natexlab{}.
\newblock \showarticletitle{Pass-Join: A Partition-Based Method for Similarity
  Joins}.
\newblock \bibinfo{journal}{\emph{Proc. VLDB Endow.}} \bibinfo{volume}{5},
  \bibinfo{number}{3} (\bibinfo{date}{Nov.} \bibinfo{year}{2011}),
  \bibinfo{pages}{253–264}.
\newblock
\showISSN{2150-8097}
\urldef\tempurl%
\url{https://doi.org/10.14778/2078331.2078340}
\showDOI{\tempurl}


\bibitem[\protect\citeauthoryear{Li, Cheng, Yang, Huang, Zhao, Yan, and
  Zhao}{Li et~al\mbox{.}}{2017}]%
        {LSH-5}
\bibfield{author}{\bibinfo{person}{Jinfeng Li}, \bibinfo{person}{James Cheng},
  \bibinfo{person}{Fan Yang}, \bibinfo{person}{Yuzhen Huang},
  \bibinfo{person}{Yunjian Zhao}, \bibinfo{person}{Xiao Yan}, {and}
  \bibinfo{person}{Ruihao Zhao}.} \bibinfo{year}{2017}\natexlab{}.
\newblock \showarticletitle{Losha: A general framework for scalable locality
  sensitive hashing}. In \bibinfo{booktitle}{\emph{Proceedings of the 40th
  International ACM SIGIR Conference on Research and Development in Information
  Retrieval}}. \bibinfo{pages}{635--644}.
\newblock


\bibitem[\protect\citeauthoryear{Li, Zhang, Sun, Wang, Li, Zhang, and Lin}{Li
  et~al\mbox{.}}{2019}]%
        {Graph-based-index-5}
\bibfield{author}{\bibinfo{person}{Wen Li}, \bibinfo{person}{Ying Zhang},
  \bibinfo{person}{Yifang Sun}, \bibinfo{person}{Wei Wang},
  \bibinfo{person}{Mingjie Li}, \bibinfo{person}{Wenjie Zhang}, {and}
  \bibinfo{person}{Xuemin Lin}.} \bibinfo{year}{2019}\natexlab{}.
\newblock \showarticletitle{Approximate nearest neighbor search on high
  dimensional data—experiments, analyses, and improvement}.
\newblock \bibinfo{journal}{\emph{IEEE Transactions on Knowledge and Data
  Engineering}} \bibinfo{volume}{32}, \bibinfo{number}{8}
  (\bibinfo{year}{2019}), \bibinfo{pages}{1475--1488}.
\newblock


\bibitem[\protect\citeauthoryear{Li, Wang, and Kang}{Li et~al\mbox{.}}{2015}]%
        {deep-l2h-li2015-pairwise-labels}
\bibfield{author}{\bibinfo{person}{Wu-Jun Li}, \bibinfo{person}{Sheng Wang},
  {and} \bibinfo{person}{Wang-Cheng Kang}.} \bibinfo{year}{2015}\natexlab{}.
\newblock \showarticletitle{Feature learning based deep supervised hashing with
  pairwise labels}.
\newblock \bibinfo{journal}{\emph{arXiv preprint arXiv:1511.03855}}
  (\bibinfo{year}{2015}).
\newblock


\bibitem[\protect\citeauthoryear{Li, Li, Suhara, Doan, and Tan}{Li
  et~al\mbox{.}}{2020}]%
        {ER-embedding-8}
\bibfield{author}{\bibinfo{person}{Yuliang Li}, \bibinfo{person}{Jinfeng Li},
  \bibinfo{person}{Yoshihiko Suhara}, \bibinfo{person}{AnHai Doan}, {and}
  \bibinfo{person}{Wang-Chiew Tan}.} \bibinfo{year}{2020}\natexlab{}.
\newblock \showarticletitle{Deep entity matching with pre-trained language
  models}.
\newblock \bibinfo{journal}{\emph{arXiv preprint arXiv:2004.00584}}
  (\bibinfo{year}{2020}).
\newblock


\bibitem[\protect\citeauthoryear{Li, Wang, et~al\mbox{.}}{Li
  et~al\mbox{.}}{2016}]%
        {Condition-based-similarity-join-3}
\bibfield{author}{\bibinfo{person}{Ye Li}, \bibinfo{person}{Jian Wang},
  {et~al\mbox{.}}} \bibinfo{year}{2016}\natexlab{}.
\newblock \showarticletitle{Multidimensional similarity join using mapreduce}.
  In \bibinfo{booktitle}{\emph{International Conference on Web-Age Information
  Management}}. Springer, \bibinfo{pages}{457--468}.
\newblock


\bibitem[\protect\citeauthoryear{Lin, Xiao, Cheng, and Bhowmick}{Lin
  et~al\mbox{.}}{2012}]%
        {general-similarity-query-graph-1}
\bibfield{author}{\bibinfo{person}{Wenqing Lin}, \bibinfo{person}{Xiaokui
  Xiao}, \bibinfo{person}{James Cheng}, {and} \bibinfo{person}{Sourav~S
  Bhowmick}.} \bibinfo{year}{2012}\natexlab{}.
\newblock \showarticletitle{Efficient algorithms for generalized subgraph query
  processing}. In \bibinfo{booktitle}{\emph{Proceedings of the 21st ACM
  international conference on Information and knowledge management}}.
  \bibinfo{pages}{325--334}.
\newblock


\bibitem[\protect\citeauthoryear{Lindenstrauss}{Lindenstrauss}{1984}]%
        {random-project-high-to-low-dim-lindenstrauss1984}
\bibfield{author}{\bibinfo{person}{W~Johnson~J Lindenstrauss}.}
  \bibinfo{year}{1984}\natexlab{}.
\newblock \showarticletitle{Extensions of Lipschitz maps into a Hilbert space}.
\newblock \bibinfo{journal}{\emph{Contemp. Math}} \bibinfo{volume}{26},
  \bibinfo{number}{189-206} (\bibinfo{year}{1984}), \bibinfo{pages}{2}.
\newblock


\bibitem[\protect\citeauthoryear{Liu, Wang, Shan, and Chen}{Liu
  et~al\mbox{.}}{2016}]%
        {deep-l2h-liu2016-dsh-for-image-retrieval}
\bibfield{author}{\bibinfo{person}{Haomiao Liu}, \bibinfo{person}{Ruiping
  Wang}, \bibinfo{person}{Shiguang Shan}, {and} \bibinfo{person}{Xilin Chen}.}
  \bibinfo{year}{2016}\natexlab{}.
\newblock \showarticletitle{Deep supervised hashing for fast image retrieval}.
  In \bibinfo{booktitle}{\emph{Proceedings of the IEEE conference on computer
  vision and pattern recognition}}. \bibinfo{pages}{2064--2072}.
\newblock


\bibitem[\protect\citeauthoryear{Liu, Dai, Bai, and Duan}{Liu
  et~al\mbox{.}}{2020}]%
        {deep-pq-module-liu2020}
\bibfield{author}{\bibinfo{person}{Meihan Liu}, \bibinfo{person}{Yongxing Dai},
  \bibinfo{person}{Yan Bai}, {and} \bibinfo{person}{Ling-Yu Duan}.}
  \bibinfo{year}{2020}\natexlab{}.
\newblock \showarticletitle{Deep Product Quantization Module for Efficient
  Image Retrieval}. In \bibinfo{booktitle}{\emph{ICASSP 2020-2020 IEEE
  International Conference on Acoustics, Speech and Signal Processing
  (ICASSP)}}. IEEE, \bibinfo{pages}{4382--4386}.
\newblock


\bibitem[\protect\citeauthoryear{Liu, Guo, Chamnongthai, and Prasetyo}{Liu
  et~al\mbox{.}}{2017}]%
        {traditional-method-image-color-feature-3}
\bibfield{author}{\bibinfo{person}{Peizhong Liu}, \bibinfo{person}{Jing-Ming
  Guo}, \bibinfo{person}{Kosin Chamnongthai}, {and} \bibinfo{person}{Heri
  Prasetyo}.} \bibinfo{year}{2017}\natexlab{}.
\newblock \showarticletitle{Fusion of color histogram and LBP-based features
  for texture image retrieval and classification}.
\newblock \bibinfo{journal}{\emph{Information Sciences}}  \bibinfo{volume}{390}
  (\bibinfo{year}{2017}), \bibinfo{pages}{95--111}.
\newblock
\showISSN{0020-0255}
\urldef\tempurl%
\url{https://doi.org/10.1016/j.ins.2017.01.025}
\showDOI{\tempurl}


\bibitem[\protect\citeauthoryear{Liu, Moore, Gray, and Yang}{Liu
  et~al\mbox{.}}{2004}]%
        {Tree-based-index-metric-spill-trees-liu2004}
\bibfield{author}{\bibinfo{person}{Ting Liu}, \bibinfo{person}{Andrew~W Moore},
  \bibinfo{person}{Alexander~G Gray}, {and} \bibinfo{person}{Ke Yang}.}
  \bibinfo{year}{2004}\natexlab{}.
\newblock \showarticletitle{An investigation of practical approximate nearest
  neighbor algorithms.}. In \bibinfo{booktitle}{\emph{NIPS}},
  Vol.~\bibinfo{volume}{12}. Citeseer, \bibinfo{pages}{2004}.
\newblock


\bibitem[\protect\citeauthoryear{Liu, Wang, Ji, Jiang, and Chang}{Liu
  et~al\mbox{.}}{2012}]%
  {l2h-multidim-features-supervised-liu2012-supervised-hashing-with-kernels}
\bibfield{author}{\bibinfo{person}{Wei Liu}, \bibinfo{person}{Jun Wang},
  \bibinfo{person}{Rongrong Ji}, \bibinfo{person}{Yu-Gang Jiang}, {and}
  \bibinfo{person}{Shih-Fu Chang}.} \bibinfo{year}{2012}\natexlab{}.
\newblock \showarticletitle{Supervised hashing with kernels}. In
  \bibinfo{booktitle}{\emph{2012 IEEE Conference on Computer Vision and Pattern
  Recognition}}. IEEE, \bibinfo{pages}{2074--2081}.
\newblock


\bibitem[\protect\citeauthoryear{Liu, Wang, Kumar, and Chang}{Liu
  et~al\mbox{.}}{2011}]%
        {l2h-multidim-features-unsupervised-liu2011-Anchor-Graph-hashing}
\bibfield{author}{\bibinfo{person}{Wei Liu}, \bibinfo{person}{Jun Wang},
  \bibinfo{person}{Sanjiv Kumar}, {and} \bibinfo{person}{Shih-Fu Chang}.}
  \bibinfo{year}{2011}\natexlab{}.
\newblock \showarticletitle{Hashing with graphs}. In
  \bibinfo{booktitle}{\emph{Icml}}.
\newblock


\bibitem[\protect\citeauthoryear{Liu, Cui, Huang, Li, and Shen}{Liu
  et~al\mbox{.}}{2014}]%
        {LSH-3}
\bibfield{author}{\bibinfo{person}{Yingfan Liu}, \bibinfo{person}{Jiangtao
  Cui}, \bibinfo{person}{Zi Huang}, \bibinfo{person}{Hui Li}, {and}
  \bibinfo{person}{Heng~Tao Shen}.} \bibinfo{year}{2014}\natexlab{}.
\newblock \showarticletitle{SK-LSH: an efficient index structure for
  approximate nearest neighbor search}.
\newblock \bibinfo{journal}{\emph{Proceedings of the VLDB Endowment}}
  \bibinfo{volume}{7}, \bibinfo{number}{9} (\bibinfo{year}{2014}),
  \bibinfo{pages}{745--756}.
\newblock


\bibitem[\protect\citeauthoryear{Lu, Fang, Farahpour, and Shannon}{Lu
  et~al\mbox{.}}{2020}]%
        {fpga-configurable-knn}
\bibfield{author}{\bibinfo{person}{Alec Lu}, \bibinfo{person}{Zhenman Fang},
  \bibinfo{person}{Nazanin Farahpour}, {and} \bibinfo{person}{Lesley Shannon}.}
  \bibinfo{year}{2020}\natexlab{}.
\newblock \showarticletitle{CHIP-KNN: A Configurable and High-Performance
  K-Nearest Neighbors Accelerator on Cloud FPGAs}. In
  \bibinfo{booktitle}{\emph{2020 International Conference on Field-Programmable
  Technology (ICFPT)}}. \bibinfo{pages}{139--147}.
\newblock
\urldef\tempurl%
\url{https://doi.org/10.1109/ICFPT51103.2020.00027}
\showDOI{\tempurl}


\bibitem[\protect\citeauthoryear{Lu, Shen, Chen, and Ooi}{Lu
  et~al\mbox{.}}{2012}]%
        {KNN-join-5}
\bibfield{author}{\bibinfo{person}{Wei Lu}, \bibinfo{person}{Yanyan Shen},
  \bibinfo{person}{Su Chen}, {and} \bibinfo{person}{Beng~Chin Ooi}.}
  \bibinfo{year}{2012}\natexlab{}.
\newblock \showarticletitle{Efficient processing of k nearest neighbor joins
  using mapreduce}.
\newblock \bibinfo{journal}{\emph{arXiv preprint arXiv:1207.0141}}
  (\bibinfo{year}{2012}).
\newblock


\bibitem[\protect\citeauthoryear{Luka{\v{c}}, {\v{Z}}alik, Cui, and
  Datcu}{Luka{\v{c}} et~al\mbox{.}}{2015}]%
        {gpu-kernel-lsh-lukavc2015gpu}
\bibfield{author}{\bibinfo{person}{Niko Luka{\v{c}}}, \bibinfo{person}{Borut
  {\v{Z}}alik}, \bibinfo{person}{Shiyong Cui}, {and} \bibinfo{person}{Mihai
  Datcu}.} \bibinfo{year}{2015}\natexlab{}.
\newblock \showarticletitle{GPU-based kernelized locality-sensitive hashing for
  satellite image retrieval}. In \bibinfo{booktitle}{\emph{2015 IEEE
  International Geoscience and Remote Sensing Symposium (IGARSS)}}. IEEE,
  \bibinfo{pages}{1468--1471}.
\newblock


\bibitem[\protect\citeauthoryear{Lv, Josephson, Wang, Charikar, and Li}{Lv
  et~al\mbox{.}}{2007a}]%
        {LSH-2}
\bibfield{author}{\bibinfo{person}{Qin Lv}, \bibinfo{person}{William
  Josephson}, \bibinfo{person}{Zhe Wang}, \bibinfo{person}{Moses Charikar},
  {and} \bibinfo{person}{Kai Li}.} \bibinfo{year}{2007}\natexlab{a}.
\newblock \showarticletitle{Multi-probe LSH: efficient indexing for
  high-dimensional similarity search}. In \bibinfo{booktitle}{\emph{33rd
  International Conference on Very Large Data Bases, VLDB 2007}}. Association
  for Computing Machinery, Inc, \bibinfo{pages}{950--961}.
\newblock


\bibitem[\protect\citeauthoryear{Lv, Josephson, Wang, Charikar, and Li}{Lv
  et~al\mbox{.}}{2007b}]%
        {Multi-probe-LSH-lv2007multi}
\bibfield{author}{\bibinfo{person}{Qin Lv}, \bibinfo{person}{William
  Josephson}, \bibinfo{person}{Zhe Wang}, \bibinfo{person}{Moses Charikar},
  {and} \bibinfo{person}{Kai Li}.} \bibinfo{year}{2007}\natexlab{b}.
\newblock \showarticletitle{Multi-probe LSH: efficient indexing for
  high-dimensional similarity search}. In \bibinfo{booktitle}{\emph{33rd
  International Conference on Very Large Data Bases, VLDB 2007}}. Association
  for Computing Machinery, Inc, \bibinfo{pages}{950--961}.
\newblock


\bibitem[\protect\citeauthoryear{Ma, Zhang, and Zhang}{Ma
  et~al\mbox{.}}{2019}]%
        {Condition-based-similarity-join-2}
\bibfield{author}{\bibinfo{person}{Youzhong Ma}, \bibinfo{person}{Ruiling
  Zhang}, {and} \bibinfo{person}{Yongxin Zhang}.}
  \bibinfo{year}{2019}\natexlab{}.
\newblock \showarticletitle{Similarity histogram estimation based top-k
  similarity join algorithm on high-dimensional data}. In
  \bibinfo{booktitle}{\emph{International Conference on Web Information Systems
  and Applications}}. Springer, \bibinfo{pages}{589--600}.
\newblock


\bibitem[\protect\citeauthoryear{Malkov, Ponomarenko, Logvinov, and
  Krylov}{Malkov et~al\mbox{.}}{2014}]%
        {Graph-based-index-nsw-malkov2014}
\bibfield{author}{\bibinfo{person}{Yury Malkov}, \bibinfo{person}{Alexander
  Ponomarenko}, \bibinfo{person}{Andrey Logvinov}, {and}
  \bibinfo{person}{Vladimir Krylov}.} \bibinfo{year}{2014}\natexlab{}.
\newblock \showarticletitle{Approximate nearest neighbor algorithm based on
  navigable small world graphs}.
\newblock \bibinfo{journal}{\emph{Information Systems}}  \bibinfo{volume}{45}
  (\bibinfo{year}{2014}), \bibinfo{pages}{61--68}.
\newblock


\bibitem[\protect\citeauthoryear{Malkov and Yashunin}{Malkov and
  Yashunin}{2016}]%
        {Graph-based-index-hnsw-malkov2016}
\bibfield{author}{\bibinfo{person}{Yu~A Malkov} {and} \bibinfo{person}{DA
  Yashunin}.} \bibinfo{year}{2016}\natexlab{}.
\newblock \showarticletitle{Efficient and robust approximate nearest neighbor
  search using Hierarchical Navigable Small World graphs}.
\newblock \bibinfo{journal}{\emph{arXiv preprint arXiv:1603.09320}}
  (\bibinfo{year}{2016}).
\newblock


\bibitem[\protect\citeauthoryear{Malkov and Yashunin}{Malkov and
  Yashunin}{2018}]%
        {Similarity-search-graph-based-1}
\bibfield{author}{\bibinfo{person}{Yu~A Malkov} {and} \bibinfo{person}{Dmitry~A
  Yashunin}.} \bibinfo{year}{2018}\natexlab{}.
\newblock \showarticletitle{Efficient and robust approximate nearest neighbor
  search using hierarchical navigable small world graphs}.
\newblock \bibinfo{journal}{\emph{IEEE transactions on pattern analysis and
  machine intelligence}} \bibinfo{volume}{42}, \bibinfo{number}{4}
  (\bibinfo{year}{2018}), \bibinfo{pages}{824--836}.
\newblock


\bibitem[\protect\citeauthoryear{Matsui, Uchida, J{\'e}gou, and Satoh}{Matsui
  et~al\mbox{.}}{2018}]%
        {Product-quantization-1}
\bibfield{author}{\bibinfo{person}{Yusuke Matsui}, \bibinfo{person}{Yusuke
  Uchida}, \bibinfo{person}{Herv{\'e} J{\'e}gou}, {and}
  \bibinfo{person}{Shin'ichi Satoh}.} \bibinfo{year}{2018}\natexlab{}.
\newblock \showarticletitle{A survey of product quantization}.
\newblock \bibinfo{journal}{\emph{ITE Transactions on Media Technology and
  Applications}} \bibinfo{volume}{6}, \bibinfo{number}{1}
  (\bibinfo{year}{2018}), \bibinfo{pages}{2--10}.
\newblock


\bibitem[\protect\citeauthoryear{Matsui, Yamasaki, and Aizawa}{Matsui
  et~al\mbox{.}}{2015}]%
        {pq-pqtable-2015}
\bibfield{author}{\bibinfo{person}{Yusuke Matsui}, \bibinfo{person}{Toshihiko
  Yamasaki}, {and} \bibinfo{person}{Kiyoharu Aizawa}.}
  \bibinfo{year}{2015}\natexlab{}.
\newblock \showarticletitle{Pqtable: Fast exact asymmetric distance neighbor
  search for product quantization using hash tables}. In
  \bibinfo{booktitle}{\emph{Proceedings of the IEEE International Conference on
  Computer Vision}}. \bibinfo{pages}{1940--1948}.
\newblock


\bibitem[\protect\citeauthoryear{Meng, Huang, Wang, Zhang, Zhuang, Kaplan, and
  Han}{Meng et~al\mbox{.}}{2019}]%
        {embedding-based-work-using-euc-distance-2}
\bibfield{author}{\bibinfo{person}{Yu Meng}, \bibinfo{person}{Jiaxin Huang},
  \bibinfo{person}{Guangyuan Wang}, \bibinfo{person}{Chao Zhang},
  \bibinfo{person}{Honglei Zhuang}, \bibinfo{person}{Lance Kaplan}, {and}
  \bibinfo{person}{Jiawei Han}.} \bibinfo{year}{2019}\natexlab{}.
\newblock \showarticletitle{Spherical text embedding}.
\newblock \bibinfo{journal}{\emph{Advances in Neural Information Processing
  Systems}}  \bibinfo{volume}{32} (\bibinfo{year}{2019}),
  \bibinfo{pages}{8208--8217}.
\newblock


\bibitem[\protect\citeauthoryear{Mikolov, Chen, Corrado, and Dean}{Mikolov
  et~al\mbox{.}}{2013a}]%
        {embedding-text-1}
\bibfield{author}{\bibinfo{person}{Tomas Mikolov}, \bibinfo{person}{Kai Chen},
  \bibinfo{person}{Greg Corrado}, {and} \bibinfo{person}{Jeffrey Dean}.}
  \bibinfo{year}{2013}\natexlab{a}.
\newblock \showarticletitle{Efficient estimation of word representations in
  vector space}.
\newblock \bibinfo{journal}{\emph{arXiv preprint arXiv:1301.3781}}
  (\bibinfo{year}{2013}).
\newblock


\bibitem[\protect\citeauthoryear{Mikolov, Chen, Corrado, and Dean}{Mikolov
  et~al\mbox{.}}{2013b}]%
        {mikolov2013efficient}
\bibfield{author}{\bibinfo{person}{Tomas Mikolov}, \bibinfo{person}{Kai Chen},
  \bibinfo{person}{Greg Corrado}, {and} \bibinfo{person}{Jeffrey Dean}.}
  \bibinfo{year}{2013}\natexlab{b}.
\newblock \showarticletitle{Efficient estimation of word representations in
  vector space}.
\newblock \bibinfo{journal}{\emph{arXiv preprint arXiv:1301.3781}}
  (\bibinfo{year}{2013}).
\newblock


\bibitem[\protect\citeauthoryear{Minati, Movsisyan, Mccormick, Gyozalyan,
  Papazyan, Makaryan, Aldrigo, Harutyunyan, Ghaltaghchyan, Mccormick, and
  Fandrich}{Minati et~al\mbox{.}}{2019}]%
        {fpga-open-source-knn}
\bibfield{author}{\bibinfo{person}{Ludovico Minati}, \bibinfo{person}{Vardan
  Movsisyan}, \bibinfo{person}{Matthew Mccormick}, \bibinfo{person}{Khachatur
  Gyozalyan}, \bibinfo{person}{Tigran Papazyan}, \bibinfo{person}{Hrach
  Makaryan}, \bibinfo{person}{Stefano Aldrigo}, \bibinfo{person}{Taron
  Harutyunyan}, \bibinfo{person}{Hayk Ghaltaghchyan}, \bibinfo{person}{Chris
  Mccormick}, {and} \bibinfo{person}{Mick Fandrich}.}
  \bibinfo{year}{2019}\natexlab{}.
\newblock \showarticletitle{iFLEX: A Fully Open-Source, High-Density
  Field-Programmable Gate Array (FPGA)-Based Hardware Co-Processor for Vector
  Similarity Searching}.
\newblock \bibinfo{journal}{\emph{IEEE Access}}  \bibinfo{volume}{7}
  (\bibinfo{year}{2019}), \bibinfo{pages}{112269--112283}.
\newblock
\urldef\tempurl%
\url{https://doi.org/10.1109/ACCESS.2019.2934715}
\showDOI{\tempurl}


\bibitem[\protect\citeauthoryear{Mithun, Li, Metze, and Roy-Chowdhury}{Mithun
  et~al\mbox{.}}{2018}]%
        {IR-embedding-joint-mithun2018learning}
\bibfield{author}{\bibinfo{person}{Niluthpol~Chowdhury Mithun},
  \bibinfo{person}{Juncheng Li}, \bibinfo{person}{Florian Metze}, {and}
  \bibinfo{person}{Amit~K Roy-Chowdhury}.} \bibinfo{year}{2018}\natexlab{}.
\newblock \showarticletitle{Learning joint embedding with multimodal cues for
  cross-modal video-text retrieval}. In \bibinfo{booktitle}{\emph{Proceedings
  of the 2018 ACM on International Conference on Multimedia Retrieval}}.
  \bibinfo{pages}{19--27}.
\newblock


\bibitem[\protect\citeauthoryear{Mudgal, Li, Rekatsinas, Doan, Park, Krishnan,
  Deep, Arcaute, and Raghavendra}{Mudgal et~al\mbox{.}}{2018}]%
        {ER-embedding-1}
\bibfield{author}{\bibinfo{person}{Sidharth Mudgal}, \bibinfo{person}{Han Li},
  \bibinfo{person}{Theodoros Rekatsinas}, \bibinfo{person}{AnHai Doan},
  \bibinfo{person}{Youngchoon Park}, \bibinfo{person}{Ganesh Krishnan},
  \bibinfo{person}{Rohit Deep}, \bibinfo{person}{Esteban Arcaute}, {and}
  \bibinfo{person}{Vijay Raghavendra}.} \bibinfo{year}{2018}\natexlab{}.
\newblock \showarticletitle{Deep learning for entity matching: A design space
  exploration}. In \bibinfo{booktitle}{\emph{Proceedings of the 2018
  International Conference on Management of Data}}. \bibinfo{pages}{19--34}.
\newblock


\bibitem[\protect\citeauthoryear{Muja and Lowe}{Muja and Lowe}{2009}]%
        {simSearch-lib-flann}
\bibfield{author}{\bibinfo{person}{Marius Muja} {and} \bibinfo{person}{David
  Lowe}.} \bibinfo{year}{2009}\natexlab{}.
\newblock \showarticletitle{Fast Approximate Nearest Neighbors with Automatic
  Algorithm Configuration.}
\newblock \bibinfo{journal}{\emph{VISAPP 2009 - Proceedings of the 4th
  International Conference on Computer Vision Theory and Applications}}
  \bibinfo{volume}{1}, \bibinfo{pages}{331--340}.
\newblock


\bibitem[\protect\citeauthoryear{Muja and Lowe}{Muja and Lowe}{2014}]%
        {simSearch-lib-flann-2}
\bibfield{author}{\bibinfo{person}{Marius Muja} {and} \bibinfo{person}{David~G
  Lowe}.} \bibinfo{year}{2014}\natexlab{}.
\newblock \showarticletitle{Scalable nearest neighbor algorithms for high
  dimensional data}.
\newblock \bibinfo{journal}{\emph{IEEE transactions on pattern analysis and
  machine intelligence}} \bibinfo{volume}{36}, \bibinfo{number}{11}
  (\bibinfo{year}{2014}), \bibinfo{pages}{2227--2240}.
\newblock


\bibitem[\protect\citeauthoryear{Norouzi and Fleet}{Norouzi and Fleet}{2011}]%
        {l2h-multidim-features-supervised-norouzi2011-minimal-loss-hashing}
\bibfield{author}{\bibinfo{person}{Mohammad Norouzi} {and}
  \bibinfo{person}{David~J Fleet}.} \bibinfo{year}{2011}\natexlab{}.
\newblock \showarticletitle{Minimal loss hashing for compact binary codes}. In
  \bibinfo{booktitle}{\emph{ICML}}.
\newblock


\bibitem[\protect\citeauthoryear{Norouzi, Punjani, and Fleet}{Norouzi
  et~al\mbox{.}}{2012}]%
        {multi-index-hashing-norouzi2012}
\bibfield{author}{\bibinfo{person}{Mohammad Norouzi}, \bibinfo{person}{Ali
  Punjani}, {and} \bibinfo{person}{David~J Fleet}.}
  \bibinfo{year}{2012}\natexlab{}.
\newblock \showarticletitle{Fast search in hamming space with multi-index
  hashing}. In \bibinfo{booktitle}{\emph{2012 IEEE conference on computer
  vision and pattern recognition}}. IEEE, \bibinfo{pages}{3108--3115}.
\newblock


\bibitem[\protect\citeauthoryear{Pan and Manocha}{Pan and Manocha}{2011}]%
        {gpu-lsh-2011}
\bibfield{author}{\bibinfo{person}{Jia Pan} {and} \bibinfo{person}{Dinesh
  Manocha}.} \bibinfo{year}{2011}\natexlab{}.
\newblock \showarticletitle{Fast GPU-Based Locality Sensitive Hashing for
  k-Nearest Neighbor Computation}. In \bibinfo{booktitle}{\emph{Proceedings of
  the 19th ACM SIGSPATIAL International Conference on Advances in Geographic
  Information Systems}} (Chicago, Illinois) \emph{(\bibinfo{series}{GIS '11})}.
  \bibinfo{publisher}{Association for Computing Machinery},
  \bibinfo{address}{New York, NY, USA}, \bibinfo{pages}{211–220}.
\newblock
\showISBNx{9781450310314}
\urldef\tempurl%
\url{https://doi.org/10.1145/2093973.2094002}
\showDOI{\tempurl}


\bibitem[\protect\citeauthoryear{Pandove, Goel, and Rani}{Pandove
  et~al\mbox{.}}{2018}]%
        {clustering-survey-pandove2018systematic}
\bibfield{author}{\bibinfo{person}{Divya Pandove}, \bibinfo{person}{Shivan
  Goel}, {and} \bibinfo{person}{Rinkl Rani}.} \bibinfo{year}{2018}\natexlab{}.
\newblock \showarticletitle{Systematic review of clustering high-dimensional
  and large datasets}.
\newblock \bibinfo{journal}{\emph{ACM Transactions on Knowledge Discovery from
  Data (TKDD)}} \bibinfo{volume}{12}, \bibinfo{number}{2}
  (\bibinfo{year}{2018}), \bibinfo{pages}{1--68}.
\newblock


\bibitem[\protect\citeauthoryear{Pavithra and Parvathi}{Pavithra and
  Parvathi}{2017}]%
        {clustering-survey-pavithra2017survey}
\bibfield{author}{\bibinfo{person}{Mudamala Pavithra} {and}
  \bibinfo{person}{RMS Parvathi}.} \bibinfo{year}{2017}\natexlab{}.
\newblock \showarticletitle{A survey on clustering high dimensional data
  techniques}.
\newblock \bibinfo{journal}{\emph{International Journal of Applied Engineering
  Research}} \bibinfo{volume}{12}, \bibinfo{number}{11} (\bibinfo{year}{2017}),
  \bibinfo{pages}{2893--2899}.
\newblock


\bibitem[\protect\citeauthoryear{Pennington, Socher, and Manning}{Pennington
  et~al\mbox{.}}{2014}]%
        {embedding-text-2}
\bibfield{author}{\bibinfo{person}{Jeffrey Pennington},
  \bibinfo{person}{Richard Socher}, {and} \bibinfo{person}{Christopher~D
  Manning}.} \bibinfo{year}{2014}\natexlab{}.
\newblock \showarticletitle{Glove: Global vectors for word representation}. In
  \bibinfo{booktitle}{\emph{Proceedings of the 2014 conference on empirical
  methods in natural language processing (EMNLP)}}.
  \bibinfo{pages}{1532--1543}.
\newblock


\bibitem[\protect\citeauthoryear{Perdacher, Plant, and B{\"o}hm}{Perdacher
  et~al\mbox{.}}{2019}]%
        {Condition-based-similarity-join-8}
\bibfield{author}{\bibinfo{person}{Martin Perdacher}, \bibinfo{person}{Claudia
  Plant}, {and} \bibinfo{person}{Christian B{\"o}hm}.}
  \bibinfo{year}{2019}\natexlab{}.
\newblock \showarticletitle{Cache-oblivious high-performance similarity join}.
  In \bibinfo{booktitle}{\emph{Proceedings of the 2019 International Conference
  on Management of Data}}. \bibinfo{pages}{87--104}.
\newblock


\bibitem[\protect\citeauthoryear{Perozzi, Al-Rfou, and Skiena}{Perozzi
  et~al\mbox{.}}{2014a}]%
        {embedding-graph-1}
\bibfield{author}{\bibinfo{person}{Bryan Perozzi}, \bibinfo{person}{Rami
  Al-Rfou}, {and} \bibinfo{person}{Steven Skiena}.}
  \bibinfo{year}{2014}\natexlab{a}.
\newblock \showarticletitle{Deepwalk: Online learning of social
  representations}. In \bibinfo{booktitle}{\emph{Proceedings of the 20th ACM
  SIGKDD international conference on Knowledge discovery and data mining}}.
  \bibinfo{pages}{701--710}.
\newblock


\bibitem[\protect\citeauthoryear{Perozzi, Al-Rfou, and Skiena}{Perozzi
  et~al\mbox{.}}{2014b}]%
        {example-learning-task-specific-embedding-for-Graph-1}
\bibfield{author}{\bibinfo{person}{Bryan Perozzi}, \bibinfo{person}{Rami
  Al-Rfou}, {and} \bibinfo{person}{Steven Skiena}.}
  \bibinfo{year}{2014}\natexlab{b}.
\newblock \showarticletitle{Deepwalk: Online learning of social
  representations}. In \bibinfo{booktitle}{\emph{Proceedings of the 20th ACM
  SIGKDD international conference on Knowledge discovery and data mining}}.
  \bibinfo{pages}{701--710}.
\newblock


\bibitem[\protect\citeauthoryear{Petri, Moffat, Mackenzie, Culpepper, and
  Beck}{Petri et~al\mbox{.}}{2019}]%
        {general-similarity-query-text-2}
\bibfield{author}{\bibinfo{person}{Matthias Petri}, \bibinfo{person}{Alistair
  Moffat}, \bibinfo{person}{Joel Mackenzie}, \bibinfo{person}{J~Shane
  Culpepper}, {and} \bibinfo{person}{Daniel Beck}.}
  \bibinfo{year}{2019}\natexlab{}.
\newblock \showarticletitle{Accelerated query processing via similarity score
  prediction}. In \bibinfo{booktitle}{\emph{Proceedings of the 42nd
  International ACM SIGIR Conference on Research and Development in Information
  Retrieval}}. \bibinfo{pages}{485--494}.
\newblock


\bibitem[\protect\citeauthoryear{Plaku and Kavraki}{Plaku and Kavraki}{2007}]%
        {KNN-join-2}
\bibfield{author}{\bibinfo{person}{Erion Plaku} {and} \bibinfo{person}{Lydia~E
  Kavraki}.} \bibinfo{year}{2007}\natexlab{}.
\newblock \showarticletitle{Distributed computation of the knn graph for large
  high-dimensional point sets}.
\newblock \bibinfo{journal}{\emph{Journal of parallel and distributed
  computing}} \bibinfo{volume}{67}, \bibinfo{number}{3} (\bibinfo{year}{2007}),
  \bibinfo{pages}{346--359}.
\newblock


\bibitem[\protect\citeauthoryear{Qin, Wang, Lu, Xiao, and Lin}{Qin
  et~al\mbox{.}}{2011}]%
        {general-similarity-query-text-1}
\bibfield{author}{\bibinfo{person}{Jianbin Qin}, \bibinfo{person}{Wei Wang},
  \bibinfo{person}{Yifei Lu}, \bibinfo{person}{Chuan Xiao}, {and}
  \bibinfo{person}{Xuemin Lin}.} \bibinfo{year}{2011}\natexlab{}.
\newblock \showarticletitle{Efficient exact edit similarity query processing
  with the asymmetric signature scheme}. In
  \bibinfo{booktitle}{\emph{Proceedings of the 2011 ACM SIGMOD International
  Conference on Management of data}}. \bibinfo{pages}{1033--1044}.
\newblock


\bibitem[\protect\citeauthoryear{Qin, Wang, Xiao, and Zhang}{Qin
  et~al\mbox{.}}{2020}]%
        {general-similarity-query-processing-2}
\bibfield{author}{\bibinfo{person}{Jianbin Qin}, \bibinfo{person}{Wei Wang},
  \bibinfo{person}{Chuan Xiao}, {and} \bibinfo{person}{Ying Zhang}.}
  \bibinfo{year}{2020}\natexlab{}.
\newblock \showarticletitle{Similarity query processing for high-dimensional
  data}.
\newblock \bibinfo{journal}{\emph{Proceedings of the VLDB Endowment}}
  (\bibinfo{year}{2020}).
\newblock


\bibitem[\protect\citeauthoryear{Ram and Gray}{Ram and Gray}{2013}]%
        {Tree-based-index-which-tree-to-use-2013}
\bibfield{author}{\bibinfo{person}{Parikshit Ram} {and}
  \bibinfo{person}{Alexander~G Gray}.} \bibinfo{year}{2013}\natexlab{}.
\newblock \showarticletitle{Which space partitioning tree to use for search?}.
  In \bibinfo{booktitle}{\emph{NIPS}}. Citeseer, \bibinfo{pages}{656--664}.
\newblock


\bibitem[\protect\citeauthoryear{Ram and Sinha}{Ram and Sinha}{2019}]%
        {Tree-based-index-revisiting-kd-tree-2019}
\bibfield{author}{\bibinfo{person}{Parikshit Ram} {and}
  \bibinfo{person}{Kaushik Sinha}.} \bibinfo{year}{2019}\natexlab{}.
\newblock \showarticletitle{Revisiting kd-tree for nearest neighbor search}. In
  \bibinfo{booktitle}{\emph{Proceedings of the 25th acm sigkdd international
  conference on knowledge discovery \& data mining}}.
  \bibinfo{pages}{1378--1388}.
\newblock


\bibitem[\protect\citeauthoryear{Roy and Mukherjee}{Roy and Mukherjee}{2013}]%
        {traditional-method-image-color-feature-2}
\bibfield{author}{\bibinfo{person}{Kalyan Roy} {and} \bibinfo{person}{Joydeep
  Mukherjee}.} \bibinfo{year}{2013}\natexlab{}.
\newblock \showarticletitle{Image similarity measure using color histogram,
  color coherence vector, and sobel method}.
\newblock \bibinfo{journal}{\emph{International Journal of Science and Research
  (IJSR)}} \bibinfo{volume}{2}, \bibinfo{number}{1} (\bibinfo{year}{2013}),
  \bibinfo{pages}{538--543}.
\newblock


\bibitem[\protect\citeauthoryear{Sadeh, Fritz, Shalev, and Oks}{Sadeh
  et~al\mbox{.}}{2019}]%
        {IR-embedding-joint-sadeh2019joint}
\bibfield{author}{\bibinfo{person}{Gil Sadeh}, \bibinfo{person}{Lior Fritz},
  \bibinfo{person}{Gabi Shalev}, {and} \bibinfo{person}{Eduard Oks}.}
  \bibinfo{year}{2019}\natexlab{}.
\newblock \bibinfo{title}{Joint Visual-Textual Embedding for Multimodal Style
  Search}.
\newblock
\newblock
\showeprint[arxiv]{1906.06620}~[cs.LG]


\bibitem[\protect\citeauthoryear{Sakurai, Yoshikawa, and Faloutsos}{Sakurai
  et~al\mbox{.}}{2005}]%
        {general-similarity-query-multimedia-3}
\bibfield{author}{\bibinfo{person}{Yasushi Sakurai}, \bibinfo{person}{Masatoshi
  Yoshikawa}, {and} \bibinfo{person}{Christos Faloutsos}.}
  \bibinfo{year}{2005}\natexlab{}.
\newblock \showarticletitle{FTW: fast similarity search under the time warping
  distance}. In \bibinfo{booktitle}{\emph{Proceedings of the twenty-fourth ACM
  SIGMOD-SIGACT-SIGART symposium on Principles of database systems}}.
  \bibinfo{pages}{326--337}.
\newblock


\bibitem[\protect\citeauthoryear{Santiago~de Araújo, da~Cruz~Ferreira,
  Figueiroa~Goldstein, Solon~Nery, Augusto Justen~Marzulo, Kundu, and Maia
  Galvão~França}{Santiago~de Araújo et~al\mbox{.}}{2019}]%
        {fpga-multi-index-hashing-2019}
\bibfield{author}{\bibinfo{person}{Leandro Santiago~de Araújo},
  \bibinfo{person}{Victor da Cruz~Ferreira}, \bibinfo{person}{Brunno
  Figueiroa~Goldstein}, \bibinfo{person}{Alexandre Solon~Nery},
  \bibinfo{person}{Leandro Augusto Justen~Marzulo}, \bibinfo{person}{Sandip
  Kundu}, {and} \bibinfo{person}{Felipe Maia Galvão~França}.}
  \bibinfo{year}{2019}\natexlab{}.
\newblock \showarticletitle{Hardware-Accelerated Similarity Search with
  Multi-Index Hashing}. In \bibinfo{booktitle}{\emph{2019 IEEE Intl Conf on
  Dependable, Autonomic and Secure Computing, Intl Conf on Pervasive
  Intelligence and Computing, Intl Conf on Cloud and Big Data Computing, Intl
  Conf on Cyber Science and Technology Congress
  (DASC/PiCom/CBDCom/CyberSciTech)}}. \bibinfo{pages}{733--740}.
\newblock
\urldef\tempurl%
\url{https://doi.org/10.1109/DASC/PiCom/CBDCom/CyberSciTech.2019.00138}
\showDOI{\tempurl}


\bibitem[\protect\citeauthoryear{Schallehn, Sattler, and Saake}{Schallehn
  et~al\mbox{.}}{2004}]%
        {Similarity-group-by-1}
\bibfield{author}{\bibinfo{person}{Eike Schallehn}, \bibinfo{person}{Kai-Uwe
  Sattler}, {and} \bibinfo{person}{Gunter Saake}.}
  \bibinfo{year}{2004}\natexlab{}.
\newblock \showarticletitle{Efficient similarity-based operations for data
  integration}.
\newblock \bibinfo{journal}{\emph{Data \& Knowledge Engineering}}
  \bibinfo{volume}{48}, \bibinfo{number}{3} (\bibinfo{year}{2004}),
  \bibinfo{pages}{361--387}.
\newblock


\bibitem[\protect\citeauthoryear{Schroff, Kalenichenko, and Philbin}{Schroff
  et~al\mbox{.}}{2015}]%
        {embedding-image-1}
\bibfield{author}{\bibinfo{person}{Florian Schroff}, \bibinfo{person}{Dmitry
  Kalenichenko}, {and} \bibinfo{person}{James Philbin}.}
  \bibinfo{year}{2015}\natexlab{}.
\newblock \showarticletitle{Facenet: A unified embedding for face recognition
  and clustering}. In \bibinfo{booktitle}{\emph{Proceedings of the IEEE
  conference on computer vision and pattern recognition}}.
  \bibinfo{pages}{815--823}.
\newblock


\bibitem[\protect\citeauthoryear{Shahvarani and Jacobsen}{Shahvarani and
  Jacobsen}{2021}]%
        {KNN-join-on-spatial-2021}
\bibfield{author}{\bibinfo{person}{Amirhesam Shahvarani} {and}
  \bibinfo{person}{Hans-Arno Jacobsen}.} \bibinfo{year}{2021}\natexlab{}.
\newblock \showarticletitle{Distributed Stream KNN Join}. In
  \bibinfo{booktitle}{\emph{Proceedings of the 2021 International Conference on
  Management of Data}} (Virtual Event, China)
  \emph{(\bibinfo{series}{SIGMOD/PODS '21})}. \bibinfo{publisher}{Association
  for Computing Machinery}, \bibinfo{address}{New York, NY, USA},
  \bibinfo{pages}{1597–1609}.
\newblock
\showISBNx{9781450383431}
\urldef\tempurl%
\url{https://doi.org/10.1145/3448016.3457269}
\showDOI{\tempurl}


\bibitem[\protect\citeauthoryear{Shen, Xu, Liu, Yang, Huang, and Shen}{Shen
  et~al\mbox{.}}{2018}]%
        {deep-lash-shen2018-unsupervised}
\bibfield{author}{\bibinfo{person}{Fumin Shen}, \bibinfo{person}{Yan Xu},
  \bibinfo{person}{Li Liu}, \bibinfo{person}{Yang Yang}, \bibinfo{person}{Zi
  Huang}, {and} \bibinfo{person}{Heng~Tao Shen}.}
  \bibinfo{year}{2018}\natexlab{}.
\newblock \showarticletitle{Unsupervised deep hashing with similarity-adaptive
  and discrete optimization}.
\newblock \bibinfo{journal}{\emph{IEEE transactions on pattern analysis and
  machine intelligence}} \bibinfo{volume}{40}, \bibinfo{number}{12}
  (\bibinfo{year}{2018}), \bibinfo{pages}{3034--3044}.
\newblock


\bibitem[\protect\citeauthoryear{Shen, Shao, Huang, and Zhou}{Shen
  et~al\mbox{.}}{2008}]%
        {general-similarity-query-multimedia-2}
\bibfield{author}{\bibinfo{person}{Heng~Tao Shen}, \bibinfo{person}{Jie Shao},
  \bibinfo{person}{Zi Huang}, {and} \bibinfo{person}{Xiaofang Zhou}.}
  \bibinfo{year}{2008}\natexlab{}.
\newblock \showarticletitle{Effective and efficient query processing for video
  subsequence identification}.
\newblock \bibinfo{journal}{\emph{IEEE Transactions on Knowledge and Data
  Engineering}} \bibinfo{volume}{21}, \bibinfo{number}{3}
  (\bibinfo{year}{2008}), \bibinfo{pages}{321--334}.
\newblock


\bibitem[\protect\citeauthoryear{Shi, Ma, Zhang, Zhang, Yu, Shan, Liu, and
  Ma}{Shi et~al\mbox{.}}{2020}]%
        {deep-l2h-shi2020-for-recommendation}
\bibfield{author}{\bibinfo{person}{Shaoyun Shi}, \bibinfo{person}{Weizhi Ma},
  \bibinfo{person}{Min Zhang}, \bibinfo{person}{Yongfeng Zhang},
  \bibinfo{person}{Xinxing Yu}, \bibinfo{person}{Houzhi Shan},
  \bibinfo{person}{Yiqun Liu}, {and} \bibinfo{person}{Shaoping Ma}.}
  \bibinfo{year}{2020}\natexlab{}.
\newblock \showarticletitle{Beyond User Embedding Matrix: Learning to Hash for
  Modeling Large-Scale Users in Recommendation}. In
  \bibinfo{booktitle}{\emph{Proceedings of the 43rd International ACM SIGIR
  Conference on Research and Development in Information Retrieval}}.
  \bibinfo{pages}{319--328}.
\newblock


\bibitem[\protect\citeauthoryear{Shi and Jain}{Shi and Jain}{2019}]%
        {shi2019probabilistic}
\bibfield{author}{\bibinfo{person}{Yichun Shi} {and} \bibinfo{person}{Anil~K
  Jain}.} \bibinfo{year}{2019}\natexlab{}.
\newblock \showarticletitle{Probabilistic face embeddings}. In
  \bibinfo{booktitle}{\emph{Proceedings of the IEEE International Conference on
  Computer Vision}}. \bibinfo{pages}{6902--6911}.
\newblock


\bibitem[\protect\citeauthoryear{Shrivastava and Li}{Shrivastava and
  Li}{2014}]%
        {LSH-4}
\bibfield{author}{\bibinfo{person}{Anshumali Shrivastava} {and}
  \bibinfo{person}{Ping Li}.} \bibinfo{year}{2014}\natexlab{}.
\newblock \showarticletitle{Asymmetric LSH (ALSH) for sublinear time maximum
  inner product search (MIPS)}.
\newblock \bibinfo{journal}{\emph{arXiv preprint arXiv:1405.5869}}
  (\bibinfo{year}{2014}).
\newblock


\bibitem[\protect\citeauthoryear{Silpa-Anan and Hartley}{Silpa-Anan and
  Hartley}{2008}]%
        {Tree-based-index-randomized-kd-tree-silpa2008}
\bibfield{author}{\bibinfo{person}{Chanop Silpa-Anan} {and}
  \bibinfo{person}{Richard Hartley}.} \bibinfo{year}{2008}\natexlab{}.
\newblock \showarticletitle{Optimised KD-trees for fast image descriptor
  matching}. In \bibinfo{booktitle}{\emph{2008 IEEE Conference on Computer
  Vision and Pattern Recognition}}. IEEE, \bibinfo{pages}{1--8}.
\newblock


\bibitem[\protect\citeauthoryear{Silva, Aly, Aref, and Larson}{Silva
  et~al\mbox{.}}{2010}]%
        {Similarity-group-by-3}
\bibfield{author}{\bibinfo{person}{Yasin~N Silva}, \bibinfo{person}{Ahmed~M
  Aly}, \bibinfo{person}{Walid~G Aref}, {and} \bibinfo{person}{Per-Ake
  Larson}.} \bibinfo{year}{2010}\natexlab{}.
\newblock \showarticletitle{Simdb: a similarity-aware database system}. In
  \bibinfo{booktitle}{\emph{Proceedings of the 2010 ACM SIGMOD International
  Conference on Management of data}}. \bibinfo{pages}{1243--1246}.
\newblock


\bibitem[\protect\citeauthoryear{Silva, Aref, and Ali}{Silva
  et~al\mbox{.}}{2009}]%
        {Similarity-group-by-5}
\bibfield{author}{\bibinfo{person}{Yasin~N Silva}, \bibinfo{person}{Walid~G
  Aref}, {and} \bibinfo{person}{Mohamed~H Ali}.}
  \bibinfo{year}{2009}\natexlab{}.
\newblock \showarticletitle{Similarity group-by}. In
  \bibinfo{booktitle}{\emph{2009 IEEE 25th International Conference on Data
  Engineering}}. IEEE, \bibinfo{pages}{904--915}.
\newblock


\bibitem[\protect\citeauthoryear{Silva, Aref, Larson, Pearson, and Ali}{Silva
  et~al\mbox{.}}{2013a}]%
        {general-similarity-query-processing-1}
\bibfield{author}{\bibinfo{person}{Yasin~N Silva}, \bibinfo{person}{Walid~G
  Aref}, \bibinfo{person}{Per-Ake Larson}, \bibinfo{person}{Spencer~S Pearson},
  {and} \bibinfo{person}{Mohamed~H Ali}.} \bibinfo{year}{2013}\natexlab{a}.
\newblock \showarticletitle{Similarity queries: their conceptual evaluation,
  transformations, and processing}.
\newblock \bibinfo{journal}{\emph{The VLDB Journal}} \bibinfo{volume}{22},
  \bibinfo{number}{3} (\bibinfo{year}{2013}), \bibinfo{pages}{395--420}.
\newblock


\bibitem[\protect\citeauthoryear{Silva, Aref, Larson, Pearson, and Ali}{Silva
  et~al\mbox{.}}{2013b}]%
        {Similarity-group-by-4}
\bibfield{author}{\bibinfo{person}{Yasin~N Silva}, \bibinfo{person}{Walid~G
  Aref}, \bibinfo{person}{Per-Ake Larson}, \bibinfo{person}{Spencer~S Pearson},
  {and} \bibinfo{person}{Mohamed~H Ali}.} \bibinfo{year}{2013}\natexlab{b}.
\newblock \showarticletitle{Similarity queries: their conceptual evaluation,
  transformations, and processing}.
\newblock \bibinfo{journal}{\emph{The VLDB Journal}} \bibinfo{volume}{22},
  \bibinfo{number}{3} (\bibinfo{year}{2013}), \bibinfo{pages}{395--420}.
\newblock


\bibitem[\protect\citeauthoryear{Song, Gu, Zhang, and Yu}{Song
  et~al\mbox{.}}{2020}]%
        {Similarity-search-tree-based-1}
\bibfield{author}{\bibinfo{person}{Yang Song}, \bibinfo{person}{Yu Gu},
  \bibinfo{person}{Rui Zhang}, {and} \bibinfo{person}{Ge Yu}.}
  \bibinfo{year}{2020}\natexlab{}.
\newblock \showarticletitle{Brepartition: Optimized high-dimensional knn search
  with bregman distances}.
\newblock \bibinfo{journal}{\emph{IEEE Transactions on Knowledge and Data
  Engineering}} (\bibinfo{year}{2020}).
\newblock


\bibitem[\protect\citeauthoryear{Stupar, Michel, and Schenkel}{Stupar
  et~al\mbox{.}}{2010}]%
        {KNN-join-Rankreduce-stupar2010}
\bibfield{author}{\bibinfo{person}{Aleksandar Stupar},
  \bibinfo{person}{Sebastian Michel}, {and} \bibinfo{person}{Ralf Schenkel}.}
  \bibinfo{year}{2010}\natexlab{}.
\newblock \showarticletitle{Rankreduce-processing k-nearest neighbor queries on
  top of mapreduce}. In \bibinfo{booktitle}{\emph{LSDS-IR@ SIGIR}}.
\newblock


\bibitem[\protect\citeauthoryear{Subramanya, Kadekodi, Krishaswamy, and
  Simhadri}{Subramanya et~al\mbox{.}}{2019}]%
        {Graph-based-index-2}
\bibfield{author}{\bibinfo{person}{Suhas~Jayaram Subramanya},
  \bibinfo{person}{Rohan Kadekodi}, \bibinfo{person}{Ravishankar Krishaswamy},
  {and} \bibinfo{person}{Harsha~Vardhan Simhadri}.}
  \bibinfo{year}{2019}\natexlab{}.
\newblock \showarticletitle{Diskann: Fast accurate billion-point nearest
  neighbor search on a single node}. In \bibinfo{booktitle}{\emph{Proceedings
  of the 33rd International Conference on Neural Information Processing
  Systems}}. \bibinfo{pages}{13766--13776}.
\newblock


\bibitem[\protect\citeauthoryear{Sun, Shang, Li, Deng, and Bao}{Sun
  et~al\mbox{.}}{2019}]%
        {general-similarity-query-text-3}
\bibfield{author}{\bibinfo{person}{Ji Sun}, \bibinfo{person}{Zeyuan Shang},
  \bibinfo{person}{Guoliang Li}, \bibinfo{person}{Dong Deng}, {and}
  \bibinfo{person}{Zhifeng Bao}.} \bibinfo{year}{2019}\natexlab{}.
\newblock \showarticletitle{Balance-aware distributed string similarity-based
  query processing system}.
\newblock \bibinfo{journal}{\emph{Proceedings of the VLDB Endowment}}
  \bibinfo{volume}{12}, \bibinfo{number}{9} (\bibinfo{year}{2019}),
  \bibinfo{pages}{961--974}.
\newblock


\bibitem[\protect\citeauthoryear{Sun, Cheng, Li, Cheung, and Han}{Sun
  et~al\mbox{.}}{2011}]%
        {general-similarity-query-graph-2}
\bibfield{author}{\bibinfo{person}{Liwen Sun}, \bibinfo{person}{Reynold Cheng},
  \bibinfo{person}{Xiang Li}, \bibinfo{person}{David~W Cheung}, {and}
  \bibinfo{person}{Jiawei Han}.} \bibinfo{year}{2011}\natexlab{}.
\newblock \showarticletitle{On link-based similarity join}.
\newblock \bibinfo{journal}{\emph{Proceedings of the VLDB Endowment}}
  \bibinfo{volume}{4}, \bibinfo{number}{11} (\bibinfo{year}{2011}),
  \bibinfo{pages}{714--725}.
\newblock


\bibitem[\protect\citeauthoryear{Sundaram, Turmukhametova, Satish, Mostak,
  Indyk, Madden, and Dubey}{Sundaram et~al\mbox{.}}{2013}]%
        {Similarity-search-hashing-based-3}
\bibfield{author}{\bibinfo{person}{Narayanan Sundaram}, \bibinfo{person}{Aizana
  Turmukhametova}, \bibinfo{person}{Nadathur Satish}, \bibinfo{person}{Todd
  Mostak}, \bibinfo{person}{Piotr Indyk}, \bibinfo{person}{Samuel Madden},
  {and} \bibinfo{person}{Pradeep Dubey}.} \bibinfo{year}{2013}\natexlab{}.
\newblock \showarticletitle{Streaming similarity search over one billion tweets
  using parallel locality-sensitive hashing}.
\newblock \bibinfo{journal}{\emph{Proceedings of the VLDB Endowment}}
  \bibinfo{volume}{6}, \bibinfo{number}{14} (\bibinfo{year}{2013}),
  \bibinfo{pages}{1930--1941}.
\newblock


\bibitem[\protect\citeauthoryear{Tan, Liu, Zhao, Yang, Zhou, and Hu}{Tan
  et~al\mbox{.}}{2020}]%
        {deep-l2h-tan2020-with-GNN-for-recommendation}
\bibfield{author}{\bibinfo{person}{Qiaoyu Tan}, \bibinfo{person}{Ninghao Liu},
  \bibinfo{person}{Xing Zhao}, \bibinfo{person}{Hongxia Yang},
  \bibinfo{person}{Jingren Zhou}, {and} \bibinfo{person}{Xia Hu}.}
  \bibinfo{year}{2020}\natexlab{}.
\newblock \showarticletitle{Learning to hash with graph neural networks for
  recommender systems}. In \bibinfo{booktitle}{\emph{Proceedings of The Web
  Conference 2020}}. \bibinfo{pages}{1988--1998}.
\newblock


\bibitem[\protect\citeauthoryear{Tang, Tahboub, Aref, Atallah, Malluhi,
  Ouzzani, and Silva}{Tang et~al\mbox{.}}{2016}]%
        {Similarity-group-by-6}
\bibfield{author}{\bibinfo{person}{Mingjie Tang}, \bibinfo{person}{Ruby~Y.
  Tahboub}, \bibinfo{person}{Walid~G. Aref}, \bibinfo{person}{Mikhail~J.
  Atallah}, \bibinfo{person}{Qutaibah~M. Malluhi}, \bibinfo{person}{Mourad
  Ouzzani}, {and} \bibinfo{person}{Yasin~N. Silva}.}
  \bibinfo{year}{2016}\natexlab{}.
\newblock \showarticletitle{Similarity Group-by Operators for Multi-Dimensional
  Relational Data}.
\newblock \bibinfo{journal}{\emph{IEEE Transactions on Knowledge and Data
  Engineering}} \bibinfo{volume}{28}, \bibinfo{number}{2}
  (\bibinfo{year}{2016}), \bibinfo{pages}{510--523}.
\newblock
\urldef\tempurl%
\url{https://doi.org/10.1109/TKDE.2015.2480400}
\showDOI{\tempurl}


\bibitem[\protect\citeauthoryear{Tang, Tahboub, Aref, Malluhi, and
  Ouzzani}{Tang et~al\mbox{.}}{2014}]%
        {Similarity-group-by-2}
\bibfield{author}{\bibinfo{person}{Mingjie Tang}, \bibinfo{person}{Ruby~Y
  Tahboub}, \bibinfo{person}{Walid~G Aref}, \bibinfo{person}{Qutaibah~M
  Malluhi}, {and} \bibinfo{person}{Mourad Ouzzani}.}
  \bibinfo{year}{2014}\natexlab{}.
\newblock \showarticletitle{On order-independent semantics of the similarity
  group-by relational database operator}.
\newblock \bibinfo{journal}{\emph{arXiv preprint arXiv:1412.4303}}
  (\bibinfo{year}{2014}).
\newblock


\bibitem[\protect\citeauthoryear{Teixeira, Teodoro, Valle, and Saltz}{Teixeira
  et~al\mbox{.}}{2013}]%
        {KNN-join-parallel-LHS-teixeira2013scalable}
\bibfield{author}{\bibinfo{person}{Thiago~SFX Teixeira},
  \bibinfo{person}{George Teodoro}, \bibinfo{person}{Eduardo Valle}, {and}
  \bibinfo{person}{Joel~H Saltz}.} \bibinfo{year}{2013}\natexlab{}.
\newblock \showarticletitle{Scalable locality-sensitive hashing for similarity
  search in high-dimensional, large-scale multimedia datasets}.
\newblock \bibinfo{journal}{\emph{arXiv preprint arXiv:1310.4136}}
  (\bibinfo{year}{2013}).
\newblock


\bibitem[\protect\citeauthoryear{Teodoro, Valle, Mariano, Torres, au2, and
  Saltz}{Teodoro et~al\mbox{.}}{2012}]%
        {teodoro2012approximate}
\bibfield{author}{\bibinfo{person}{George Teodoro}, \bibinfo{person}{Eduardo
  Valle}, \bibinfo{person}{Nathan Mariano}, \bibinfo{person}{Ricardo Torres},
  \bibinfo{person}{Wagner Meira~Jr au2}, {and} \bibinfo{person}{Joel~H.
  Saltz}.} \bibinfo{year}{2012}\natexlab{}.
\newblock \bibinfo{title}{Approximate Similarity Search for Online Multimedia
  Services on Distributed CPU-GPU Platforms}.
\newblock
\newblock
\showeprint[arxiv]{1209.0410}~[cs.MM]


\bibitem[\protect\citeauthoryear{Tuxen}{Tuxen}{2016}]%
        {range-search-using-hashing-Tuxen2016RangeQO}
\bibfield{author}{\bibinfo{person}{Clarissa~Bruno Tuxen}.}
  \bibinfo{year}{2016}\natexlab{}.
\newblock \showarticletitle{Range Queries over Hashing}.
\newblock


\bibitem[\protect\citeauthoryear{Vempala}{Vempala}{2012}]%
        {Tree-based-index-randomly-oriented-kd-tree-vempala2012}
\bibfield{author}{\bibinfo{person}{Santosh~S Vempala}.}
  \bibinfo{year}{2012}\natexlab{}.
\newblock \showarticletitle{Randomly-oriented kd trees adapt to intrinsic
  dimension}. In \bibinfo{booktitle}{\emph{IARCS Annual Conference on
  Foundations of Software Technology and Theoretical Computer Science (FSTTCS
  2012)}}. Schloss Dagstuhl-Leibniz-Zentrum fuer Informatik.
\newblock


\bibitem[\protect\citeauthoryear{Vukoti\'{c}, Raymond, and Gravier}{Vukoti\'{c}
  et~al\mbox{.}}{2016}]%
        {IR-embedding-joint-2016}
\bibfield{author}{\bibinfo{person}{Vedran Vukoti\'{c}},
  \bibinfo{person}{Christian Raymond}, {and} \bibinfo{person}{Guillaume
  Gravier}.} \bibinfo{year}{2016}\natexlab{}.
\newblock \showarticletitle{Bidirectional Joint Representation Learning with
  Symmetrical Deep Neural Networks for Multimodal and Crossmodal Applications}.
  In \bibinfo{booktitle}{\emph{Proceedings of the 2016 ACM on International
  Conference on Multimedia Retrieval}} (New York, New York, USA)
  \emph{(\bibinfo{series}{ICMR '16})}. \bibinfo{publisher}{Association for
  Computing Machinery}, \bibinfo{address}{New York, NY, USA},
  \bibinfo{pages}{343–346}.
\newblock
\showISBNx{9781450343596}
\urldef\tempurl%
\url{https://doi.org/10.1145/2911996.2912064}
\showDOI{\tempurl}


\bibitem[\protect\citeauthoryear{Wang, Li, and Fe}{Wang et~al\mbox{.}}{2011}]%
        {general-similarity-query-text-4}
\bibfield{author}{\bibinfo{person}{Jiannan Wang}, \bibinfo{person}{Guoliang
  Li}, {and} \bibinfo{person}{Jianhua Fe}.} \bibinfo{year}{2011}\natexlab{}.
\newblock \showarticletitle{Fast-join: An efficient method for fuzzy token
  matching based string similarity join}. In \bibinfo{booktitle}{\emph{2011
  IEEE 27th International Conference on Data Engineering}}. IEEE,
  \bibinfo{pages}{458--469}.
\newblock


\bibitem[\protect\citeauthoryear{Wang and Li}{Wang and Li}{2012}]%
        {Graph-based-index-query-driven-iterated-wang2012}
\bibfield{author}{\bibinfo{person}{Jingdong Wang} {and}
  \bibinfo{person}{Shipeng Li}.} \bibinfo{year}{2012}\natexlab{}.
\newblock \showarticletitle{Query-driven iterated neighborhood graph search for
  large scale indexing}. In \bibinfo{booktitle}{\emph{Proceedings of the 20th
  ACM international conference on Multimedia}}. \bibinfo{pages}{179--188}.
\newblock


\bibitem[\protect\citeauthoryear{Wang, Liu, Kumar, and Chang}{Wang
  et~al\mbox{.}}{2015}]%
        {learning-to-hash-good-survey-2}
\bibfield{author}{\bibinfo{person}{Jun Wang}, \bibinfo{person}{Wei Liu},
  \bibinfo{person}{Sanjiv Kumar}, {and} \bibinfo{person}{Shih-Fu Chang}.}
  \bibinfo{year}{2015}\natexlab{}.
\newblock \showarticletitle{Learning to hash for indexing big data—A survey}.
\newblock \bibinfo{journal}{\emph{Proc. IEEE}} \bibinfo{volume}{104},
  \bibinfo{number}{1} (\bibinfo{year}{2015}), \bibinfo{pages}{34--57}.
\newblock


\bibitem[\protect\citeauthoryear{Wang, Shen, Song, and Ji}{Wang
  et~al\mbox{.}}{2014}]%
        {Index-survey-hashing-1}
\bibfield{author}{\bibinfo{person}{Jingdong Wang}, \bibinfo{person}{Heng~Tao
  Shen}, \bibinfo{person}{Jingkuan Song}, {and} \bibinfo{person}{Jianqiu Ji}.}
  \bibinfo{year}{2014}\natexlab{}.
\newblock \showarticletitle{Hashing for similarity search: A survey}.
\newblock \bibinfo{journal}{\emph{arXiv preprint arXiv:1408.2927}}
  (\bibinfo{year}{2014}).
\newblock


\bibitem[\protect\citeauthoryear{Wang, Wang, Zeng, Tu, Gan, and Li}{Wang
  et~al\mbox{.}}{2012}]%
        {Graph-based-index-partition-construct-wang2012}
\bibfield{author}{\bibinfo{person}{Jing Wang}, \bibinfo{person}{Jingdong Wang},
  \bibinfo{person}{Gang Zeng}, \bibinfo{person}{Zhuowen Tu},
  \bibinfo{person}{Rui Gan}, {and} \bibinfo{person}{Shipeng Li}.}
  \bibinfo{year}{2012}\natexlab{}.
\newblock \showarticletitle{Scalable k-nn graph construction for visual
  descriptors}. In \bibinfo{booktitle}{\emph{2012 IEEE Conference on Computer
  Vision and Pattern Recognition}}. IEEE, \bibinfo{pages}{1106--1113}.
\newblock


\bibitem[\protect\citeauthoryear{Wang, Wang, Jia, Li, Zeng, Zha, and Hua}{Wang
  et~al\mbox{.}}{2013}]%
        {Tree-based-index-trinary-projection-trees}
\bibfield{author}{\bibinfo{person}{Jingdong Wang}, \bibinfo{person}{Naiyan
  Wang}, \bibinfo{person}{You Jia}, \bibinfo{person}{Jian Li},
  \bibinfo{person}{Gang Zeng}, \bibinfo{person}{Hongbin Zha}, {and}
  \bibinfo{person}{Xian-Sheng Hua}.} \bibinfo{year}{2013}\natexlab{}.
\newblock \showarticletitle{Trinary-projection trees for approximate nearest
  neighbor search}.
\newblock \bibinfo{journal}{\emph{IEEE transactions on pattern analysis and
  machine intelligence}} \bibinfo{volume}{36}, \bibinfo{number}{2}
  (\bibinfo{year}{2013}), \bibinfo{pages}{388--403}.
\newblock


\bibitem[\protect\citeauthoryear{Wang, Zhang, Sebe, Shen, et~al\mbox{.}}{Wang
  et~al\mbox{.}}{2017}]%
        {Learning-to-hash-1}
\bibfield{author}{\bibinfo{person}{Jingdong Wang}, \bibinfo{person}{Ting
  Zhang}, \bibinfo{person}{Nicu Sebe}, \bibinfo{person}{Heng~Tao Shen},
  {et~al\mbox{.}}} \bibinfo{year}{2017}\natexlab{}.
\newblock \showarticletitle{A survey on learning to hash}.
\newblock \bibinfo{journal}{\emph{IEEE transactions on pattern analysis and
  machine intelligence}} \bibinfo{volume}{40}, \bibinfo{number}{4}
  (\bibinfo{year}{2017}), \bibinfo{pages}{769--790}.
\newblock


\bibitem[\protect\citeauthoryear{Wang, Shen, Wang, Yao, Jiang, Qi, and
  Chen}{Wang et~al\mbox{.}}{2019}]%
        {deep-l2h-wang2019-hash-for-imcomplete-kg-search}
\bibfield{author}{\bibinfo{person}{Meng Wang}, \bibinfo{person}{Haomin Shen},
  \bibinfo{person}{Sen Wang}, \bibinfo{person}{Lina Yao},
  \bibinfo{person}{Yinlin Jiang}, \bibinfo{person}{Guilin Qi}, {and}
  \bibinfo{person}{Yang Chen}.} \bibinfo{year}{2019}\natexlab{}.
\newblock \showarticletitle{Learning to hash for efficient search over
  incomplete knowledge graphs}. In \bibinfo{booktitle}{\emph{2019 IEEE
  International Conference on Data Mining (ICDM)}}. IEEE,
  \bibinfo{pages}{1360--1365}.
\newblock


\bibitem[\protect\citeauthoryear{Wang, Wu, and Qi}{Wang et~al\mbox{.}}{2020}]%
        {deep-l2h-wang2020-hash-for-KG-embed}
\bibfield{author}{\bibinfo{person}{Meng Wang}, \bibinfo{person}{Tongtong Wu},
  {and} \bibinfo{person}{Guilin Qi}.} \bibinfo{year}{2020}\natexlab{}.
\newblock \showarticletitle{A Hash Learning Framework for Search-Oriented
  Knowledge Graph Embedding}.
\newblock In \bibinfo{booktitle}{\emph{ECAI 2020}}. \bibinfo{publisher}{IOS
  Press}, \bibinfo{pages}{921--928}.
\newblock


\bibitem[\protect\citeauthoryear{Wang, Xu, Yue, and Wang}{Wang
  et~al\mbox{.}}{2021}]%
        {graph-based-index-comprehensive-survey-vldb2021}
\bibfield{author}{\bibinfo{person}{Mengzhao Wang}, \bibinfo{person}{Xiaoliang
  Xu}, \bibinfo{person}{Qiang Yue}, {and} \bibinfo{person}{Yuxiang Wang}.}
  \bibinfo{year}{2021}\natexlab{}.
\newblock \showarticletitle{A Comprehensive Survey and Experimental Comparison
  of Graph-Based Approximate Nearest Neighbor Search}.
\newblock \bibinfo{journal}{\emph{Proc. {VLDB} Endow.}} \bibinfo{volume}{14},
  \bibinfo{number}{11} (\bibinfo{year}{2021}), \bibinfo{pages}{1964--1978}.
\newblock
\urldef\tempurl%
\url{http://www.vldb.org/pvldb/vol14/p1964-wang.pdf}
\showURL{%
\tempurl}


\bibitem[\protect\citeauthoryear{Wang and Deng}{Wang and Deng}{2020}]%
        {Product-quantization-9}
\bibfield{author}{\bibinfo{person}{Runhui Wang} {and} \bibinfo{person}{Dong
  Deng}.} \bibinfo{year}{2020}\natexlab{}.
\newblock \showarticletitle{DeltaPQ: lossless product quantization code
  compression for high dimensional similarity search}.
\newblock \bibinfo{journal}{\emph{Proceedings of the VLDB Endowment}}
  \bibinfo{volume}{13}, \bibinfo{number}{13} (\bibinfo{year}{2020}),
  \bibinfo{pages}{3603--3616}.
\newblock


\bibitem[\protect\citeauthoryear{Wang, Shrivastava, and Ryu}{Wang
  et~al\mbox{.}}{2018}]%
        {gpu-fast-lsh-ultra-high-dim-Wang2018RandomizedAA}
\bibfield{author}{\bibinfo{person}{Yiqiu Wang}, \bibinfo{person}{Anshumali
  Shrivastava}, {and} \bibinfo{person}{Junghee Ryu}.}
  \bibinfo{year}{2018}\natexlab{}.
\newblock \showarticletitle{Randomized Algorithms Accelerated over CPU-GPU for
  Ultra-High Dimensional Similarity Search}.
\newblock \bibinfo{journal}{\emph{Proceedings of the 2018 International
  Conference on Management of Data}} (\bibinfo{year}{2018}).
\newblock


\bibitem[\protect\citeauthoryear{WANG, CHEN, QIAN, and CHEN}{WANG
  et~al\mbox{.}}{2016}]%
        {Similarity-search-hashing-based-2}
\bibfield{author}{\bibinfo{person}{Zhong-wei WANG}, \bibinfo{person}{Ye-fang
  CHEN}, \bibinfo{person}{Jiang-bo QIAN}, {and} \bibinfo{person}{Hua-hui
  CHEN}.} \bibinfo{year}{2016}\natexlab{}.
\newblock \showarticletitle{LSH-based algorithm for k nearest neighbor search
  on big data}.
\newblock \bibinfo{journal}{\emph{ACTA ELECTONICA SINICA}}
  \bibinfo{volume}{44}, \bibinfo{number}{4} (\bibinfo{year}{2016}),
  \bibinfo{pages}{906}.
\newblock


\bibitem[\protect\citeauthoryear{Weiss, Torralba, Fergus, et~al\mbox{.}}{Weiss
  et~al\mbox{.}}{2008}]%
        {l2h-multidim-features-unsupervised-weiss2008-spectral-hashing}
\bibfield{author}{\bibinfo{person}{Yair Weiss}, \bibinfo{person}{Antonio
  Torralba}, \bibinfo{person}{Robert Fergus}, {et~al\mbox{.}}}
  \bibinfo{year}{2008}\natexlab{}.
\newblock \showarticletitle{Spectral hashing.}. In
  \bibinfo{booktitle}{\emph{Nips}}, Vol.~\bibinfo{volume}{1}. Citeseer,
  \bibinfo{pages}{4}.
\newblock


\bibitem[\protect\citeauthoryear{White and Jain}{White and Jain}{1996}]%
        {range-search-using-tree-white1996similarity}
\bibfield{author}{\bibinfo{person}{David~A White} {and}
  \bibinfo{person}{Ramesh~C Jain}.} \bibinfo{year}{1996}\natexlab{}.
\newblock \showarticletitle{Similarity indexing: Algorithms and performance}.
  In \bibinfo{booktitle}{\emph{Storage and Retrieval for Still Image and Video
  Databases IV}}, Vol.~\bibinfo{volume}{2670}. International Society for Optics
  and Photonics, \bibinfo{pages}{62--73}.
\newblock


\bibitem[\protect\citeauthoryear{Wieschollek, Wang, Sorkine-Hornung, and
  Lensch}{Wieschollek et~al\mbox{.}}{2016}]%
        {gpu-PQ-Wieschollek_2016_CVPR}
\bibfield{author}{\bibinfo{person}{Patrick Wieschollek},
  \bibinfo{person}{Oliver Wang}, \bibinfo{person}{Alexander Sorkine-Hornung},
  {and} \bibinfo{person}{Hendrik P.~A. Lensch}.}
  \bibinfo{year}{2016}\natexlab{}.
\newblock \showarticletitle{Efficient Large-Scale Approximate Nearest Neighbor
  Search on the GPU}. In \bibinfo{booktitle}{\emph{Proceedings of the IEEE
  Conference on Computer Vision and Pattern Recognition (CVPR)}}.
\newblock


\bibitem[\protect\citeauthoryear{Xia, Lu, Ooi, and Hu}{Xia
  et~al\mbox{.}}{2004}]%
        {KNN-join-Gorder-2004}
\bibfield{author}{\bibinfo{person}{Chenyi Xia}, \bibinfo{person}{Hongjun Lu},
  \bibinfo{person}{Beng~Chin Ooi}, {and} \bibinfo{person}{Jing Hu}.}
  \bibinfo{year}{2004}\natexlab{}.
\newblock \showarticletitle{Gorder: an efficient method for knn join
  processing}. In \bibinfo{booktitle}{\emph{Proceedings of the Thirtieth
  international conference on Very large data bases-Volume 30}}.
  \bibinfo{pages}{756--767}.
\newblock


\bibitem[\protect\citeauthoryear{Xia, Pan, Lai, Liu, and Yan}{Xia
  et~al\mbox{.}}{2014}]%
        {deep-l2h-xia2014-supervised-hashing-image-retrieval}
\bibfield{author}{\bibinfo{person}{Rongkai Xia}, \bibinfo{person}{Yan Pan},
  \bibinfo{person}{Hanjiang Lai}, \bibinfo{person}{Cong Liu}, {and}
  \bibinfo{person}{Shuicheng Yan}.} \bibinfo{year}{2014}\natexlab{}.
\newblock \showarticletitle{Supervised hashing for image retrieval via image
  representation learning}. In \bibinfo{booktitle}{\emph{Twenty-eighth AAAI
  conference on artificial intelligence}}.
\newblock


\bibitem[\protect\citeauthoryear{Xiao, Wang, Lin, Yu, and Wang}{Xiao
  et~al\mbox{.}}{2011a}]%
        {Condition-based-join-text-token-xiao2011efficient}
\bibfield{author}{\bibinfo{person}{Chuan Xiao}, \bibinfo{person}{Wei Wang},
  \bibinfo{person}{Xuemin Lin}, \bibinfo{person}{Jeffrey~Xu Yu}, {and}
  \bibinfo{person}{Guoren Wang}.} \bibinfo{year}{2011}\natexlab{a}.
\newblock \showarticletitle{Efficient similarity joins for near-duplicate
  detection}.
\newblock \bibinfo{journal}{\emph{ACM Transactions on Database Systems (TODS)}}
  \bibinfo{volume}{36}, \bibinfo{number}{3} (\bibinfo{year}{2011}),
  \bibinfo{pages}{1--41}.
\newblock


\bibitem[\protect\citeauthoryear{Xiao, Wang, Lin, Yu, and Wang}{Xiao
  et~al\mbox{.}}{2011b}]%
        {ER-traditional-1}
\bibfield{author}{\bibinfo{person}{Chuan Xiao}, \bibinfo{person}{Wei Wang},
  \bibinfo{person}{Xuemin Lin}, \bibinfo{person}{Jeffrey~Xu Yu}, {and}
  \bibinfo{person}{Guoren Wang}.} \bibinfo{year}{2011}\natexlab{b}.
\newblock \showarticletitle{Efficient similarity joins for near-duplicate
  detection}.
\newblock \bibinfo{journal}{\emph{ACM Transactions on Database Systems (TODS)}}
  \bibinfo{volume}{36}, \bibinfo{number}{3} (\bibinfo{year}{2011}),
  \bibinfo{pages}{1--41}.
\newblock


\bibitem[\protect\citeauthoryear{Xiao, Guo, Lan, Xu, and Cheng}{Xiao
  et~al\mbox{.}}{2018}]%
        {Graph-based-index-kDNNG-xiao2018}
\bibfield{author}{\bibinfo{person}{Yan Xiao}, \bibinfo{person}{Jiafeng Guo},
  \bibinfo{person}{Yanyan Lan}, \bibinfo{person}{Jun Xu}, {and}
  \bibinfo{person}{Xueqi Cheng}.} \bibinfo{year}{2018}\natexlab{}.
\newblock \showarticletitle{Fast Approximate Nearest Neighbor Search via
  k-Diverse Nearest Neighbor Graph}. In \bibinfo{booktitle}{\emph{Thirty-Second
  AAAI Conference on Artificial Intelligence}}.
\newblock


\bibitem[\protect\citeauthoryear{Xiong, Zhu, and Philip}{Xiong
  et~al\mbox{.}}{2014}]%
        {general-similarity-query-graph-3}
\bibfield{author}{\bibinfo{person}{Yun Xiong}, \bibinfo{person}{Yangyong Zhu},
  {and} \bibinfo{person}{S~Yu Philip}.} \bibinfo{year}{2014}\natexlab{}.
\newblock \showarticletitle{Top-k similarity join in heterogeneous information
  networks}.
\newblock \bibinfo{journal}{\emph{IEEE Transactions on Knowledge and Data
  Engineering}} \bibinfo{volume}{27}, \bibinfo{number}{6}
  (\bibinfo{year}{2014}), \bibinfo{pages}{1710--1723}.
\newblock


\bibitem[\protect\citeauthoryear{Xu and Tian}{Xu and Tian}{2015}]%
        {clustering-survey-xu2015comprehensive}
\bibfield{author}{\bibinfo{person}{Dongkuan Xu} {and} \bibinfo{person}{Yingjie
  Tian}.} \bibinfo{year}{2015}\natexlab{}.
\newblock \showarticletitle{A comprehensive survey of clustering algorithms}.
\newblock \bibinfo{journal}{\emph{Annals of Data Science}} \bibinfo{volume}{2},
  \bibinfo{number}{2} (\bibinfo{year}{2015}), \bibinfo{pages}{165--193}.
\newblock


\bibitem[\protect\citeauthoryear{Yan, Wang, Wang, Wang, and Li}{Yan
  et~al\mbox{.}}{2019}]%
        {Tree-based-index-rpForests-2019}
\bibfield{author}{\bibinfo{person}{Donghui Yan}, \bibinfo{person}{Yingjie
  Wang}, \bibinfo{person}{Jin Wang}, \bibinfo{person}{Honggang Wang}, {and}
  \bibinfo{person}{Zhenpeng Li}.} \bibinfo{year}{2019}\natexlab{}.
\newblock \showarticletitle{K-nearest neighbor search by random projection
  forests}.
\newblock \bibinfo{journal}{\emph{IEEE Transactions on Big Data}}
  \bibinfo{volume}{7}, \bibinfo{number}{1} (\bibinfo{year}{2019}),
  \bibinfo{pages}{147--157}.
\newblock


\bibitem[\protect\citeauthoryear{Yang, Yu, and Liu}{Yang et~al\mbox{.}}{2014}]%
        {KNN-join-dynamic-HDR-tree-2014}
\bibfield{author}{\bibinfo{person}{Chong Yang}, \bibinfo{person}{Xiaohui Yu},
  {and} \bibinfo{person}{Yang Liu}.} \bibinfo{year}{2014}\natexlab{}.
\newblock \showarticletitle{Continuous KNN join processing for real-time
  recommendation}. In \bibinfo{booktitle}{\emph{2014 IEEE International
  Conference on Data Mining}}. IEEE, \bibinfo{pages}{640--649}.
\newblock


\bibitem[\protect\citeauthoryear{Yao, Li, and Kumar}{Yao et~al\mbox{.}}{2010}]%
        {KNN-join-z-value-2010}
\bibfield{author}{\bibinfo{person}{Bin Yao}, \bibinfo{person}{Feifei Li}, {and}
  \bibinfo{person}{Piyush Kumar}.} \bibinfo{year}{2010}\natexlab{}.
\newblock \showarticletitle{K nearest neighbor queries and knn-joins in large
  relational databases (almost) for free}. In \bibinfo{booktitle}{\emph{2010
  IEEE 26th International Conference on Data Engineering (ICDE 2010)}}. IEEE,
  \bibinfo{pages}{4--15}.
\newblock


\bibitem[\protect\citeauthoryear{Yu, Cui, Wang, and Su}{Yu
  et~al\mbox{.}}{2007a}]%
        {KNN-join-3}
\bibfield{author}{\bibinfo{person}{Cui Yu}, \bibinfo{person}{Bin Cui},
  \bibinfo{person}{Shuguang Wang}, {and} \bibinfo{person}{Jianwen Su}.}
  \bibinfo{year}{2007}\natexlab{a}.
\newblock \showarticletitle{Efficient index-based KNN join processing for
  high-dimensional data}.
\newblock \bibinfo{journal}{\emph{Information and Software Technology}}
  \bibinfo{volume}{49}, \bibinfo{number}{4} (\bibinfo{year}{2007}),
  \bibinfo{pages}{332--344}.
\newblock


\bibitem[\protect\citeauthoryear{Yu, Cui, Wang, and Su}{Yu
  et~al\mbox{.}}{2007b}]%
        {KNN-join-iJoin-yu2007}
\bibfield{author}{\bibinfo{person}{Cui Yu}, \bibinfo{person}{Bin Cui},
  \bibinfo{person}{Shuguang Wang}, {and} \bibinfo{person}{Jianwen Su}.}
  \bibinfo{year}{2007}\natexlab{b}.
\newblock \showarticletitle{Efficient index-based KNN join processing for
  high-dimensional data}.
\newblock \bibinfo{journal}{\emph{Information and Software Technology}}
  \bibinfo{volume}{49}, \bibinfo{number}{4} (\bibinfo{year}{2007}),
  \bibinfo{pages}{332--344}.
\newblock


\bibitem[\protect\citeauthoryear{Yu, Nutanong, Li, Wang, and Yuan}{Yu
  et~al\mbox{.}}{2016}]%
        {Condition-based-similarity-join-1}
\bibfield{author}{\bibinfo{person}{Chenyun Yu}, \bibinfo{person}{Sarana
  Nutanong}, \bibinfo{person}{Hangyu Li}, \bibinfo{person}{Cong Wang}, {and}
  \bibinfo{person}{Xingliang Yuan}.} \bibinfo{year}{2016}\natexlab{}.
\newblock \showarticletitle{A generic method for accelerating LSH-based
  similarity join processing}.
\newblock \bibinfo{journal}{\emph{IEEE Transactions on Knowledge and Data
  Engineering}} \bibinfo{volume}{29}, \bibinfo{number}{4}
  (\bibinfo{year}{2016}), \bibinfo{pages}{712--726}.
\newblock


\bibitem[\protect\citeauthoryear{Yu, Zhang, Huang, and Xiong}{Yu
  et~al\mbox{.}}{2010}]%
        {KNN-join-dynamic-incremental-Yu2010HighdimensionalKJ}
\bibfield{author}{\bibinfo{person}{Cui Yu}, \bibinfo{person}{Rui Zhang},
  \bibinfo{person}{Yaochun Huang}, {and} \bibinfo{person}{Hui Xiong}.}
  \bibinfo{year}{2010}\natexlab{}.
\newblock \showarticletitle{High-dimensional kNN joins with incremental
  updates}.
\newblock \bibinfo{journal}{\emph{GeoInformatica}}  \bibinfo{volume}{14}
  (\bibinfo{year}{2010}), \bibinfo{pages}{55--82}.
\newblock


\bibitem[\protect\citeauthoryear{Yu, Nie, Shen, and Kou}{Yu
  et~al\mbox{.}}{2020}]%
        {Heterogeneous-hardware-4}
\bibfield{author}{\bibinfo{person}{Lining Yu}, \bibinfo{person}{Tiezheng Nie},
  \bibinfo{person}{Derong Shen}, {and} \bibinfo{person}{Yue Kou}.}
  \bibinfo{year}{2020}\natexlab{}.
\newblock \showarticletitle{An Approach for Progressive Set Similarity Join
  with GPU Accelerating}. In \bibinfo{booktitle}{\emph{International Conference
  on Web Information Systems and Applications}}. Springer,
  \bibinfo{pages}{155--167}.
\newblock


\bibitem[\protect\citeauthoryear{Yu, Yuan, Fang, and Jin}{Yu
  et~al\mbox{.}}{2018a}]%
        {Product-quantization-3}
\bibfield{author}{\bibinfo{person}{Tan Yu}, \bibinfo{person}{Junsong Yuan},
  \bibinfo{person}{Chen Fang}, {and} \bibinfo{person}{Hailin Jin}.}
  \bibinfo{year}{2018}\natexlab{a}.
\newblock \showarticletitle{Product quantization network for fast image
  retrieval}. In \bibinfo{booktitle}{\emph{Proceedings of the European
  Conference on Computer Vision (ECCV)}}. \bibinfo{pages}{186--201}.
\newblock


\bibitem[\protect\citeauthoryear{Yu, Yuan, Fang, and Jin}{Yu
  et~al\mbox{.}}{2018b}]%
        {pq-pqnet-yu2018}
\bibfield{author}{\bibinfo{person}{Tan Yu}, \bibinfo{person}{Junsong Yuan},
  \bibinfo{person}{Chen Fang}, {and} \bibinfo{person}{Hailin Jin}.}
  \bibinfo{year}{2018}\natexlab{b}.
\newblock \showarticletitle{Product quantization network for fast image
  retrieval}. In \bibinfo{booktitle}{\emph{Proceedings of the European
  Conference on Computer Vision (ECCV)}}. \bibinfo{pages}{186--201}.
\newblock


\bibitem[\protect\citeauthoryear{Zhang, Li, and Jestes}{Zhang
  et~al\mbox{.}}{2012}]%
        {KNN-join-z-value-2012}
\bibfield{author}{\bibinfo{person}{Chi Zhang}, \bibinfo{person}{Feifei Li},
  {and} \bibinfo{person}{Jeffrey Jestes}.} \bibinfo{year}{2012}\natexlab{}.
\newblock \showarticletitle{Efficient parallel kNN joins for large data in
  MapReduce}. In \bibinfo{booktitle}{\emph{Proceedings of the 15th
  international conference on extending database technology}}.
  \bibinfo{pages}{38--49}.
\newblock


\bibitem[\protect\citeauthoryear{Zhang and Zhang}{Zhang and Zhang}{2017a}]%
        {Condition-based-similarity-join-5}
\bibfield{author}{\bibinfo{person}{Haoyu Zhang} {and} \bibinfo{person}{Qin
  Zhang}.} \bibinfo{year}{2017}\natexlab{a}.
\newblock \showarticletitle{Embedjoin: Efficient edit similarity joins via
  embeddings}. In \bibinfo{booktitle}{\emph{Proceedings of the 23rd ACM SIGKDD
  international conference on knowledge discovery and data mining}}.
  \bibinfo{pages}{585--594}.
\newblock


\bibitem[\protect\citeauthoryear{Zhang and Zhang}{Zhang and Zhang}{2017b}]%
        {Condition-based-join-string-sim-join-edit-dist-2017}
\bibfield{author}{\bibinfo{person}{Haoyu Zhang} {and} \bibinfo{person}{Qin
  Zhang}.} \bibinfo{year}{2017}\natexlab{b}.
\newblock \showarticletitle{EmbedJoin: Efficient Edit Similarity Joins via
  Embeddings} \emph{(\bibinfo{series}{KDD '17})}.
  \bibinfo{publisher}{Association for Computing Machinery},
  \bibinfo{address}{New York, NY, USA}, \bibinfo{pages}{585–594}.
\newblock
\showISBNx{9781450348874}
\urldef\tempurl%
\url{https://doi.org/10.1145/3097983.3098003}
\showDOI{\tempurl}


\bibitem[\protect\citeauthoryear{Zhang, Khoram, and Li}{Zhang
  et~al\mbox{.}}{2018}]%
        {fpga-PQ-Zhang_2018_CVPR}
\bibfield{author}{\bibinfo{person}{Jialiang Zhang}, \bibinfo{person}{Soroosh
  Khoram}, {and} \bibinfo{person}{Jing Li}.} \bibinfo{year}{2018}\natexlab{}.
\newblock \showarticletitle{Efficient Large-Scale Approximate Nearest Neighbor
  Search on OpenCL FPGA}. In \bibinfo{booktitle}{\emph{Proceedings of the IEEE
  Conference on Computer Vision and Pattern Recognition (CVPR)}}.
\newblock


\bibitem[\protect\citeauthoryear{Zhang, Gao, Zhang, and Li}{Zhang
  et~al\mbox{.}}{2010}]%
        {Data-oriented-lsh-zhang2010data}
\bibfield{author}{\bibinfo{person}{Wei Zhang}, \bibinfo{person}{Ke Gao},
  \bibinfo{person}{Yong-dong Zhang}, {and} \bibinfo{person}{Jin-tao Li}.}
  \bibinfo{year}{2010}\natexlab{}.
\newblock \showarticletitle{Data-oriented locality sensitive hashing}. In
  \bibinfo{booktitle}{\emph{Proceedings of the 18th ACM international
  conference on Multimedia}}. \bibinfo{pages}{1131--1134}.
\newblock


\bibitem[\protect\citeauthoryear{Zhou, Guo, Jagadish, Krcal, Liu, Luan, Tung,
  Yang, and Zheng}{Zhou et~al\mbox{.}}{2018}]%
        {inverted-index-gpu}
\bibfield{author}{\bibinfo{person}{Jingbo Zhou}, \bibinfo{person}{Qi Guo},
  \bibinfo{person}{H.~V. Jagadish}, \bibinfo{person}{Lubos Krcal},
  \bibinfo{person}{Siyuan Liu}, \bibinfo{person}{Wenhao Luan},
  \bibinfo{person}{Anthony K.~H. Tung}, \bibinfo{person}{Yueji Yang}, {and}
  \bibinfo{person}{Yuxin Zheng}.} \bibinfo{year}{2018}\natexlab{}.
\newblock \showarticletitle{A Generic Inverted Index Framework for Similarity
  Search on the GPU}. In \bibinfo{booktitle}{\emph{2018 IEEE 34th International
  Conference on Data Engineering (ICDE)}}. \bibinfo{pages}{893--904}.
\newblock
\urldef\tempurl%
\url{https://doi.org/10.1109/ICDE.2018.00085}
\showDOI{\tempurl}


\bibitem[\protect\citeauthoryear{Zhou, Han, Zhang, Dai, and Xu}{Zhou
  et~al\mbox{.}}{2013}]%
        {KNN-join-LSH-zhou2013hdkv}
\bibfield{author}{\bibinfo{person}{Wei Zhou}, \bibinfo{person}{Jizhong Han},
  \bibinfo{person}{Zhang Zhang}, \bibinfo{person}{Jiao Dai}, {and}
  \bibinfo{person}{Zhiyong Xu}.} \bibinfo{year}{2013}\natexlab{}.
\newblock \showarticletitle{HDKV: supporting efficient high-dimensional
  similarity search in key-value stores}.
\newblock \bibinfo{journal}{\emph{Concurrency and Computation: Practice and
  Experience}} \bibinfo{volume}{25}, \bibinfo{number}{12}
  (\bibinfo{year}{2013}), \bibinfo{pages}{1675--1698}.
\newblock


\bibitem[\protect\citeauthoryear{Zhu, Long, Wang, and Cao}{Zhu
  et~al\mbox{.}}{2016}]%
        {deep-l2h-zhu2016-deep-hashing-net}
\bibfield{author}{\bibinfo{person}{Han Zhu}, \bibinfo{person}{Mingsheng Long},
  \bibinfo{person}{Jianmin Wang}, {and} \bibinfo{person}{Yue Cao}.}
  \bibinfo{year}{2016}\natexlab{}.
\newblock \showarticletitle{Deep hashing network for efficient similarity
  retrieval}. In \bibinfo{booktitle}{\emph{Proceedings of the AAAI Conference
  on Artificial Intelligence}}, Vol.~\bibinfo{volume}{30}.
\newblock


\bibitem[\protect\citeauthoryear{Zhu, Zhan, and Qiu}{Zhu et~al\mbox{.}}{2015}]%
        {KNN-join-inverted-index-LSH-2015}
\bibfield{author}{\bibinfo{person}{Pingfei Zhu}, \bibinfo{person}{Xiangwen
  Zhan}, {and} \bibinfo{person}{Wenming Qiu}.} \bibinfo{year}{2015}\natexlab{}.
\newblock \showarticletitle{Efficient k-Nearest Neighbors Search in High
  Dimensions Using MapReduce}. In \bibinfo{booktitle}{\emph{2015 IEEE Fifth
  International Conference on Big Data and Cloud Computing}}.
  \bibinfo{pages}{23--30}.
\newblock
\urldef\tempurl%
\url{https://doi.org/10.1109/BDCloud.2015.51}
\showDOI{\tempurl}


\bibitem[\protect\citeauthoryear{Zimek}{Zimek}{2018}]%
        {clustering-survey-zimek2018clustering}
\bibfield{author}{\bibinfo{person}{Arthur Zimek}.}
  \bibinfo{year}{2018}\natexlab{}.
\newblock \showarticletitle{Clustering high-dimensional data}.
\newblock In \bibinfo{booktitle}{\emph{Data Clustering}}.
  \bibinfo{publisher}{Chapman and Hall/CRC}, \bibinfo{pages}{201--230}.
\newblock


\end{thebibliography}

\end{document}